\documentclass[manuscript]{aastex}
\usepackage{multirow}

\shorttitle{Exploring Hot Gas at Filamentary Junctions}
\shortauthors{Mitsuishi et al.}

\begin{document}


\title{Exploring Hot Gas at Junctions of Galaxy Filaments with Suzaku}


\author{Ikuyuki Mitsuishi\altaffilmark{1}, Hajime Kawahara\altaffilmark{2}, Norio Sekiya\altaffilmark{3}, Shin Sasaki\altaffilmark{1}, \\Thierry Sousbie\altaffilmark{4} \& Noriko Y. Yamasaki\altaffilmark{3}}

\affil{$^1$Department of Physics, Tokyo Metropolitan University, Hachioji, Tokyo 192-0397, Japan}

%\author{H. Kawahara\altaffilmark{2}}
\affil{$^2$Department of Earth and Planetary Science, The University of Tokyo, \\
7-3-1 Hongo, Bunkyo-ku, Tokyo 113-0033, Japan}

%\author{N. Sekiya\altaffilmark{3}}
\affil{$^3$Institute of Space and Astronautical Science, Japan Aerospace Exploration Agency, \\
3-1-1 Yoshinodai, Chuo, Sagamihara, Kanagawa 252-5210, Japan}

%\author{S. Sasaki\altaffilmark{1}}

%\author{T. Sousbie\altaffilmark{4}}
\affil{$^4$Institut d'Astrophysique de Paris - 98 bis boulevard Arago, F-75014 Paris, France}
\email{mitsuisi@phys.se.tmu.ac.jp}

%\and
%
%\author{N. Y Yamasaki\altaffilmark{3}}


%\altaffiltext{1}{Visiting Astronomer, Cerro Tololo Inter-American Observatory.
%CTIO is operated by AURA, Inc.\ under contract to the National Science
%\altaffiltext{2}{Society of Fellows, Harvard University.}
%\altaffiltext{3}{present address: Center for Astrophysics,
%    60 Garden Street, Cambridge, MA 02138}
%\altaffiltext{4}{Visiting Programmer, Space Telescope Science Institute}
%\altaffiltext{5}{Patron, Alonso's Bar and Grill}


\begin{abstract}
We performed five pointing observations with {\it Suzaku} 
to search for hot gases associated with the junctions of galaxy filaments 
where no significant diffuse X-ray sources were detected so far. 
%
%Observation fields were selected by a filament extractor, 
%DisPerSE based on the SDSS galaxy distribution. 
%%
We discovered X-ray sources successfully in all five regions and 
analyzed two bright sources in each field. 
Spectral analysis indicates that 
three sources originate from X-ray diffuse halos associated with 
optically bright galaxies or groups of galaxies with $kT$$\sim$0.6--0.8 keV.
Other three sources are possibly group- and cluster-scale X-ray halos 
with temperatures of $\sim$1 keV and $\sim$4 keV, respectively 
while the others are compact object origins such as AGNs.
%
%X-ray energy spectra indicate that two and one of them originate from 
%X-ray diffuse halos associated with groups of galaxies and a cluster of galaxies 
%with temperatures of $kT$$\sim$1--1.3 keV and $\sim$3.4--3.6 keV, respectively 
%while four sources are compact object origins.
%
%Other three sources are diffuse X-ray halos 
%associated with optically-bright elliptical galaxies and/or 
%groups of galaxies.
%
All the three observed intracluster media within the junctions of the galaxy filaments 
previously found are involved in ongoing mergers.
Thus, 
we demonstrate that deep X-ray observations at the filament junctions identified 
by galaxy surveys are a powerful mean  
to explore growing halos in a hierarchical structure undetected so far.
%
%Thus, our results suggest that 
%a galaxy filamentary junction tends to have an X-ray emitting diffuse halo 
%associated with the large scale structure and 
%a merger phenomenon is more likely to occur in such an active field 
%in structure formation.
%
%This work demonstrate the validity of the filament extractor DisPerSE and 
%future X-ray surveys, such as Spectrum-Roentgen-Gamma observatory and $DIOS$
%will map the large scale structure with X-ray emission.
\end{abstract}





\section{Introduction}

%Diffuse intergalactic gas associated with filamentary large-scale structure 
%has been considered as a key component of missing baryon of the present Universe 
%\citep{1998ApJ...503..518F,1999ApJ...514....1C,2001ApJ...552..473D}
%and therefore studied actively so far 
%\citep[e.g.,][]{2007ApJ...655..831T,2009ApJ...695.1351B,2012PASJ...64...18M}. 
%%
%Observational evidences of the diffuse and low temperature gases have been obtained 
%through the metal absorption lines imprinted in the QSO spectra 
%\citep[e.g.,][]{2005ApJ...624..555D,2008ApJS..177...39T} and 
%diffuse emission between two massive clusters 
%\citep[e.g.,][]{2008A&A...482L..29W}.
%
In the structure formation theory of the Universe, 
the filamentary structure is evolved by the density field, 
while the rare density peaks within the primordial field latterly collapse into halos. 
Filaments form the highways along which matter and gases are transported 
from the lower density regions that will latterly form the voids to those high density peaks, 
located at their junctions. 
Since the galaxy filaments are considered to trace the large scale structure of matter and 
gases behind, one can use them to search for undetected gas components. 
For instance, X-ray absorption lines across large galaxy filaments were interpreted 
as the signal from the missing baryons 
\citep{2009ApJ...695.1351B,2010ApJ...724L..25W}. 
Massive X-ray clusters are also often discovered at the intersection of two or more filaments 
\citep{2000AA...355..461A,2004AA...416..839B,2004AA...425..429C,2007AA...470..425B,2008AA...491..379G}.

\citet{2011ApJ...727L..38K} (hereafter Paper I) found a new X-ray halo, 
which can be interpreted as a merging group, 
by observing with {\it Suzaku} satellite at a filamentary junction of galaxies. 
This filamentary junction was visually identified within the Sloan Digital Sky Survey (SDSS) 
and no X-ray detection was previously achieved. 
The detection of a new merging group by X-ray can be performed 
due to {\it Suzaku'}s high sensitivity coming from low background capability. 
The evaluation of hot X-ray emitting gas associated with the large scale structure 
namely an intracluster medium (ICM) in the low-z Universe will enable us 
to estimate the actual population of baryons. 
In this paper, we report the further detection of ten new X-ray signals  
with {\it Suzaku} by a similar method to Paper I, however, 
with more sophisticated identification of galaxy filaments 
by the filament extractor, DisPerSE \citep{2011MNRAS.414..350S,2011MNRAS.414..384S}. 
Throughout this paper, we assume a $\Lambda$CDM universe 
with $\Omega_{\rm{m}}$ = 0.27, $\Omega_{\Lambda}$ = 0.73, and {\it h$_{\rm{0}}$} = 0.7. 
Unless otherwise specified, all errors in text and tables are 
at 90 \% confidence level, and those in figures are at 1 $\sigma$ level.
\section{Field Selection and Observations}
\label{sec:field-selection}
\begin{figure*}[h!]
\vspace{-0.8cm}
\begin{tabular}{ccc}
\begin{minipage}{0.45\hsize}
\hspace{0.8cm}
(a) FJ1
\vspace{-3.5cm}
\begin{center}
    \includegraphics[bb=0 0 1424 1067, width=11.5cm]{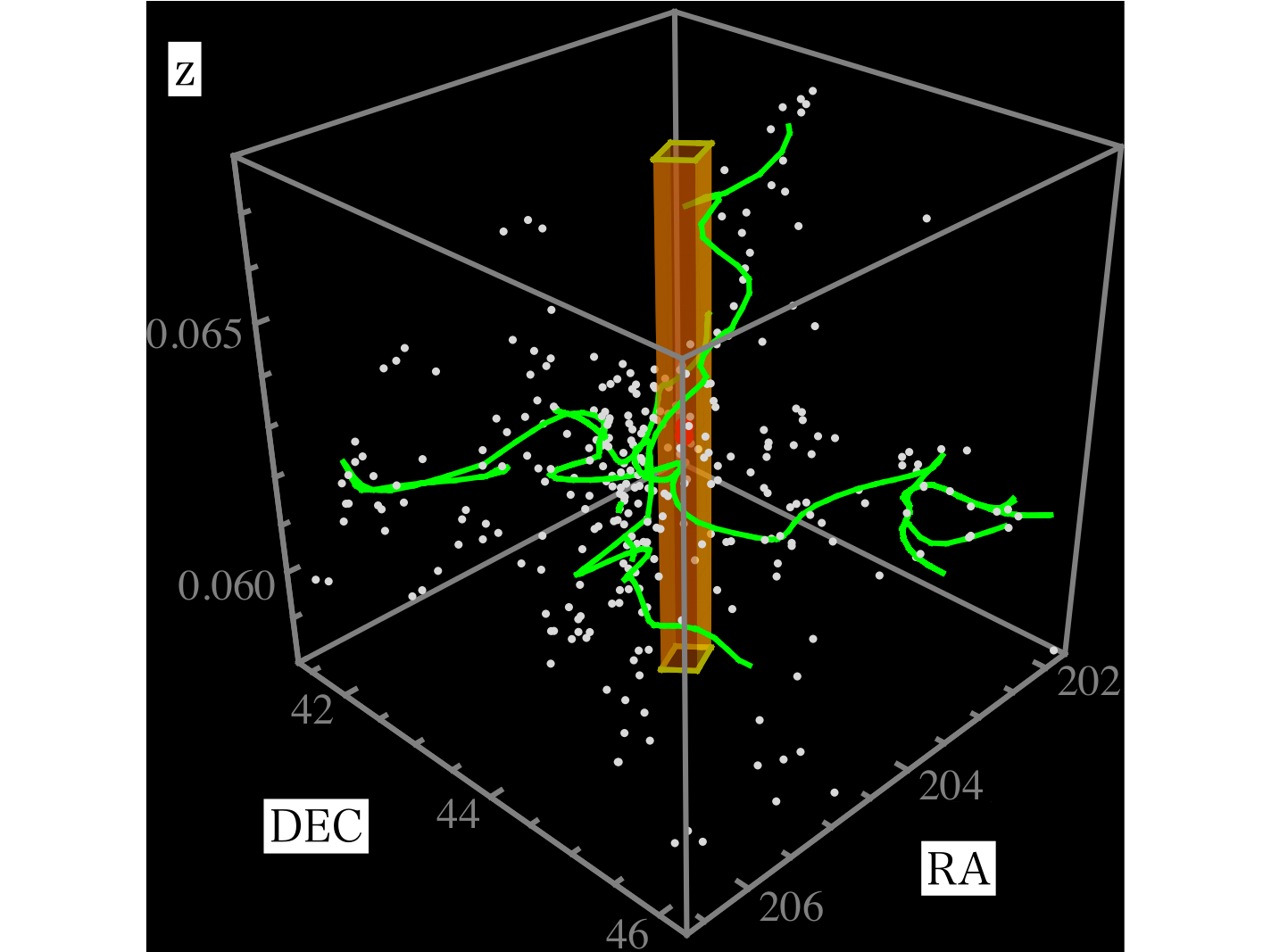}
\end{center}
\end{minipage}
\begin{minipage}{0.45\hsize}
\hspace{0.8cm}
(b) FJ2
\begin{center}
\vspace{-3.5cm}
    \includegraphics[bb=0 0 1424 1067, width=11.5cm]{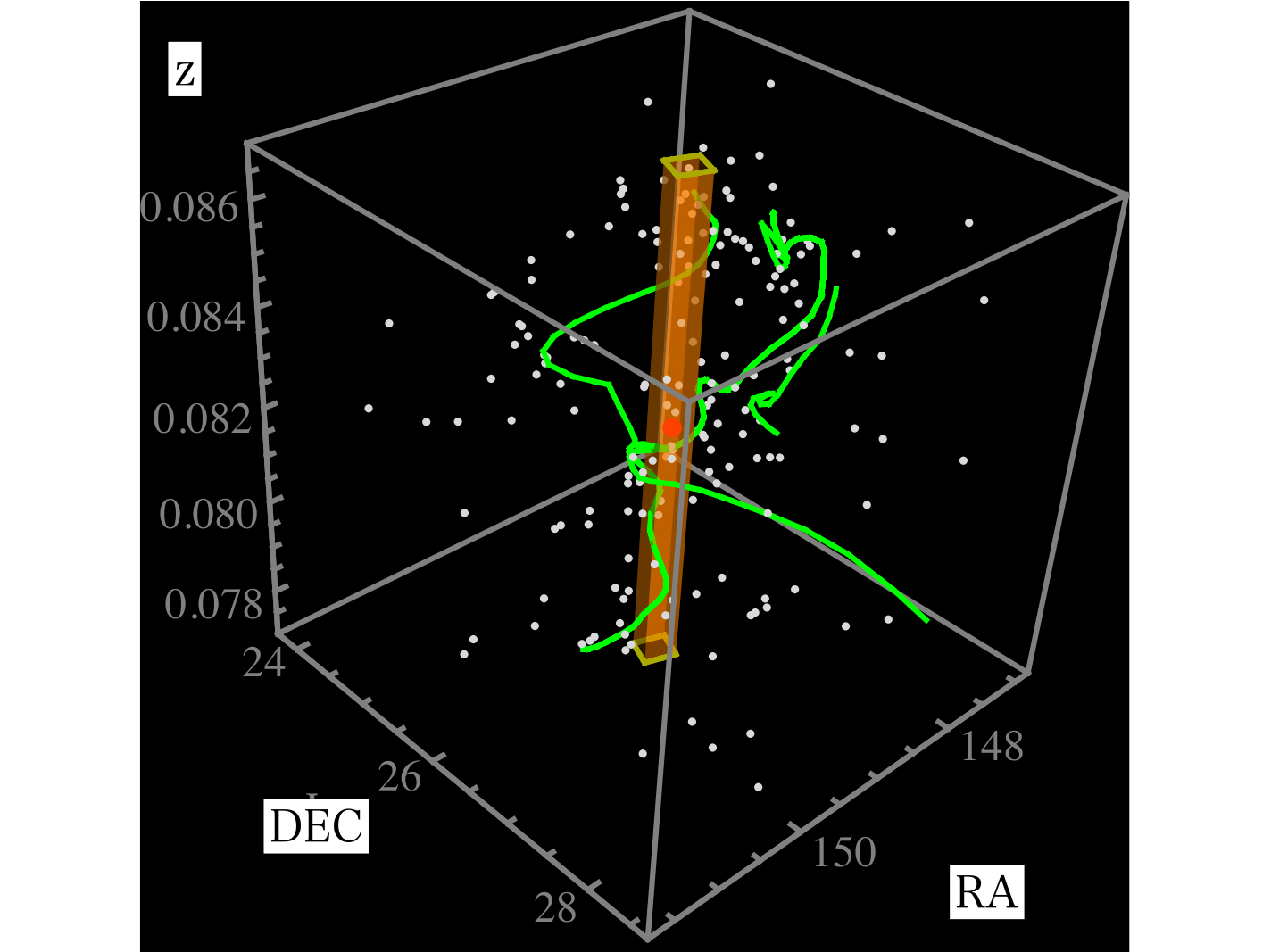}
\end{center}
\end{minipage}
\vspace{0.2cm}
\\
\begin{minipage}{0.45\hsize}
\hspace{0.8cm}
(c) FJ3
\begin{center}
\vspace{-3.5cm}
    \includegraphics[bb=0 0 1424 1067, width=11.5cm]{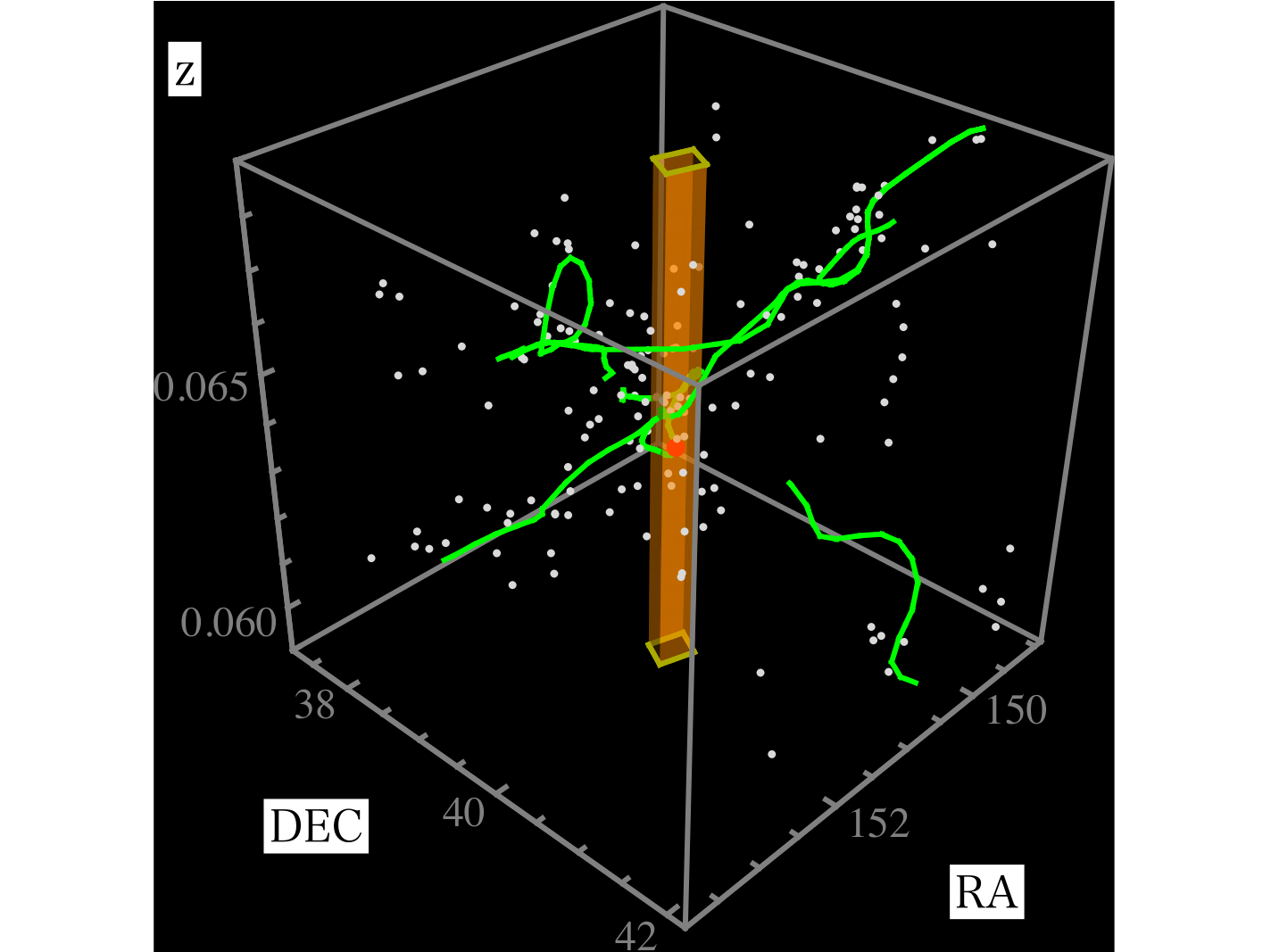}
\end{center}
\end{minipage}
\vspace{0.2cm}
\begin{minipage}{0.45\hsize}
\hspace{0.8cm}
(d) FJ4
\begin{center}
\vspace{-3.5cm}
    \includegraphics[bb=0 0 1424 1067, width=11.5cm]{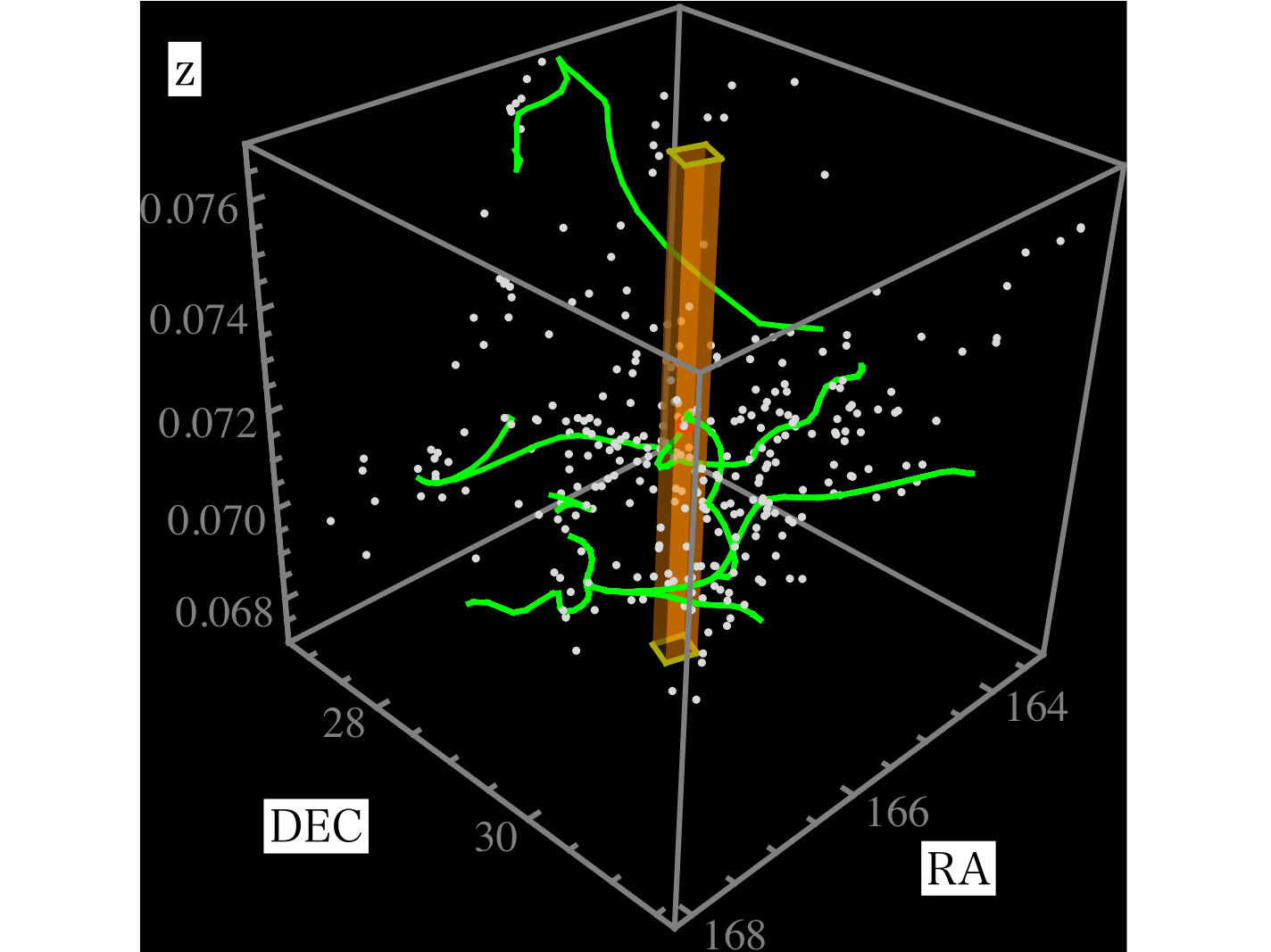}
\end{center}
\end{minipage}
\\
\begin{minipage}{0.45\hsize}
\hspace{0.8cm}
(e) FJ5
\begin{center}
\vspace{-3.5cm}
    \includegraphics[bb=0 0 1424 1067, width=11.5cm]{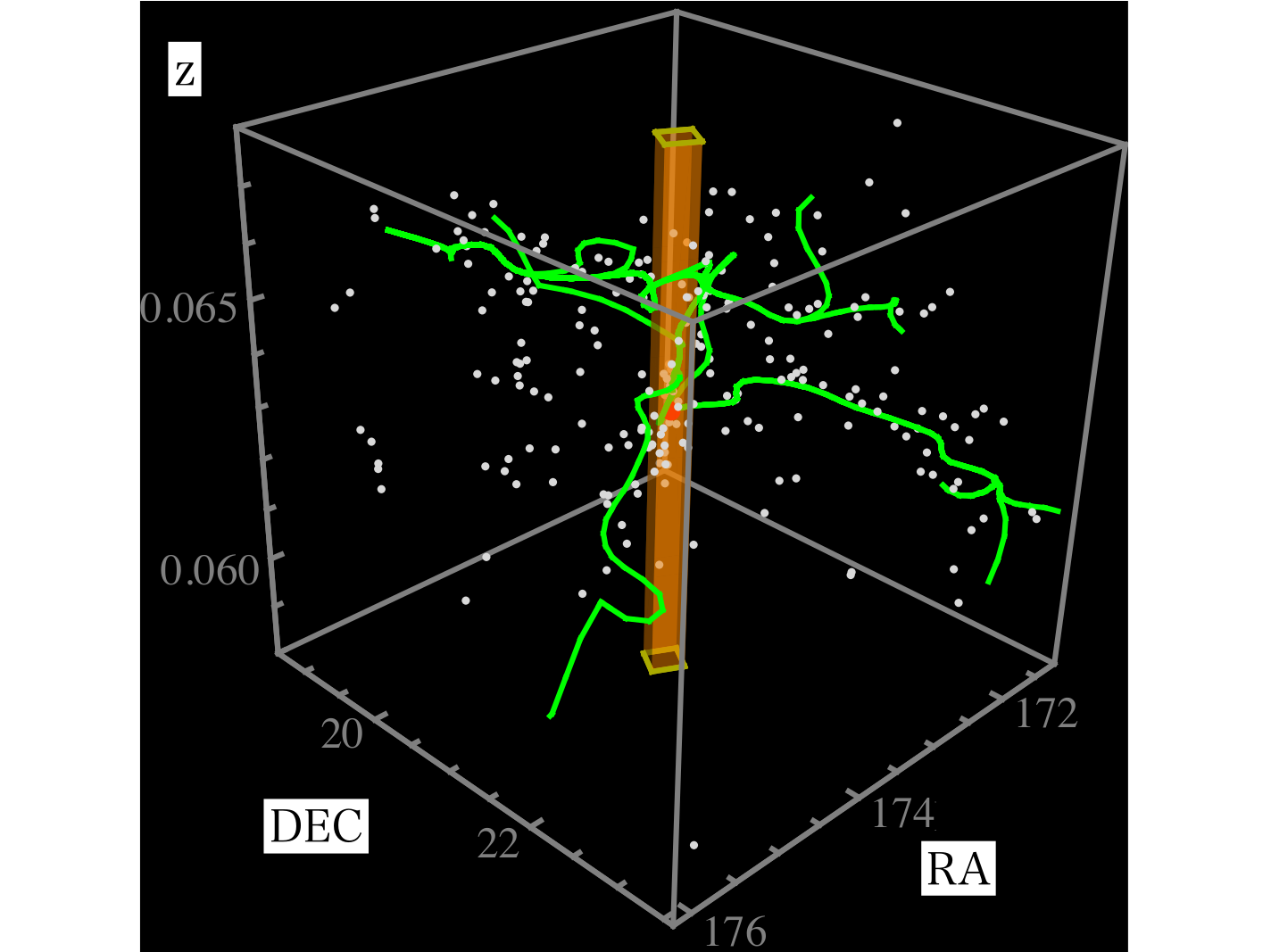}
\end{center}
\end{minipage}
\end{tabular}
\vspace{-0.2cm}
  \caption{The SDSS spectroscopic galaxy distribution around five observed regions, 
   i.e., FJ1 (top left), FJ2 (top right), FJ3 (middle left), FJ4 (middle right) and FJ5 (bottom).
   White and red dots indicate galaxies and the optically-bright galaxies, respectively.
   Filamentary structures computed by DisPerSE are represented by green solid curves. 
   The {\it Suzaku} FOVs are shown by yellow cylinders.}
  \label{fig:FJ-disperse}
\end{figure*}

Three-dimensional filamentary structures were extracted 
based on the spectroscopic data of the SDSS Data Release 7.2 
and the filament extractor, DisPerSE 
which visually allows for the coherent multiscale identification of all types of astrophysical structures, 
i.e., the voids, walls, filaments and clusters, utilizing the discrete Morse theory 
\citep{2011MNRAS.414..350S,2011MNRAS.414..384S}. 
Many filamentary junctions were identified and some of them already have been known 
as clusters of galaxies and groups of galaxies surrounded by X-ray halos. 
Among unidentified filamentary junctions, we selected five fields  
(hereafter FJ1, FJ2, FJ3, FJ4 and FJ5) satisfying the following criterion in the nearby Universe 
with the redshift between 0.05 and 0.1, where the field of view (hereafter FOV) of {\it Suzaku} can cover 1--2 Mpc:  
(1) no specific X-ray halo associated with the large scale structure 
was identified by the NASA Extragalactic Database (NED), 
(2) no strong X-ray signal was detected in the {\it ROSAT} all-sky map and 
(3) at least one optically bright galaxy exists. 
Figure \ref{fig:FJ-disperse} shows resultant three-dimensional filamentary structures 
computed by DisPerSE.
Filament Junction 1 (FJ1) has two optically bright galaxies  
(SDSS J133646.63+435028.9 \& SDSS J133646.14+435038.1) 
with $z$ $\sim$ 0.063 at the center, which seem almost colliding. 
The filaments run almost in the same plane perpendicular to the line of sight. 
Filament Junction 2 (FJ2) has optically bright galaxy (SDSS J095701.27+261027.7) at $z$ $\sim$ 0.082 
located in the center of Shakhbazian compact group (ShCG) 188 at $z$ = 0.080 \citep{2005RMxAA..41....3T} 
and two filaments run perpendicular to and along the line of sight, respectively. 
Filament Junctions 3, 4 and 5 (FJ3, FJ4 and FJ5) locate at several-filament intersections.
They possess optically bright spectroscopic identified galaxies 
with redshifts of 0.064 (SDSS J100540.53+394508.6), 0.072 (SDSS J110238.41+291525.3) and 
0.063 (SDSS J113459.83+210546.2) lying in their center positions.
%
%These central galaxies except for FJ5 are classified as elliptical galaxies based on optical data.
%
We observed five fields with {\it Suzaku} \citep{mitsuda-Suzaku} 
which has low and stable background and therefore is optimum 
to search for a faint emission from a halo and three co-aligned CCD chips, 
i.e., X-ray Imaging Spectrometer (XIS) \citep{2007PASJ...59S..23K}, 
with 18$'$$\times$18$'$ squared FOV were used in this work.
Neutral hydrogen column densities of the Galaxy 
toward the FOV and observation logs are summarized 
in Table \ref{table:junc_info}.
\begin{table*}[h!]
\footnotesize{
 \caption{{\it Suzaku} observation logs of five targets.}
\label{table:junc_info}
  \begin{center}
    \begin{tabular}{lcccccc}
\hline\hline
Field name                          & Obs. ID        & Date                         &  Exposure$^{\ast}$             &  (R.A., Dec.)      & N$_H$$^{\dagger}$ \\ 
                                              &                       &                                  & [ksec]                                     & [deg]                   & [$\times$10$^{20}$~cm$^{-2}$] \\ \hline
Filament Junction 1 (FJ1) & 806003010 & May 25-26, 2011  & 45                                           & (204.2, 43.8)    & 1.4\\
FJ2                                        & 806004010 & May 18-20, 2011  & 57                                           & (149.3, 26.2)    & 2.8\\
FJ3                                        & 806005010 & Apr 21-22, 2011   & 40                                           & (151.4, 39.7)     & 1.2 \\
FJ4                                        & 807038010 & June 8-10, 2012   & 21                                           & (165.7, 29.2)    & 1.8  \\
FJ5                                        & 807039010 & June 10-11, 2012 & 34                                          & (173.7, 21.1)     & 2.0 \\
\hline

    \end{tabular}
\begin{flushleft} 
\footnotesize{
\hspace{0.1cm}$^\ast$ Exposure time after the COR screening.\\
\hspace{0.1cm}$^\dagger$ Neutral hydrogen column densities of the Galaxy based on LAB survey \citep{2005AA...440..775K}.}
\end{flushleft}
  \end{center}}
\end{table*}
%\begin{table*}[htbp!]
%\footnotesize{
%  \caption{{\it Suzaku} observation logs of five targets.}
%\label{junc_info}
%  \begin{center}
%    \begin{tabular}{lccccr}
%\hline\hline
%Field name                          & Obs. ID        & Date                        &  Exposure$^{\ast}$ [ksec]      & \multicolumn{2}{c}{Aim point}                                   \\
%                                               &                       &                                 &       & (R.A., Dec.)                 &     \multicolumn{1}{c}{($\ell$, b)}   \\ \hline
%Filament Junction 1 (FJ1) & 806003010 & May 25-26, 2011  & 46 & (204.2, 43.8)               & (97.2, 71.0) \\
%FJ2                                        & 806004010 & May 18-20, 2011  & 57 & (149.3, 26.2)               & (204.1, 51.2) \\
%FJ3                                        & 806005010 & Apr 21-22, 2011   & 40 & (151.4, 39.7)               & (181.9, 53.0) \\ 
%FJ4                                        & 807038010 & June 8-10, 2012   & 69 & (165.7, 29.2)               & (201.5, 66.0) \\
%FJ5                                        & 807039010 & June 10-11, 2012 & 50 & (173.7, 21.1)              & (227.9, 71.6) \\
%\hline
%
%    \end{tabular}
%\begin{flushleft} 
%\footnotesize{
%$^\ast$ Exposure time after the COR screening.}
%\end{flushleft}
%  \end{center}}
%\end{table*}
%%
%%
%%
%%
%%

%%
%%
%%
\section{Data Reduction}
We applied standard data reduction processing 
(e.g., ANG$\_$DIST $<$ 1.5, SAA$\_$HXD == 0, T$\_$SAA$\_$HXD $>$ 436, ELV $>$ 5 and DYE$\_$ELV $>$ 20) 
and additional cut-off-rigidity (COR) selection 
(COR2 $>$ 8 GV for FJ1, FJ2, FJ3 and FJ5 and COR2 $>$ 12 GV for FJ4) 
for all $Suzaku$ data before analyzing energy spectra 
to subtract the non X-ray background (NXB) effectively. 
The higher COR threshold of 12 GV is needed only for FJ4 
to suppress higher background fluctuations.
The redistribution matrix files were produced by the xisrmfgen ftool and 
ancillary response files were prepared by using the xissimarfgen ftool 
\citep{2007PASJ...59S.113I}.
We assumed an input image to xissimarfgen to be a uniform flat sky with a radius of 20$'$.
The NXB was estimated from an accumulated night Earth observation 
with the ftool xisnxbgen \citep{2008PASJ...60S..11T}. 
In all analysis and data reduction, HEAsoft version 6.12 
and XSPEC version 12.7.0 were utilized.
\section{Analysis \& Results}
\subsection{X-ray Images for the Filamentary Junctions}
\begin{figure*}[h!]
\vspace{-0.8cm}
\begin{tabular}{ccc}
\begin{minipage}{0.45\hsize}
\hspace{0.05cm}
\vspace{-0.2cm}
(a) FJ1
\begin{center}
\vspace{-0.2cm}
    \includegraphics[bb=0 0 1024 768, width=7.2cm]{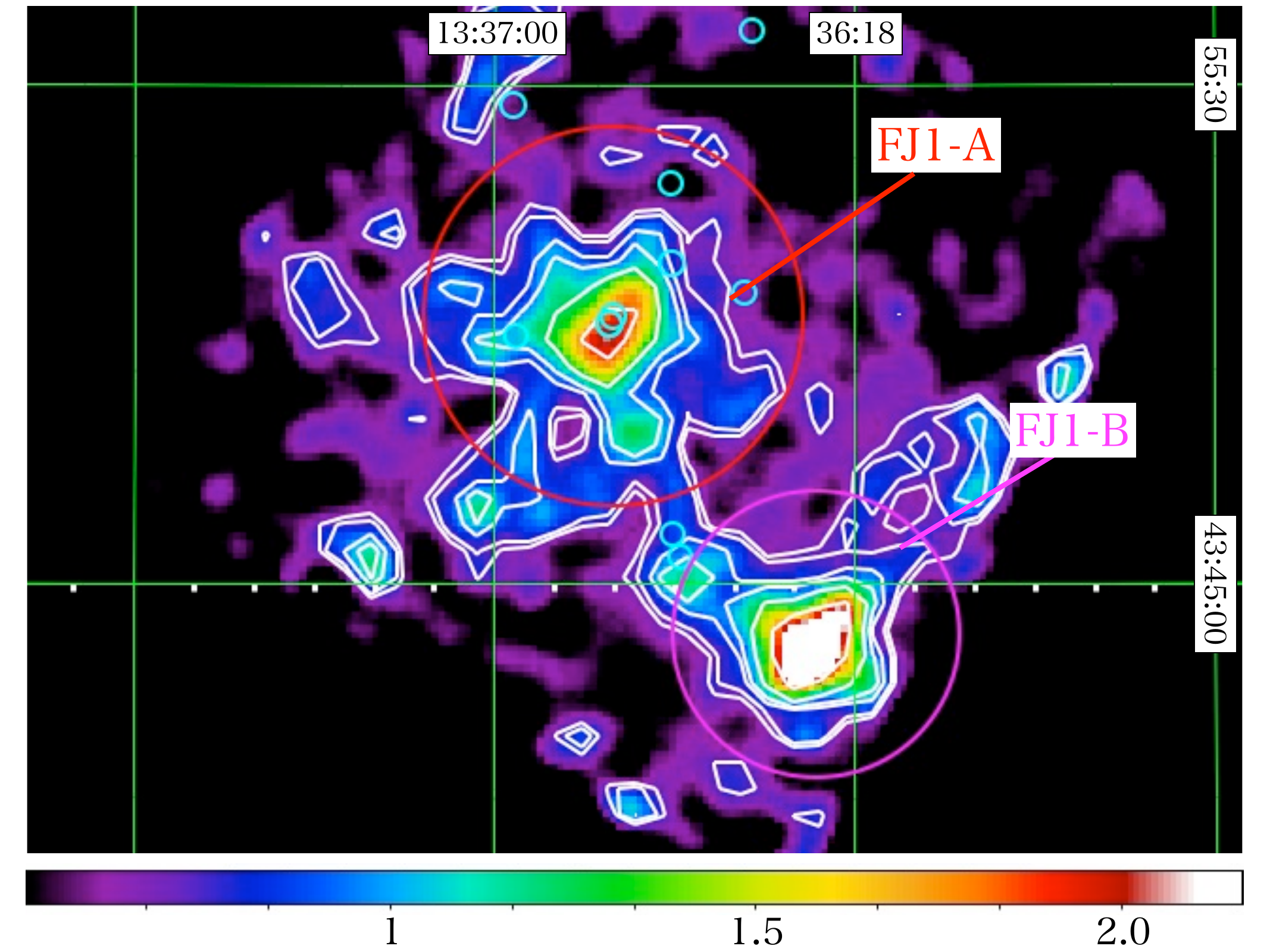}
\end{center}
\end{minipage}
\begin{minipage}{0.45\hsize}
\hspace{0.05cm}
\vspace{-0.2cm}
(b) FJ2
\begin{center}
\vspace{-0.2cm}
    \includegraphics[bb=0 0 1024 768, width=7.2cm]{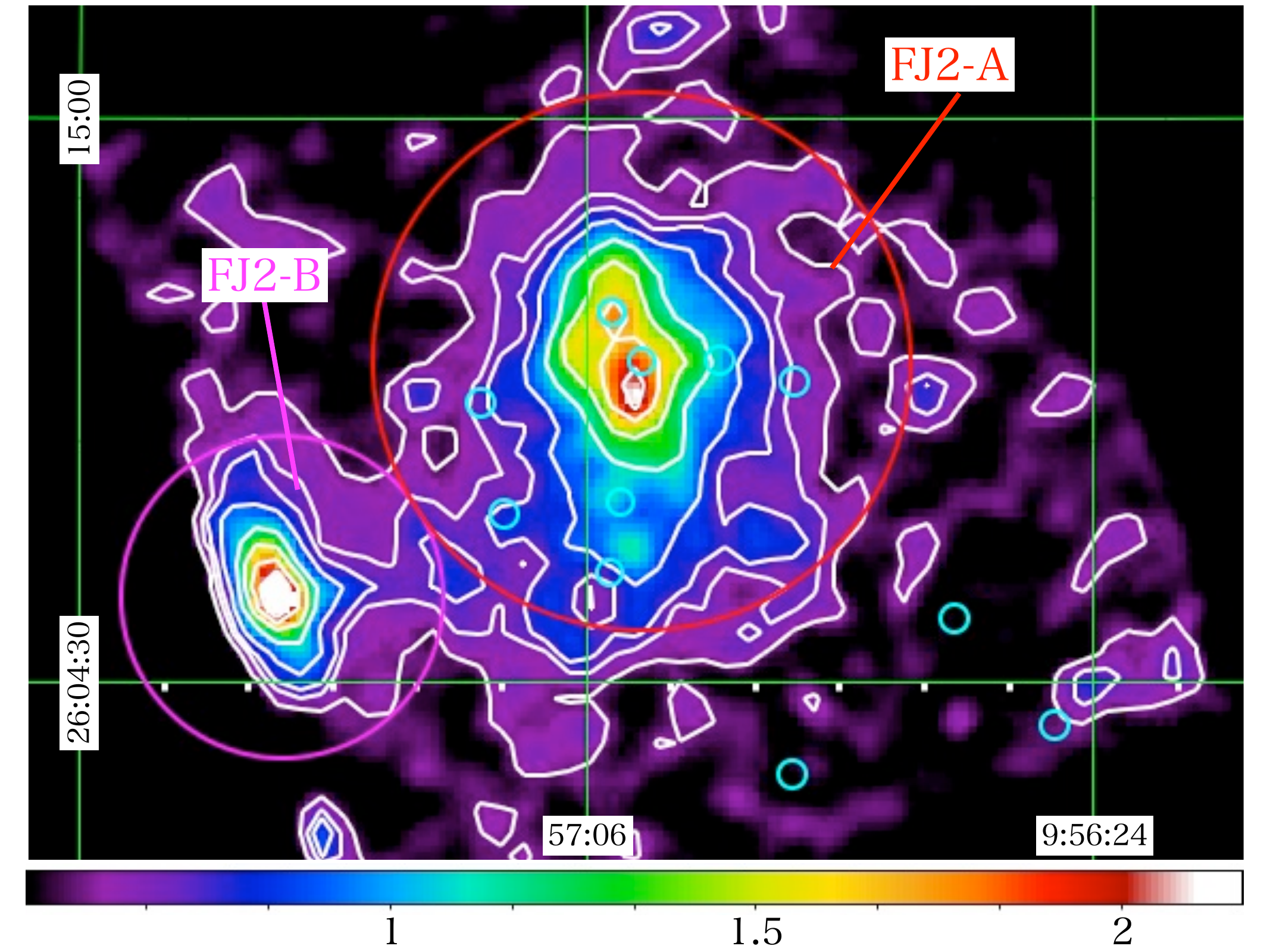}
\end{center}
\end{minipage}
\vspace{0.2cm}
\\
\begin{minipage}{0.45\hsize}
\hspace{0.05cm}
\vspace{-0.2cm}
(c) FJ3
\begin{center}
\vspace{-0.2cm}
    \includegraphics[bb=0 0 1024 768, width=7.2cm]{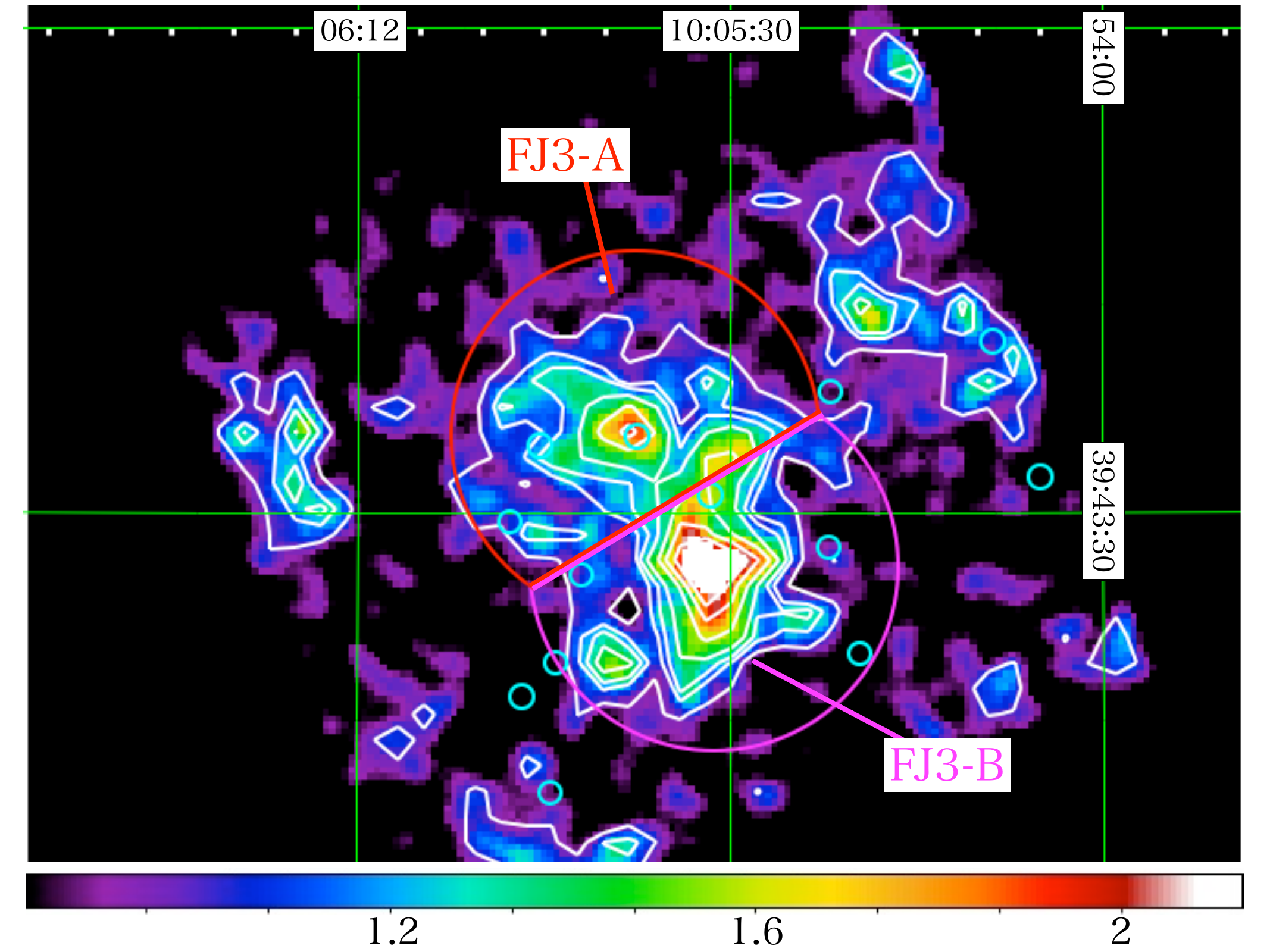}
\end{center}
\end{minipage}
\vspace{0.2cm}
\begin{minipage}{0.45\hsize}
\hspace{0.05cm}
\vspace{-0.2cm}
(d) FJ4
\begin{center}
\vspace{-0.2cm}
    \includegraphics[bb=0 0 1024 768, width=7.2cm]{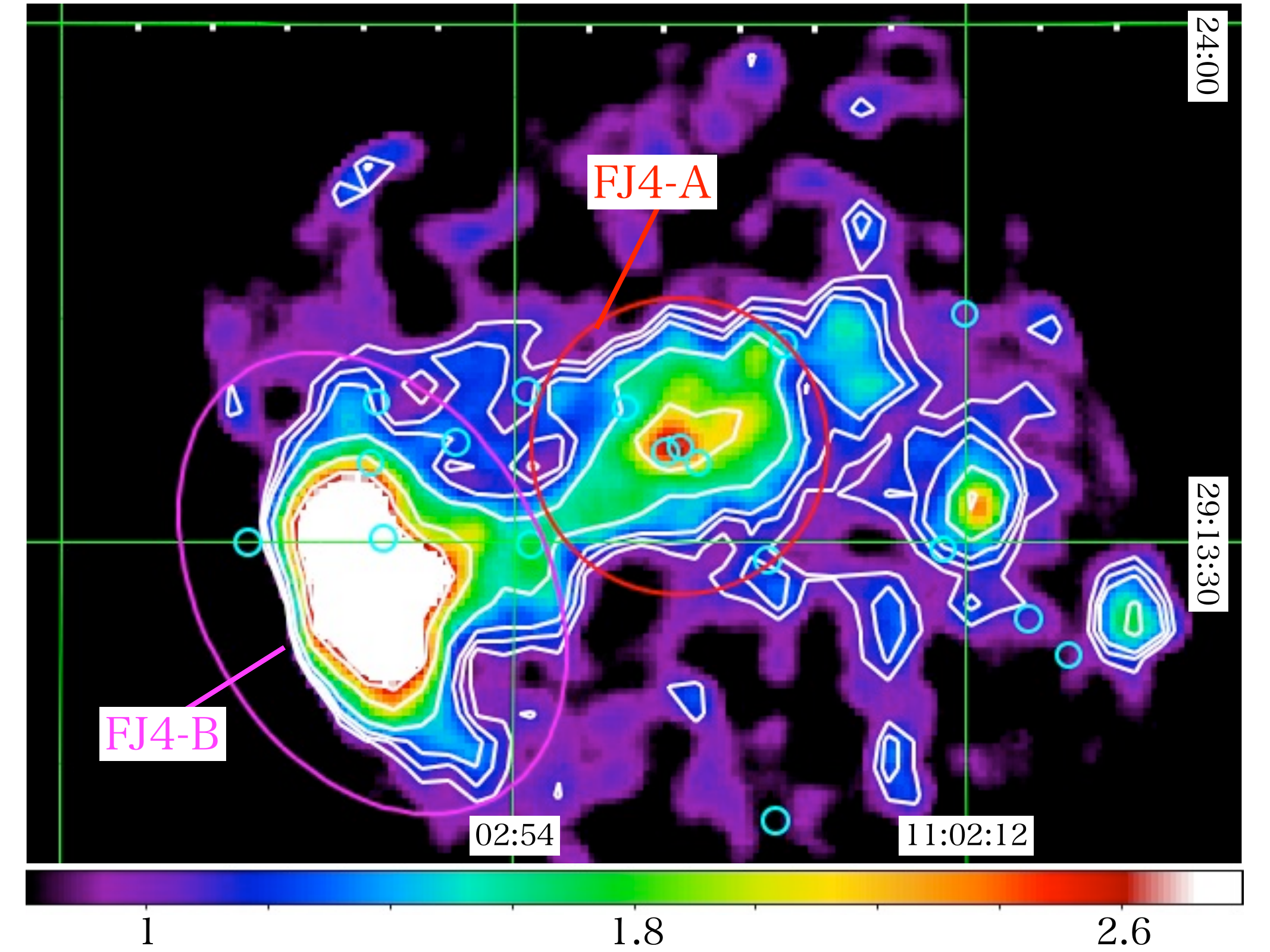}
\end{center}
\end{minipage}
\\
\begin{minipage}{0.45\hsize}
\hspace{0.05cm}
\vspace{-0.2cm}
(e) FJ5
\begin{center}
\vspace{-0.2cm}
    \includegraphics[bb=0 0 1024 768, width=7.2cm]{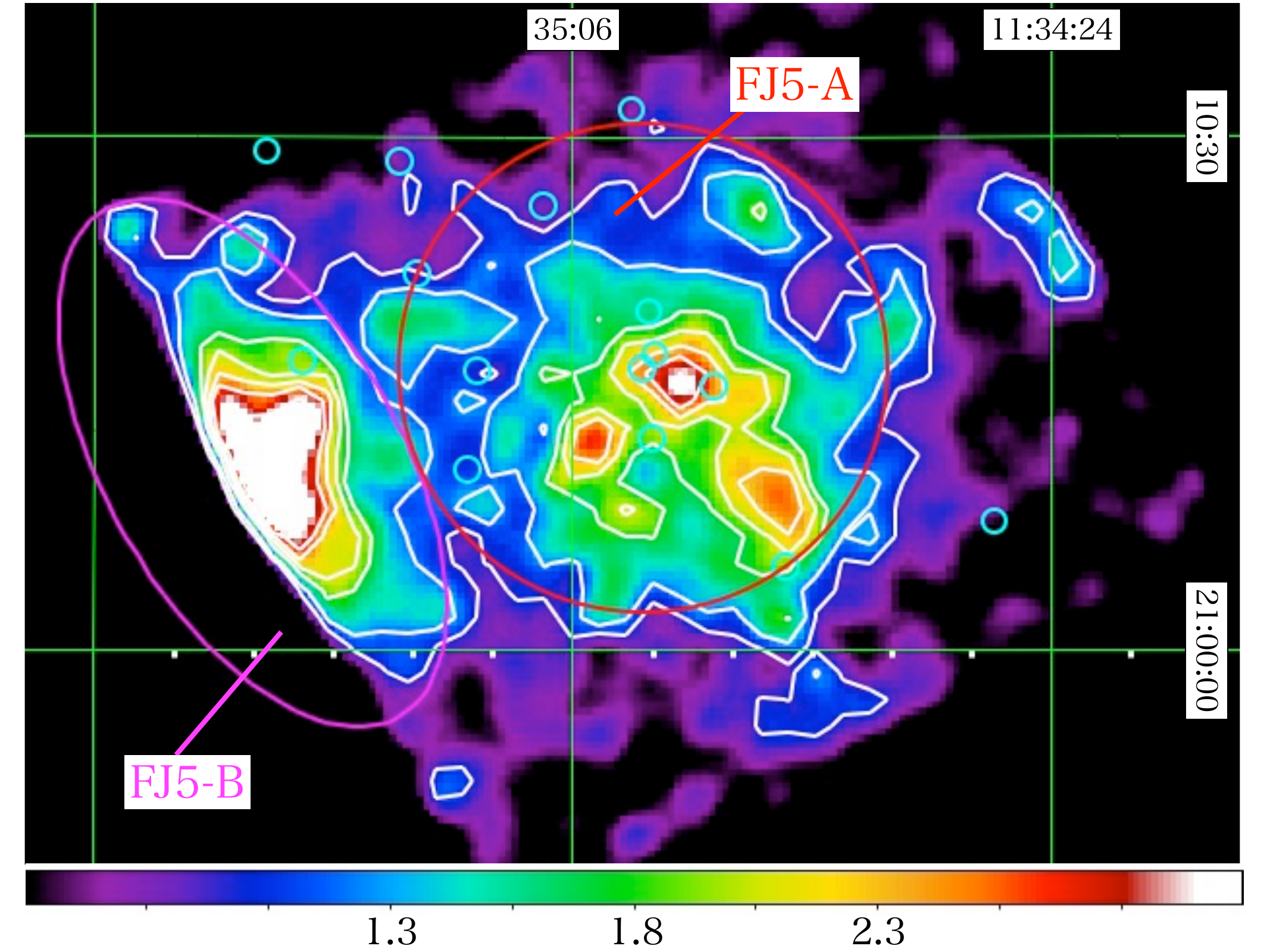}
\end{center}
\end{minipage}
\end{tabular}
\vspace{-0.2cm}
  \caption{The combined images of XIS0, 1 and 3 detectors at the energy range between 0.5 and 2.0 keV 
  in the unit of cts (exposure)$^{-1}$ (64 pixel)$^{-1}$ 
  for all five fields, i.e., FJ1 (top left), FJ2 (top right), FJ3 (middle left), FJ4 (middle right) and FJ5 (bottom).
  Vignetting correction is not applied. 
  The SDSS spectroscopic identified galaxies (cyan circles) and 
  X-ray contours (white solid lines) are also exhibited.
  Regions enclosed by red and magenta correspond to the central X-ray emitting regions (FJn-A) 
  where optically-bright galaxies are placed and 
  the brightest regions (FJn-B) other than FJn-A and are used for spectral analysis. 
  To emphasize FJn-A, some pixels in FJn-B are saturated.
  }
  \label{fig:FJ-xis1-05-2keV-images}
\end{figure*}

Figure \ref{fig:FJ-xis1-05-2keV-images} shows combined images of XIS0, 1 and 3 detectors 
with galaxy filaments calculated by DisPerSE 
having almost the same redshift as the central optically bright galaxies.
We adopted the energy band of 0.5--2.0 keV 
where an excess component is expected to be dominant as is the case of Paper I. 
Anomalous pixels in the XIS0 detector were excluded and 
complemented by the XIS3 detector.
Brighter regions are found at the center of all the images 
where optically bright galaxies are present. 
These regions are denoted as FJn-A (n = 1, 2, 3, 4, 5), 
enclosed with red lines in Figure \ref{fig:FJ-xis1-05-2keV-images}. 
FJ2-A and FJ5-A show non-spherically symmetric distributions and 
especially FJ5-A possesses multiple peaks in its surface brightness.
There are also several source candidates in all the images excluding the FJn-A regions. 
We selected the brightest region other than FJn-A in each field as FJn-B, 
enclosed with magenta lines in Figure \ref{fig:FJ-xis1-05-2keV-images}. 
Due to the poor statistics and/or their locations near the edge of CCD chips, 
most sources are hardly distinguishable from point sources 
based on their surface brightness profiles 
considering the spatial resolution of the {\it Suzaku} X-ray telescope (HPD$\sim$1.8$'$).
However, especially FJ2-A and FJ5-A show asymmetries and extended features.
Hence, 
we proceeded to spectroscopic analysis only for FJn-A and FJn-B 
because other candidates do not emit enough photons to discuss the origin 
through spectral analysis.
%
%Note that a typical sensitivity for point sources by {\it Suzaku} is 
%about several 10$^{-14}$ erg cm$^{-2}$ sec$^{-1}$ with these exposure times. 
%

%%
%%
%%
%%
%%
\clearpage
\subsection{Spectral Analysis}
We created energy spectra from the filament junction fields as shown in Figure \ref{fig:FJ-xis1-05-2keV-images} 
to examine the emission origin. 
The FJn-A regions including bright sources around optically bright galaxies were extracted 
by circles with radii of 4 arcmin for FJ1-A, 3 arcmin for FJ4-A and 5 arcmin for FJ2-A and FJ5-A.
The center positions of FJn-A correspond to the positions of optically bright galaxies 
described in \S \ref{sec:field-selection}.
The FJn-B regions were extracted by a circle  
with a radius of 3 arcmin for FJ1-B and FJ2-B and ellipses  
with major/minor axes of 5/3.5 arcmin for FJ4-B and 6/3 arcmin for FJ5-B, respectively. 
For FJ3, because the FJ3-B region is partly overlapped 
with the FJ3-A region the extracted region was not complete circle 
with a radius of 4 arcmin. 
In spectral analysis, 
the energy ranges of 0.8--5.0 keV for the XIS0 and XIS3 detectors and 
0.4--5.0 keV for the XIS1 detector were used.
\subsubsection{Evaluation of X-ray Background}
For spectral analysis of a faint X-ray emission, 
a careful approach on an X-ray background estimation is important. 
We use a surrounding region in each FOV to evaluate the energy spectrum of the X-ray background. 
We defined the FJn-BGD region excluding the FJn-A and FJn-B regions in each FOV. 
We fitted the spectrum in the FJn-BGD regions with the typical X-ray background emission 
consisting of (1) an unabsorbed thin thermal collisionally-ionized equilibrium (CIE) plasma, 
(2) an absorbed thin thermal CIE plasma and (3) an absorbed power law \citep{2009PASJ...61..805Y}. 
The first two components represent emissions from the Solar-system neighborhood 
(hereafter solar wind charge exchange: SWCX and Local Hot Bubble: LHB) and 
Galactic halo (Milky Way Halo: MWH). 
On the other hand, the last component corresponds to the accumulation of unresolved extragalactic point sources 
(cosmic X-ray background: CXB) described by an absorbed power-law model with a photon index of 1.4 
as shown in \citet{2002PASJ...54..327K}. 
The following model was adopted 
in the XSPEC software: $apec$$_\mathrm{SWCX+LHB}$ + $phabs$$_\mathrm{Galactic}$~ ($apec$$_\mathrm{MWH}$+ $power$-$law$$_\mathrm{CXB}$). 
To consider the systematic error in the different sensors, 
a constant factor ({\it f}) was multiplied for the XIS1 and XIS3 detectors \citep{2013PASJ...65...44M}.
Resulting temperatures of the Galactic foreground emission 
except for the MWH component in FJ2-BGD, FJ3-BGD and FJ5-BGD and 
surface brightness of the CXB component in all FJn-BGD regions are consistent 
with typical values reported in previous studies \citep{2009PASJ...61..805Y} within the statistical error. 
The temperatures of MWH in FJ3-BGD and FJ4-BGD are 
significantly higher than the typical value at a 90 \% significance level 
but consistent within 4 $\sigma$ significance levels.
However, the temperature of MWH in FJ5-BGD is significantly higher than the typical value 
even at a 4 $\sigma$ significance level.
Thus, we conducted further spectral analysis for FJ5-BGD in different ways.
As the first trial, the higher COR threshold value of 12 GV was imposed to reduce background contributions.
However, no difference was found within the statistical error in their best fit values.
Although the area of 2 arcmin from the chip edges was removed 
to diminish systematic uncertainties and 
the area within 2 arcmin around FJ5-A and FJ5-B was excluded 
to reduce contaminations from brighter regions 
considering the {\it Suzaku} angular resolution (HPD$\sim$1.8$'$) 
any changes were not confirmed within the statistical error.
Thus, we concluded that the temperature of MWH in FJ5 is really high.
This sort of hot foreground is known to be present 
especially at low Galactic latitudes \citep[e.g.,][]{2009PASJ...61..805Y,2009PASJ...61S.115M}.
Resultant parameters are summarized in Table \ref{table:spec-fj-bgd}.
As stated in the next section, 
statistical and systematic uncertainties on the X-ray background are taken into account 
through a simultaneous fitting using FJn-BGD.
\begin{figure*}[h!]
\begin{tabular}{ccc}
\begin{minipage}{0.45\hsize}
\hspace{0.5cm}
(a) FJ1-BGD
\begin{center}
\vspace{-3.2cm}
    \includegraphics[bb=0 0 1424 1067, width=10cm]{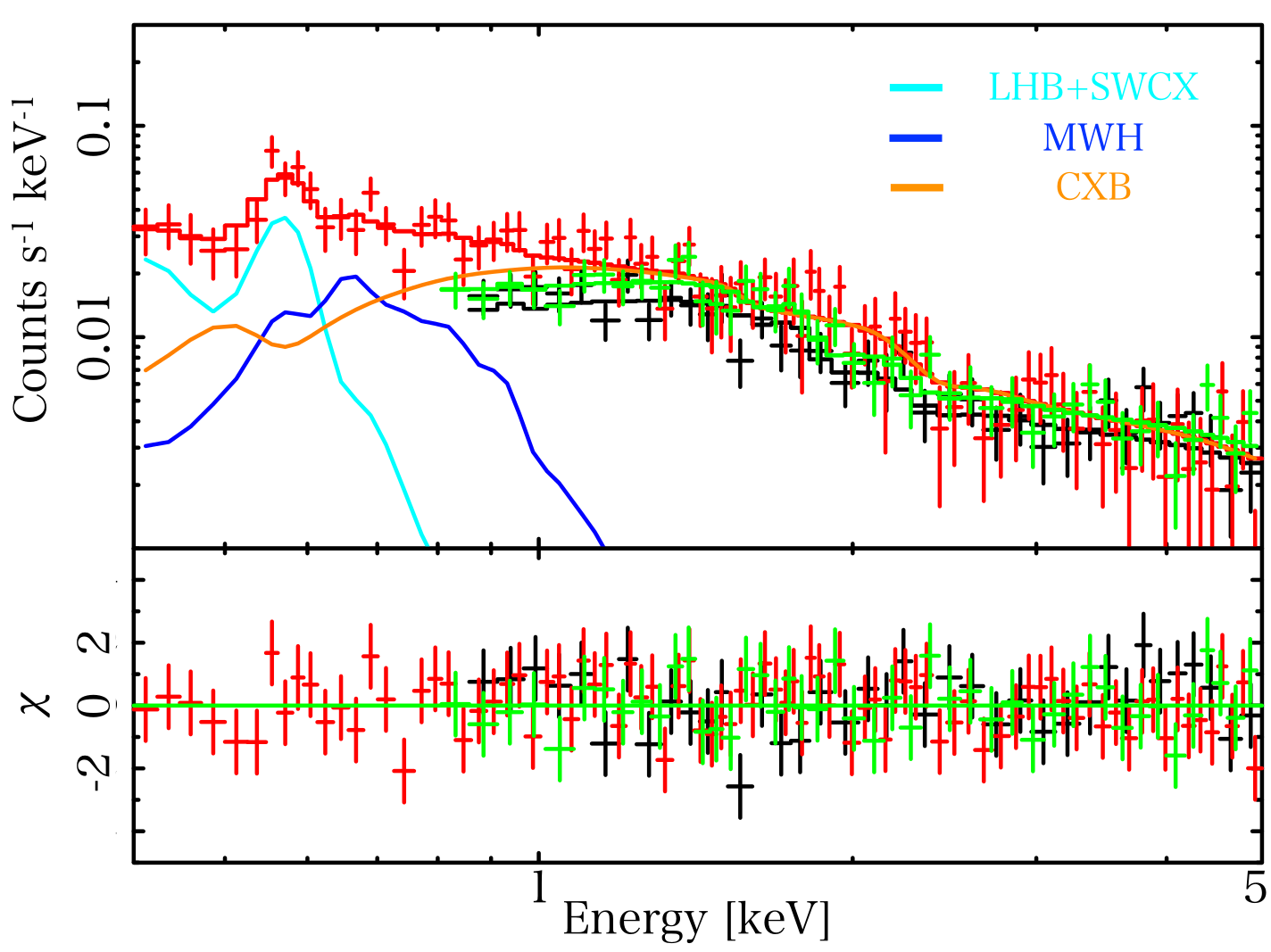}
\end{center}
\end{minipage}
\begin{minipage}{0.45\hsize}
\hspace{0.5cm}
(b) FJ2-BGD
\begin{center}
\vspace{-3.2cm}
    \includegraphics[bb=0 0 1424 1067, width=10cm]{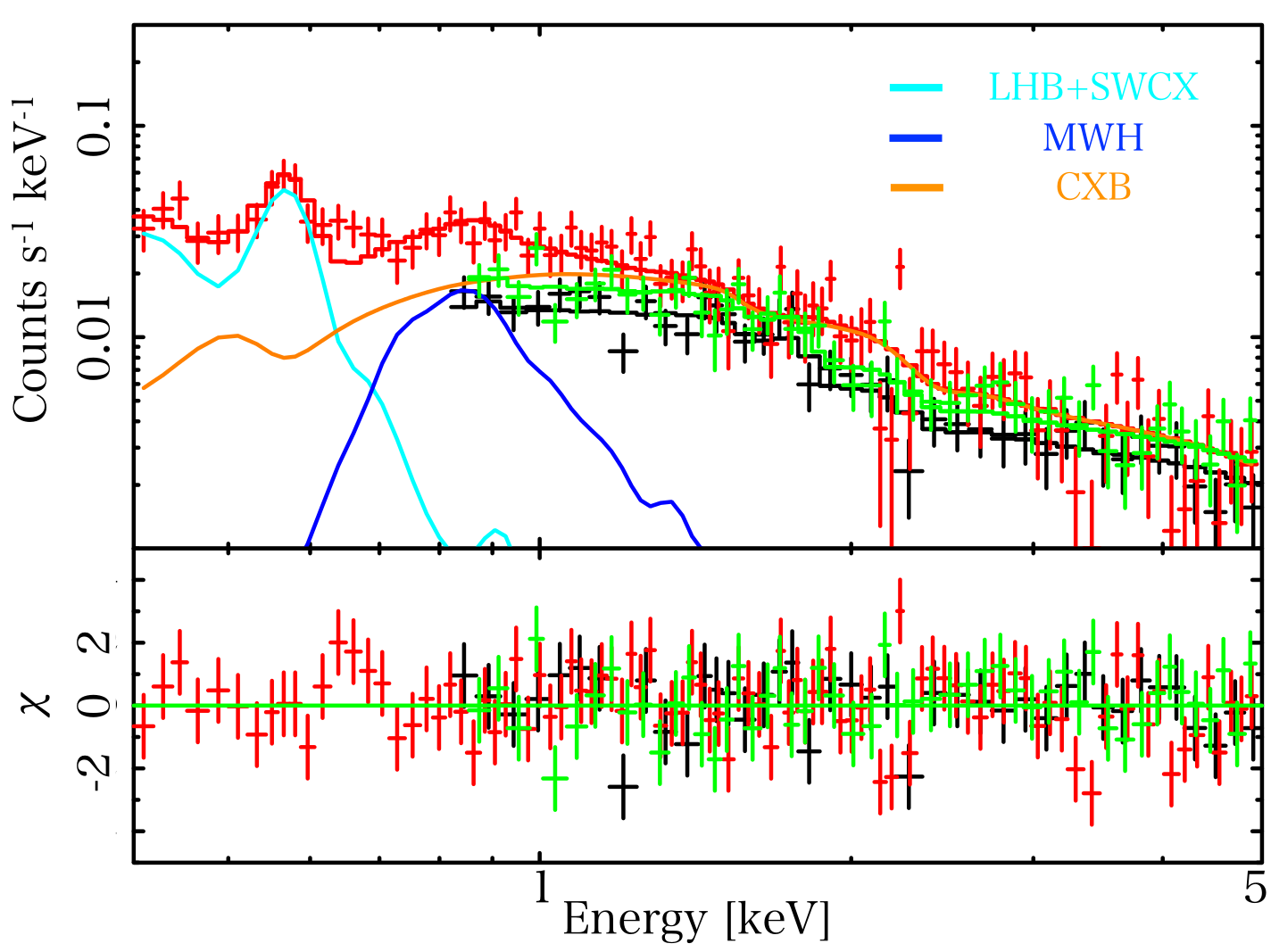}
\end{center}
\end{minipage}
\vspace{0.2cm}
\\
\begin{minipage}{0.45\hsize}
\hspace{0.5cm}
(c) FJ3-BGD
\begin{center}
\vspace{-3.2cm}
    \includegraphics[bb=0 0 1424 1067, width=10cm]{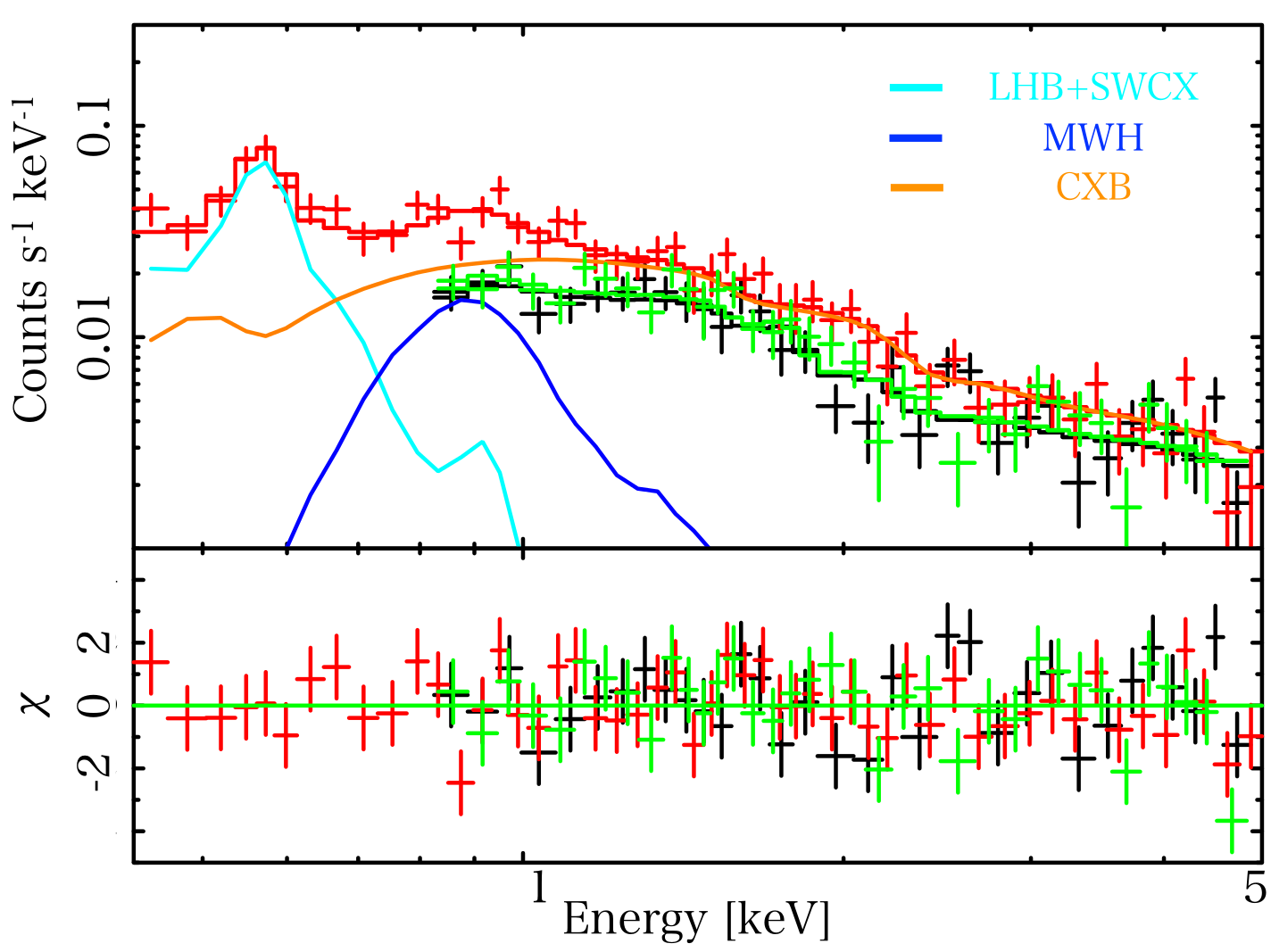}
\end{center}
\end{minipage}
\vspace{0.2cm}
\begin{minipage}{0.45\hsize}
\hspace{0.5cm}
(d) FJ4-BGD
\begin{center}
\vspace{-3.2cm}
    \includegraphics[bb=0 0 1424 1067, width=10cm]{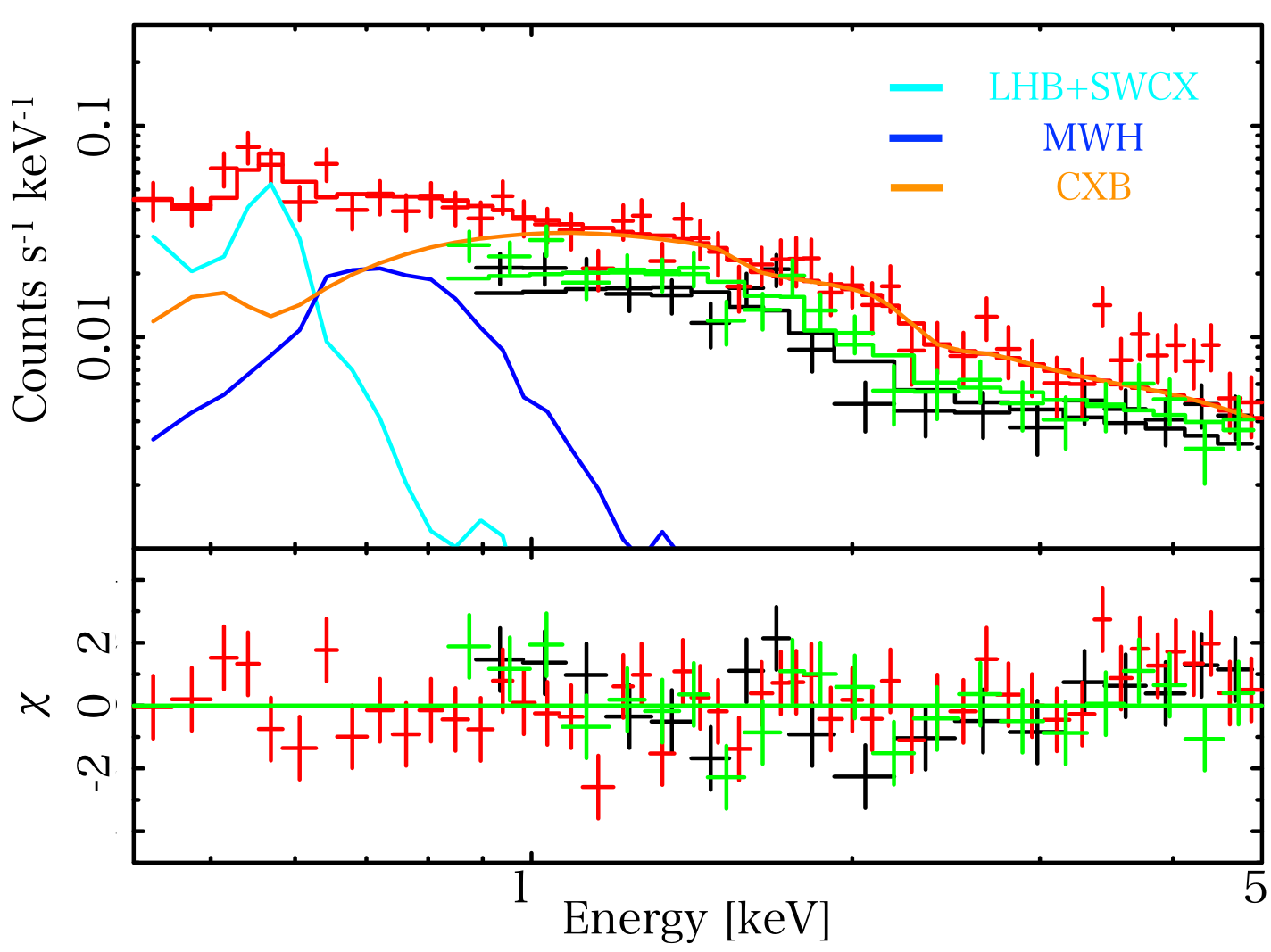}
\end{center}
\end{minipage}
\\
\begin{minipage}{0.45\hsize}
\hspace{0.5cm}
(e) FJ5-BGD
\begin{center}
\vspace{-3.2cm}
    \includegraphics[bb=0 0 1424 1067, width=10cm]{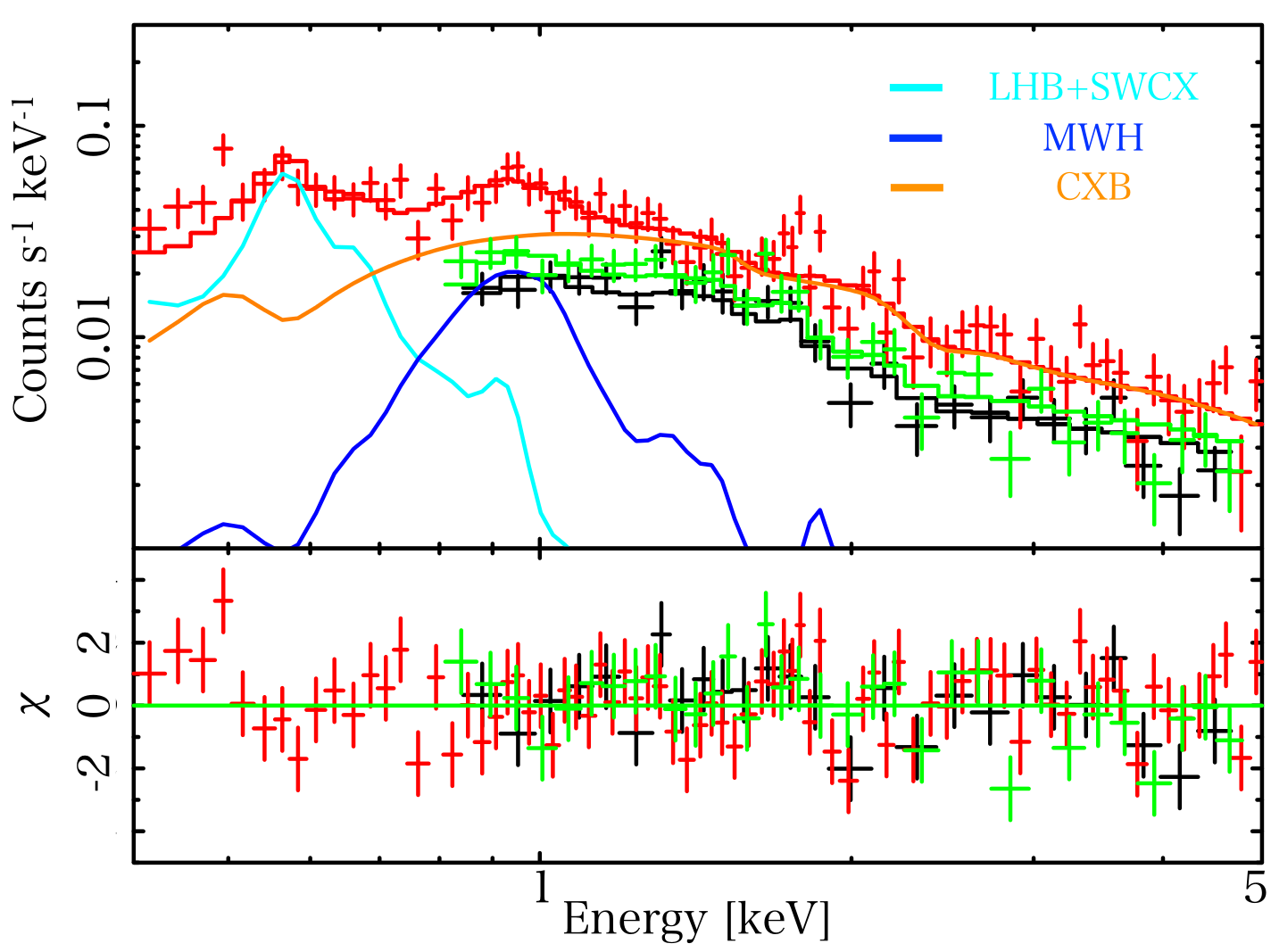}
\end{center}
\end{minipage}
\end{tabular}
  \caption{XIS0 (black), XIS1 (red) and XIS3 (green) spectra obtained 
  from the FJn-BGD regions, i.e., FJ1-BGD (top left), FJ2-BGD (top right), 
  FJ3-BGD (middle left), FJ4-BGD (middle right) and FJ5-BGD (bottom).
  Resultant spectra are fitted with the X-ray background emission model, 
  i.e., $apec$$_\mathrm{SWCX+LHB}$ (cyan) + $phabs$$_\mathrm{Galactic}$$\times$($apec$$_\mathrm{MWH}$ (blue) 
  + $power$-$law$$_\mathrm{CXB}$ (orange)), 
  and the best fit models only for the XIS1 detector are shown for simplicity.}
  \label{fig:FJ-bgd}
\end{figure*}
%%
%%
%%
%%
%%

%%%
%%%
%%%
%%%
%%%
%\begin{figure*}[h!]
%\begin{minipage}{0.5\hsize}
%\begin{center}
%    \includegraphics[width=0.9\linewidth]{figures/fj1-bgd_arf_ea_cor2-8GV.png}
%\end{center}
%\end{minipage}
%\begin{minipage}{0.5\hsize}
%\begin{center}
%    \includegraphics[width=0.9\linewidth]{figures/fj2-bgd_arf_ea_cor2-8GV.png}
%\end{center}
%\end{minipage}
%\begin{minipage}{0.5\hsize}
%\begin{center}
%    \includegraphics[width=0.9\linewidth]{figures/fj3-bgd_arf_ea_cor2-8GV.png}
%\end{center}
%\end{minipage}
%\begin{minipage}{0.5\hsize}
%\begin{center}
%    \includegraphics[width=0.9\linewidth]{figures/fj4-bgd_cor2-12GV.png}
%\end{center}
%\end{minipage}
%\begin{minipage}{0.5\hsize}
%\begin{center}
%    \includegraphics[width=0.9\linewidth]{figures/fj5-bgd_arf_att_cor2-8GV.png}
%\end{center}
%\end{minipage}
%  \caption{test}
%  \label{fig:FJ-bgd}
%\end{figure*}
%%%
%%%
%%%
%%%
%%%

%%
%%
%%
%%
%%
\begin{table*}[h!]
\scriptsize{
  \caption{The results of the model fitting of the five FJn-BGD regions.}
\label{table:spec-fj-bgd}
  \begin{center}
    \begin{tabular}{cccccccc}
\hline\hline
Region      & $f^{\ast}$                                                                          & $kT_\mathrm{SWCX+LHB}$ & Norm$_\mathrm{SWCX+LHB}$$^\dagger$ & $kT_\mathrm{MWH}$     & Norm$_\mathrm{MWH}$$^\dagger$ & $S_\mathrm{CXB}$$^\ddagger$ & $\chi^2/$d.o.f \\
                   &                                                                                            &  [keV]                                          &                                                                               & [keV]                                   &                                                                   &                                                             & \\ \hline \hline
 FJ1-BGD & (1.05$^{+0.10}_{-0.09}$, 1.12$^{+0.10}_{-0.09}$)  & 0.09$^{+0.03}_{-0.06}$          & 55$^{+1100}_{-35}$                                          & 0.23$^{+0.09}_{-0.03}$  & 4.1$^{+8.3}_{-2.2}$                               & 7.6$\pm$0.5                                     & 131/163  \\
 FJ2-BGD & (1.14$^{+0.08}_{-0.09}$, 1.11$^{+0.09}_{-0.08}$)  & 0.09$\pm$0.01                         & 69$^{+31}_{-30}$                                               & 0.67$^{+0.10}_{-0.09}$ & 1.4$\pm$0.3                                            & 7.5$\pm$0.5                                     & 187/181 \\
 FJ3-BGD & (1.15$^{+0.10}_{-0.09}$, 0.93$^{+0.09}_{-0.08}$)  & 0.12$\pm$0.03                         & 29$^{+50}_{-14}$                                               & 0.81$^{+0.14}_{-0.12}$ & 1.2$\pm$0.3                                            & 7.4$\pm$0.5                                    & 136/114 \\
 FJ4-BGD & (1.30$^{+0.15}_{-0.13}$, 1.06$^{+0.12}_{-0.11}$) & 0.09$^{+0.04}_{-0.03}$           & 60$^{+720}_{-43}$                                             & 0.28$^{+0.12}_{-0.09}$ & 3.2$^{+5.8}_{-1.6}$                                & 8.7$\pm$0.7                                   & 106/83 \\
 FJ5-BGD & (1.42$^{+0.12}_{-0.11}$, 1.04$^{+0.10}_{-0.09}$) & 0.15$^{+0.02}_{-0.01}$           & 18$^{+7}_{-4}$                                                    & 0.97$^{+0.10}_{-0.11}$ & 2.1$\pm$0.5                                            & 11$\pm$1                                        & 171/130 \\
\hline
    \end{tabular}
  \end{center}
\begin{flushleft}
\footnotesize{
% \hspace{1.2cm}$^\ast$ The absorption column density of the Galaxy in the unit of $10^{20}$ cm$^{-2}$.\\
$\ast$ Constant factors of XIS1 (left) and XIS3 (right) detectors relative to the XIS0 detector. \\
$\dagger$ Normalization of the $apec$ models divided by a solid angle $\Omega$, 
 assumed in a uniform-sky ARF calculation (20$'$ radius), 
 i.e. $\it{Norm} = (1/\Omega) \int n_e n_H dV / (4\pi(1+z)^2)D^2_A)$ cm$^{-5}$ sr$^{-1}$ 
in unit of 10$^{-14}$, where $D_A$ is the angular diameter distance.\\
$\ddagger$ Surface brightness of CXB 
 in the unit of photons s$^{-1}$ cm$^{-2}$ sr$^{-1}$ keV$^{-1}$ at 1 keV\@.
}
\end{flushleft}}
\end{table*}
\clearpage
\subsubsection{Spectral fitting for FJ sources}
\label{sec:spectral-analysis}
We fitted the spectra of both FJn-A and FJn-B with FJn-BGD regions as a template of X-ray background. 
A simultaneous fitting using the FJn-BGD regions was carried out with common parameters 
for the background emission 
($kT_{{\rm SWCX+LHB}}$, $Norm_{{\rm SWCX+LHB}}$, $kT_{{\rm MWH}}$, $Norm_{{\rm MWH}}$ and $S_{{\rm CXB}}$) 
to consider the statistical and systematic errors. 
Because the FJn-BGD spectra cannot reproduce all the spectra obtained from FJn-A and FJn-B at all 
firstly we assumed a thin thermal emission as the excess emissions as with the case of Paper I 
and then tried a non-thermal emission model. 
%
%In all cases, the background emission parameters are consistent 
%with those obtained from FJn-BGD alone tabulated in Table \ref{table:spec-fj-bgd}. 
%

For FJn-A, we let the abundance of the excess emission vary and 
then fix it to be 0.3 or 1 solar if the abundance cannot be constrained due to poor statistics.
The former value corresponds to the typical abundance 
in groups of galaxies and clusters of galaxies \citep[e.g.,][]{2003ApJS..145...39M,2005ApJ...620..680B}. 
The redshift of the excess emission in FJn-A is fixed at that of the central optically bright galaxy. 
Resultant temperatures of the additional plasmas are $\sim$0.6, 1.3, 0.8, 0.7 and 1.0 keV 
for FJ1-A, FJ2-A, FJ3-A, FJ4-A and FJ5-A, respectively.
The abundance values were constrained only for FJ2-A and FJ5-A and 
resultant metal abundances are 0.2$\pm$0.1 and $<$0.1 solar, respectively.
Temperatures with the abundance of 0.3 or 1 solar are consistent with each other in all cases.
No additional absorption is required for FJn-A.
Next, a power-law component was added instead of the thermal component. 
The goodness of the fit were worse than those in the thermal plasma emission cases for all regions 
because the power-law component does not make up for the residual around 0.8 keV 
as shown in Figure \ref{fig:FJn-A1}.
Resulting photon indices are $\sim$3.3 and 3.6 for FJ2-A and FJ5-A and 
only lower limits of $>$4.7, 2.5 and 2.7 are obtained for FJ1-A, FJ3-A and FJ4-A, respectively.
These values are much higher than the typical value of AGNs 
\citep[e.g.,][]{2008ApJ...674..686W}.
Additional absorption on the order of 10$^{21}$ cm$^{-2}$ is needed 
for FJ1-A and FJ2-A.
Spectra are shown in Figure \ref{fig:FJn-A1} and 
the best fit parameters are summarized in Table \ref{table:FJn-Aenter-result}.

As is the case with FJn-A, 
the same spectral analysis was performed also for FJn-B.
Although the redshift and the abundance are both regarded as free parameters at first,   
no reasonable constraint is extracted on the redshift for all the FJn-B sources.
Thus, their redshifts of the excess thermal plasmas in FJn-B are set to be 0 and 
we exhibit only this case.
We confirmed that temperatures are consistent with each other 
in cases of the redshift of free or 0.
However, the abundances are well constrained 
when the redshift of the excess thermal plasma is fixed at 0.
The best fit values of temperatures and abundances are $\sim$1.4, 2.1, 2.6, 1.9 and 3.2 keV 
with upper limits on the abundance of $<$0.4, 0.4, 1.4, 0.1 and 0.3 solar 
for FJ1-B, FJ2-B, FJ3-B, FJ4-B and FJ5-B, respectively.
Significant extra absorption is not required for all FJn-B.
Then, the thin thermal plasma model was replaced by a power-law component.
Photon indices are $\sim$2.9, 2.3, 2.1, 2.4 and 2.0 
for FJ1-B, FJ2-B, FJ3-B, FJ4-B and FJ5-B, respectively.
Additional absorption is not requested for the power-law components.
Statistically, 
the thermal and the non-thermal emissions are both acceptable except for FJ4-B 
while there still remains the residual below 1 keV in the fit 
with the thin thermal plasma model for FJ4-B.
Resultant spectra are shown in Figure \ref{fig:FJn-B1} 
and the best fit parameters are summarized in Table \ref{table:FJn-B-result}.

In summary, 
spectral analysis shows that the thin thermal plasma is preferable 
in both statistical and physical points of view as the origin for FJn-A
since the power-law components cannot make up for the residual around 0.8 keV 
in the energy spectrum and resulting photon indices are much higher than 
the typical values of AGNs.
For FJn-B, 
the thin thermal and the non-thermal components are both acceptable statistically.
Thus, the origin and corresponding objects utilizing catalogues 
are discussed in the next section.
\begin{figure*}[h!]
\begin{tabular}{cc}
\begin{minipage}{0.5\hsize}
\hspace{0.6cm}
FJ1-A: + absorbed thin-thermal plasma 
\begin{center}
\vspace{-3.2cm}
    \includegraphics[bb=0 0 1424 1067, width=10.cm]{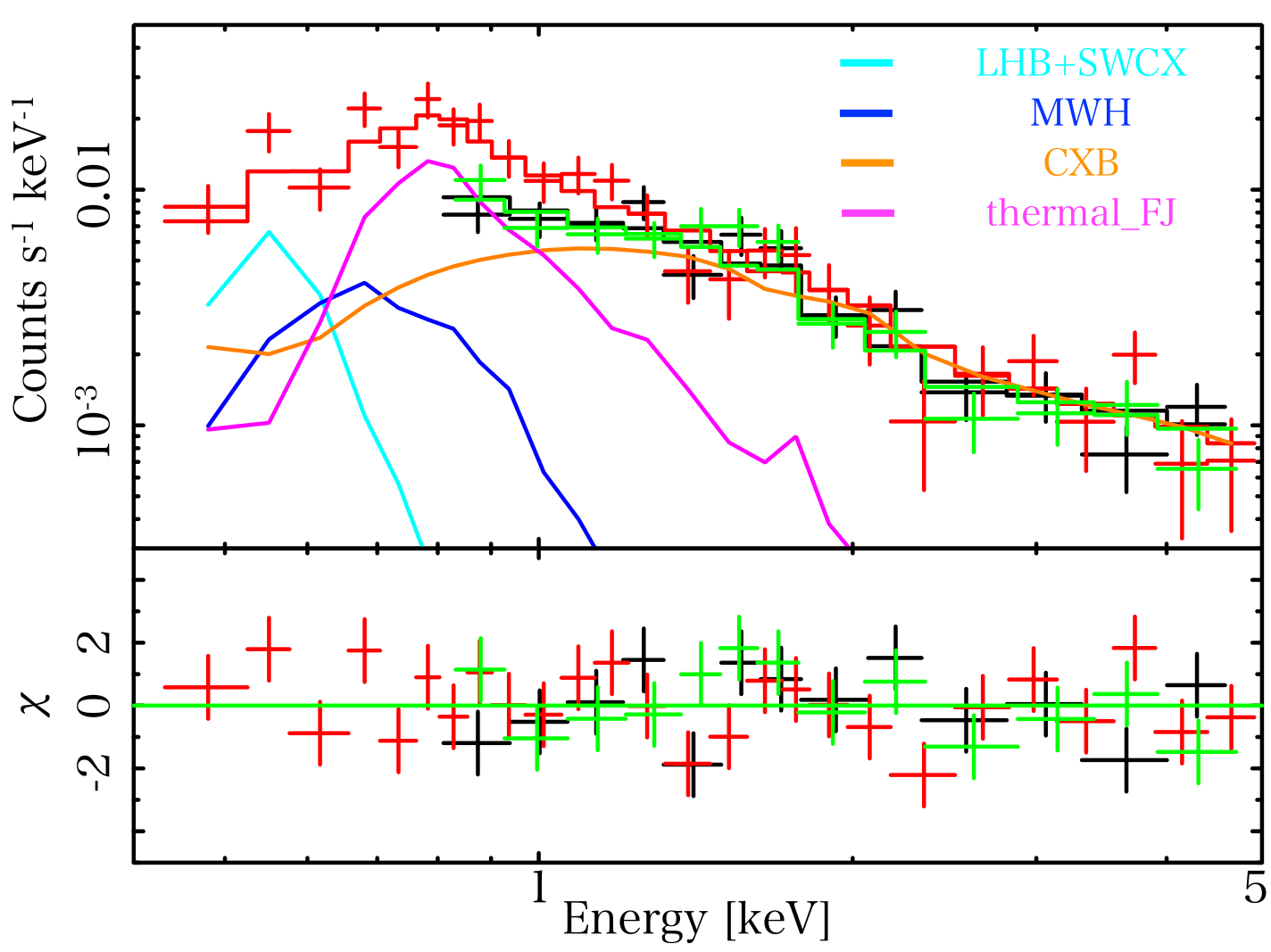}
\vspace{0.2cm}
\end{center}
\end{minipage}
\begin{minipage}{0.5\hsize}
\hspace{0.6cm}
FJ1-A: + absorbed non-thermal plasma 
\begin{center}
\vspace{-3.2cm}
    \includegraphics[bb=0 0 1424 1067, width=10.cm]{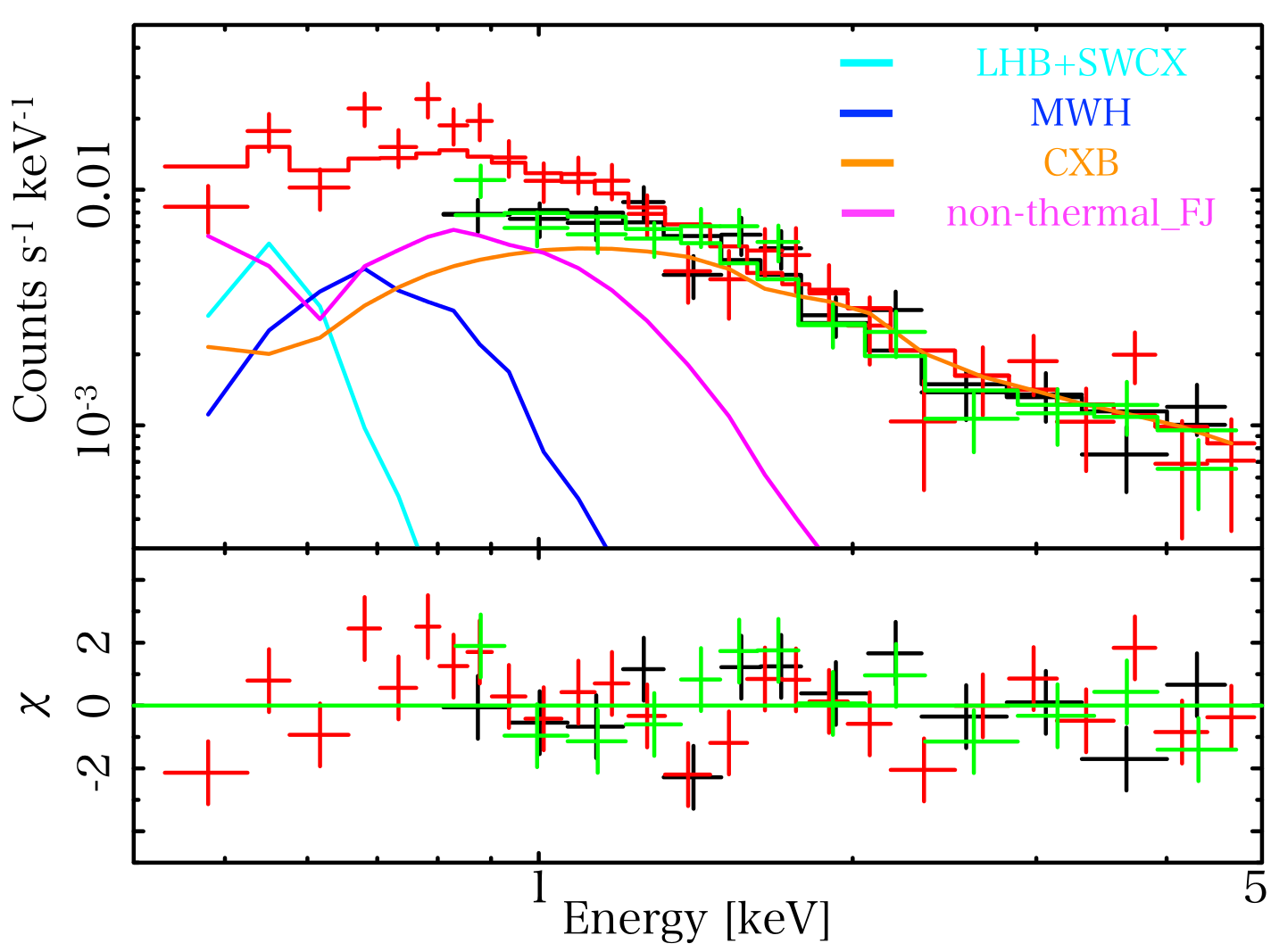}
\vspace{0.2cm}
\end{center}
\end{minipage}
\\
\begin{minipage}{0.5\hsize}
\hspace{0.6cm}
FJ2-A: + absorbed thin-thermal plasma 
\begin{center}
\vspace{-3.2cm}
    \includegraphics[bb=0 0 1424 1067, width=10cm]{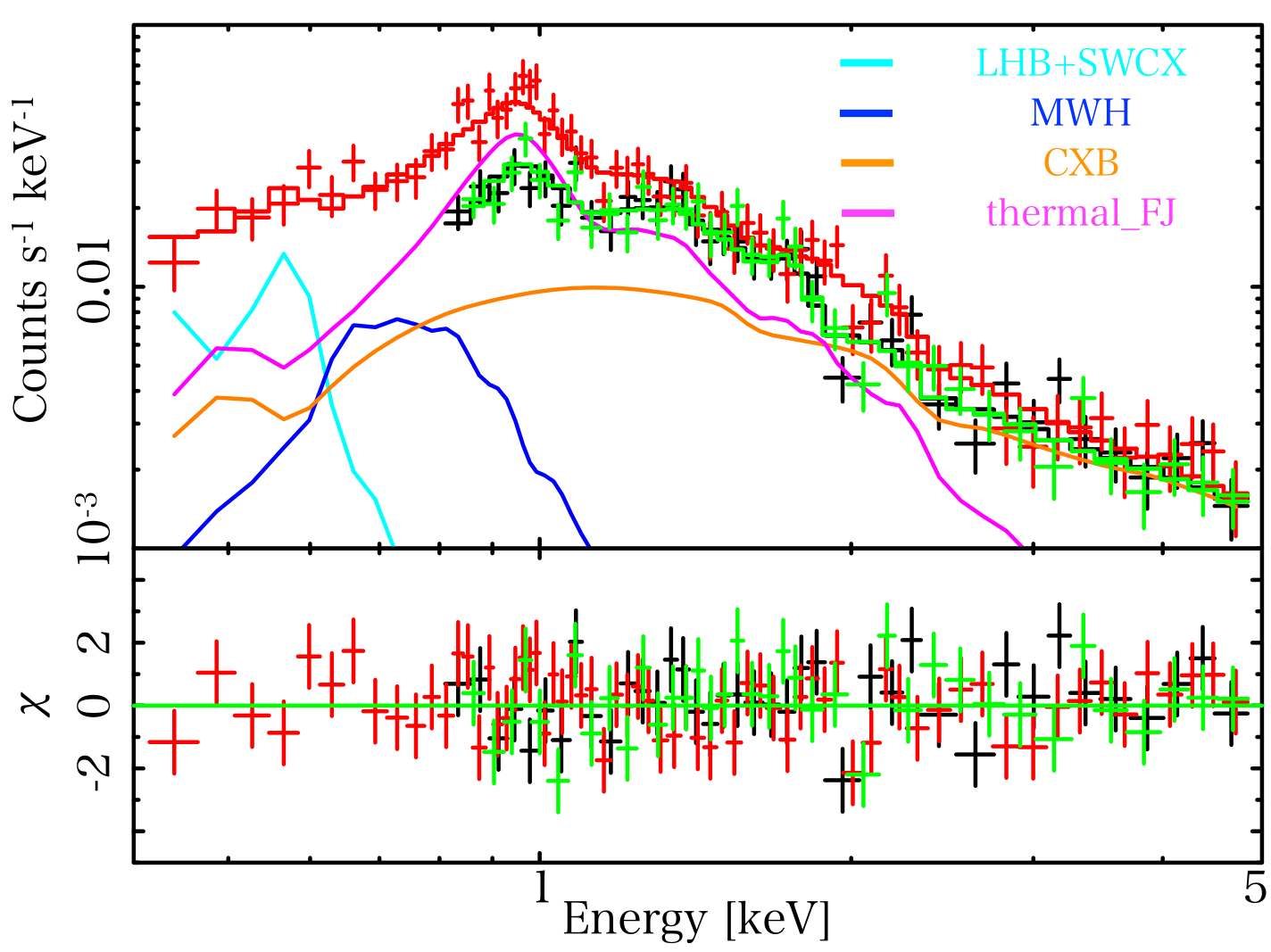}
\vspace{0.2cm}
\end{center}
\end{minipage}
\begin{minipage}{0.5\hsize}
\hspace{0.6cm}
FJ2-A: + absorbed non-thermal plasma 
\begin{center}
\vspace{-3.2cm}
    \includegraphics[bb=0 0 1424 1067, width=10cm]{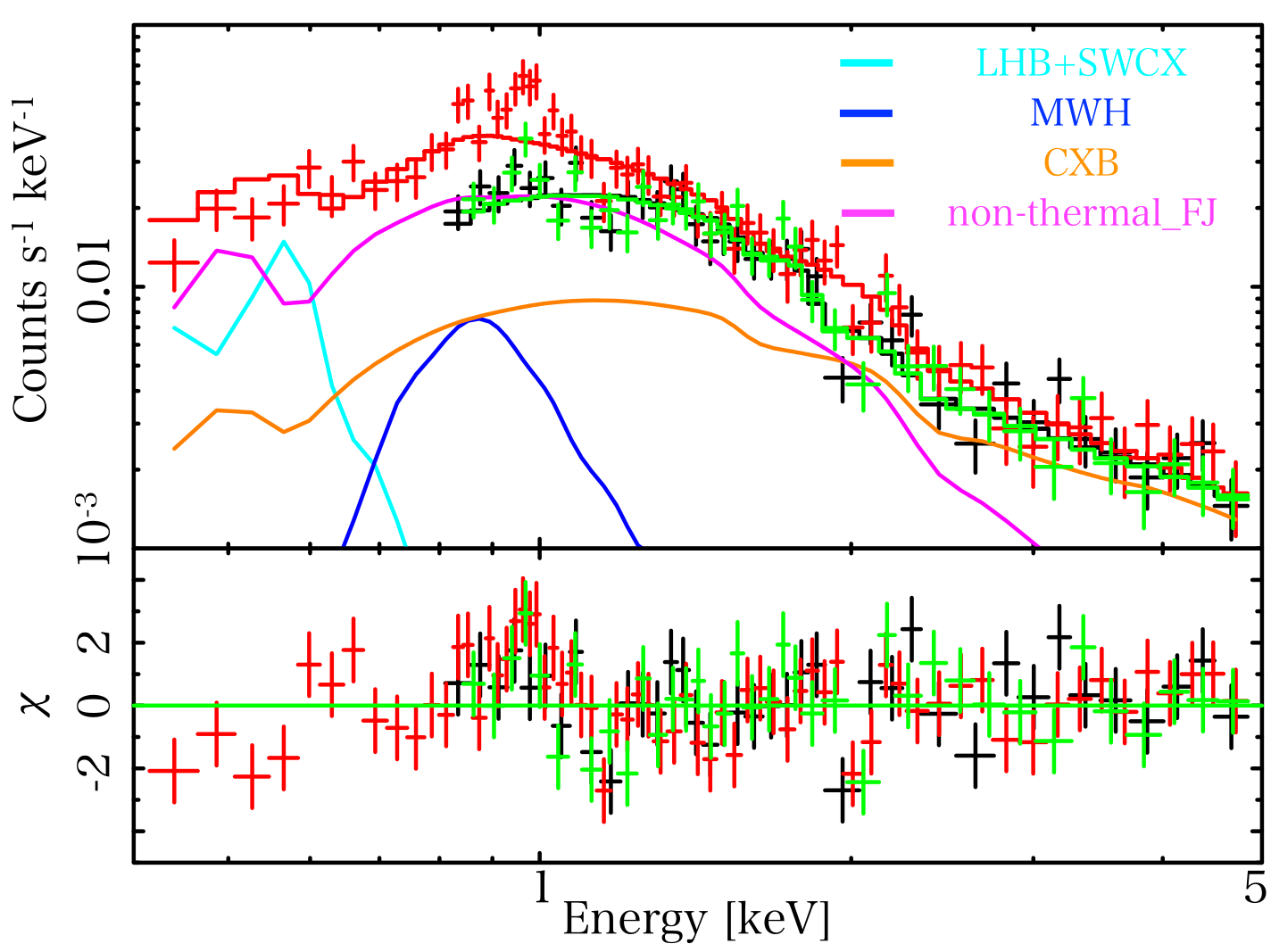}
\vspace{0.2cm}
\end{center}
\end{minipage}
\\
\begin{minipage}{0.5\hsize}
\hspace{0.6cm}
FJ3-A: + absorbed thin-thermal plasma 
\begin{center}
\vspace{-3.2cm}
    \includegraphics[bb=0 0 1424 1067, width=10cm]{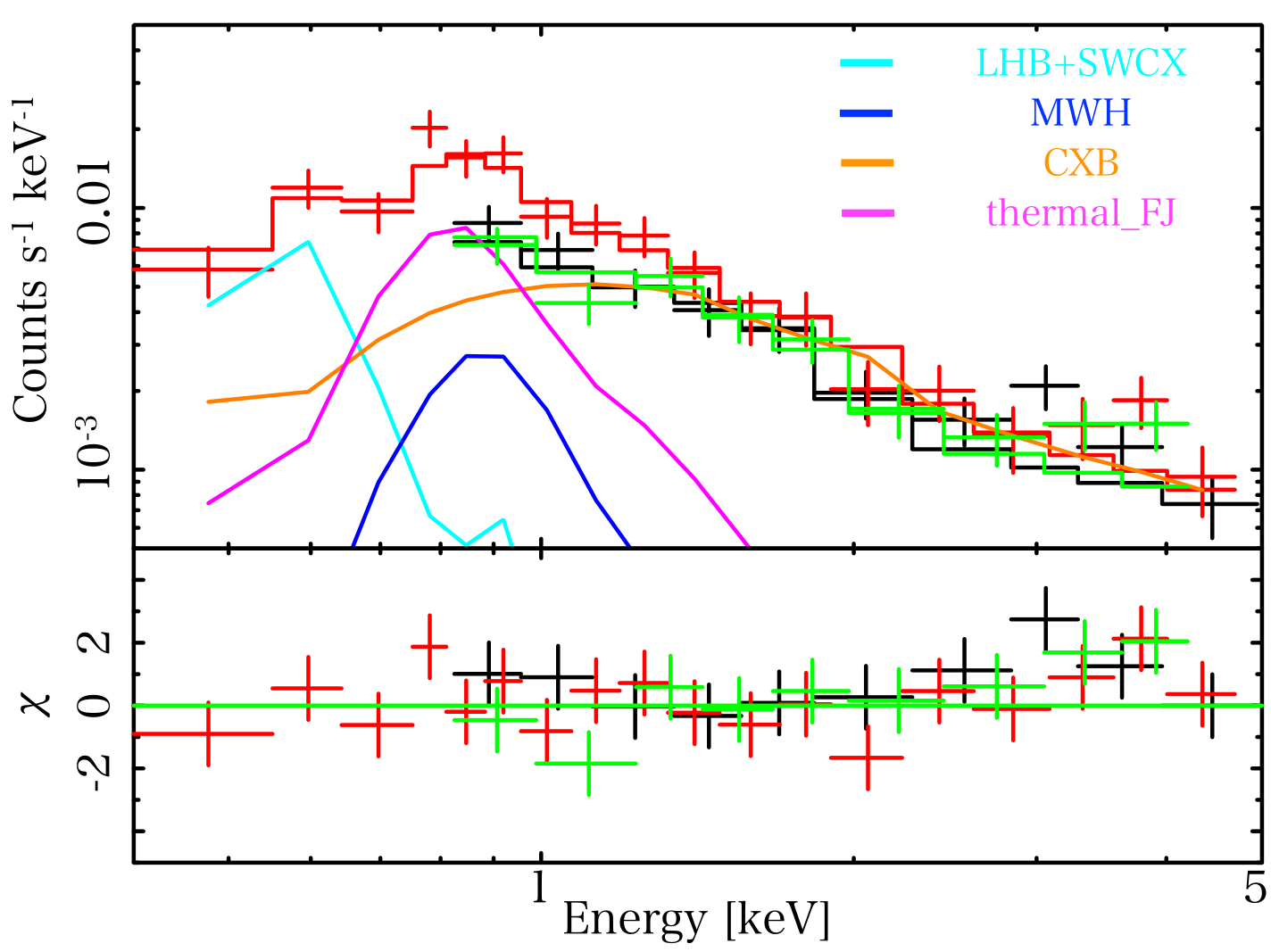}
\vspace{0.2cm}
\end{center}
\end{minipage}
\begin{minipage}{0.5\hsize}
\hspace{0.6cm}
FJ3-A: + absorbed non-thermal plasma 
\begin{center}
\vspace{-3.2cm}
    \includegraphics[bb=0 0 1424 1067, width=10cm]{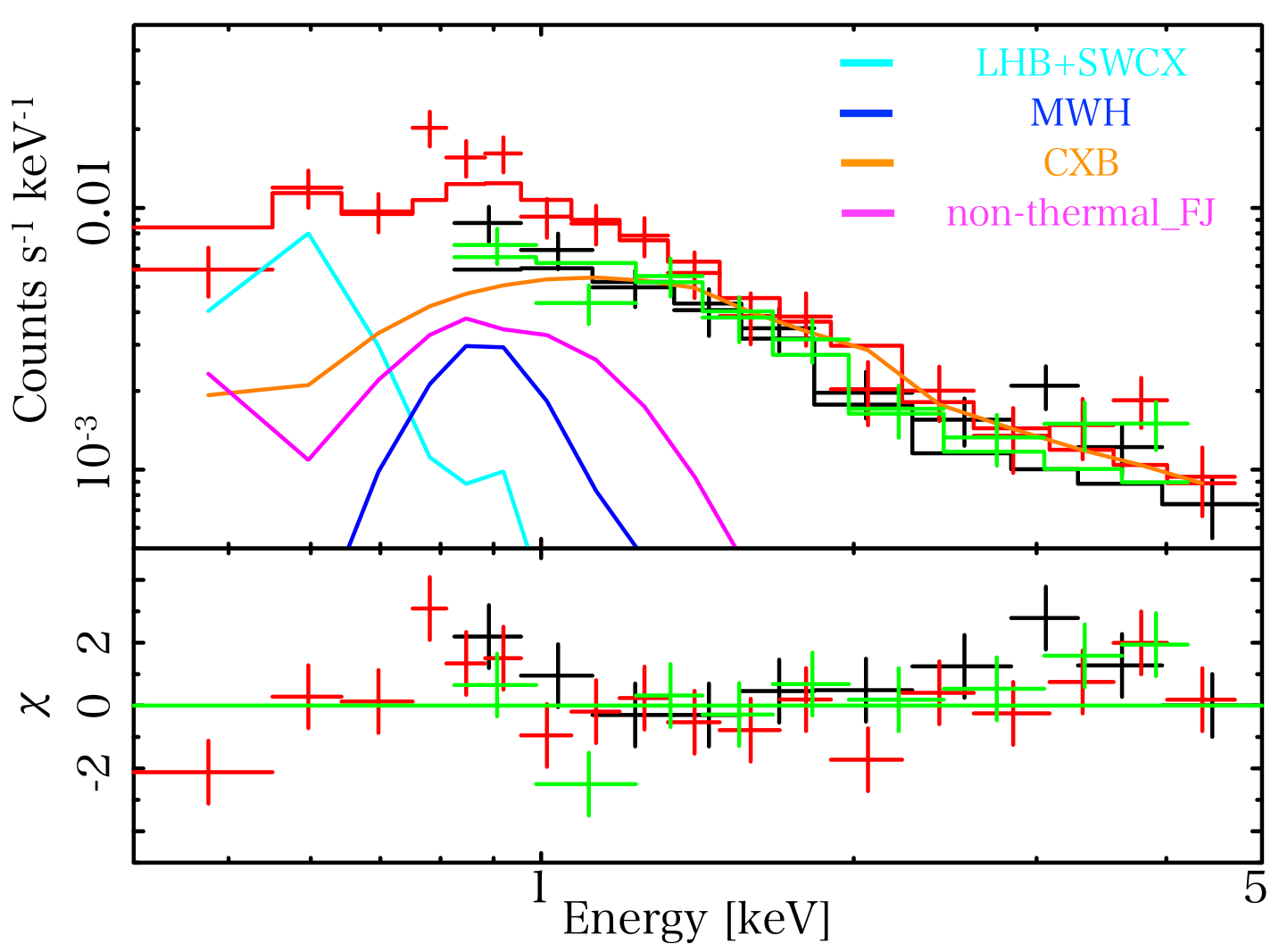}
\vspace{0.2cm}
\end{center}
\end{minipage}
\vspace{-1cm}
\end{tabular}
  \caption{The spectra obtained from FJn-A regions, i.e., FJ1-A, FJ2-A, FJ3-A, FJ4-A and FJ5-A, 
  are shown in increasing order from top to bottom.
  As the excess component, absorbed thin thermal plasma (left) or 
  absorbed power-law (right) component was added for the X-ray background emission.}
  \label{fig:FJn-A1}
\end{figure*}
\begin{figure*}[h!]
\begin{tabular}{cc}
\begin{minipage}{0.5\hsize}
\hspace{0.6cm}
FJ4-A: + absorbed thin-thermal plasma 
\begin{center}
\vspace{-3.2cm}
    \includegraphics[bb=0 0 1424 1067, width=10cm]{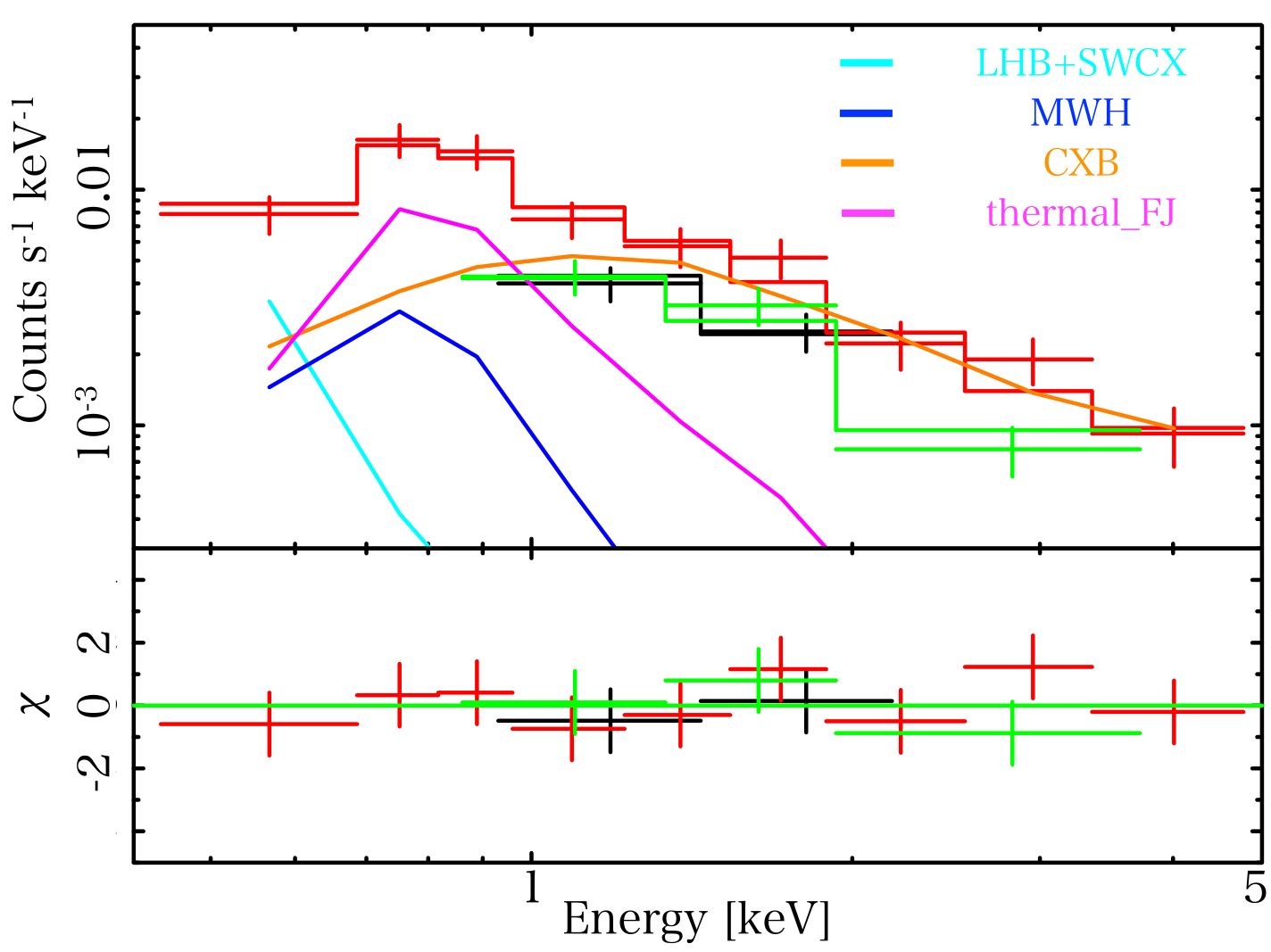}
\vspace{0.2cm}
\end{center}
\end{minipage}
\begin{minipage}{0.5\hsize}
\hspace{0.6cm}
FJ4-A: + absorbed non-thermal plasma 
\begin{center}
\vspace{-3.2cm}
    \includegraphics[bb=0 0 1424 1067, width=10cm]{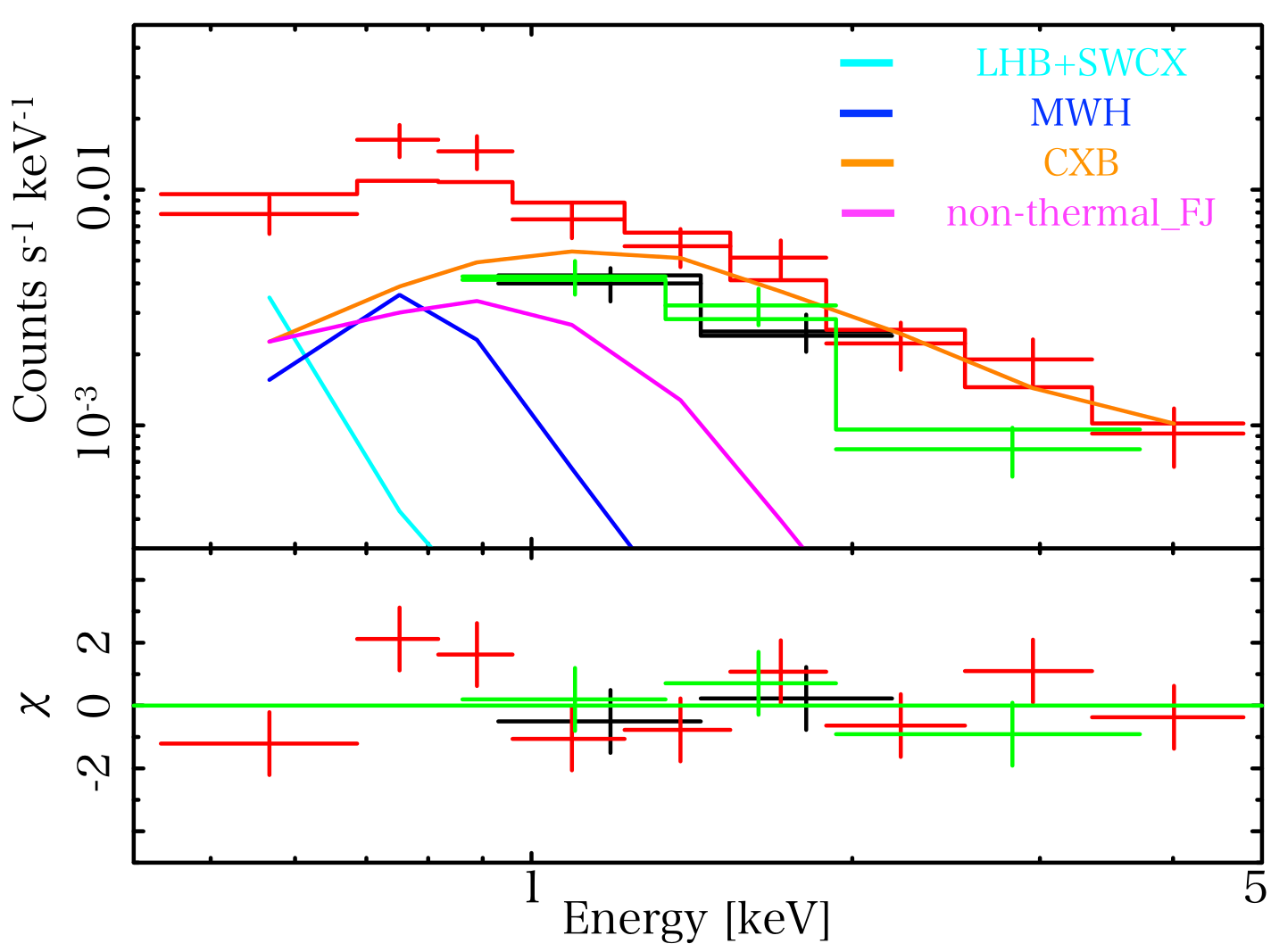}
\vspace{0.2cm}
\end{center}
\end{minipage}
\\
\begin{minipage}{0.5\hsize}
\hspace{0.6cm}
FJ5-A: + absorbed thin-thermal plasma 
\begin{center}
\vspace{-3.2cm}
    \includegraphics[bb=0 0 1424 1067, width=10cm]{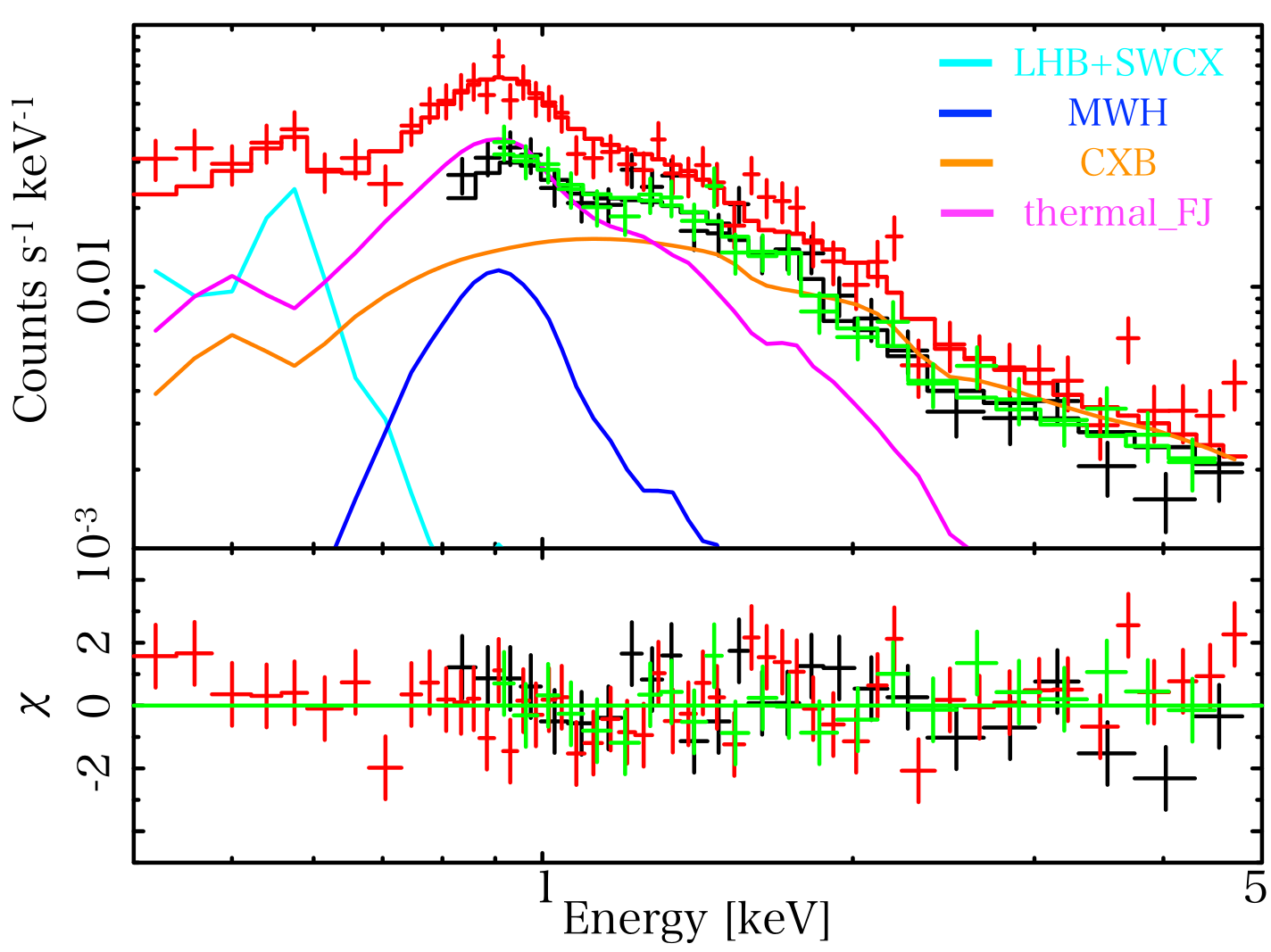}
\vspace{0.2cm}
\end{center}
\end{minipage}
\begin{minipage}{0.5\hsize}
\hspace{0.6cm}
FJ5-A: + absorbed non-thermal plasma 
\begin{center}
\vspace{-3.2cm}
    \includegraphics[bb=0 0 1424 1067, width=10cm]{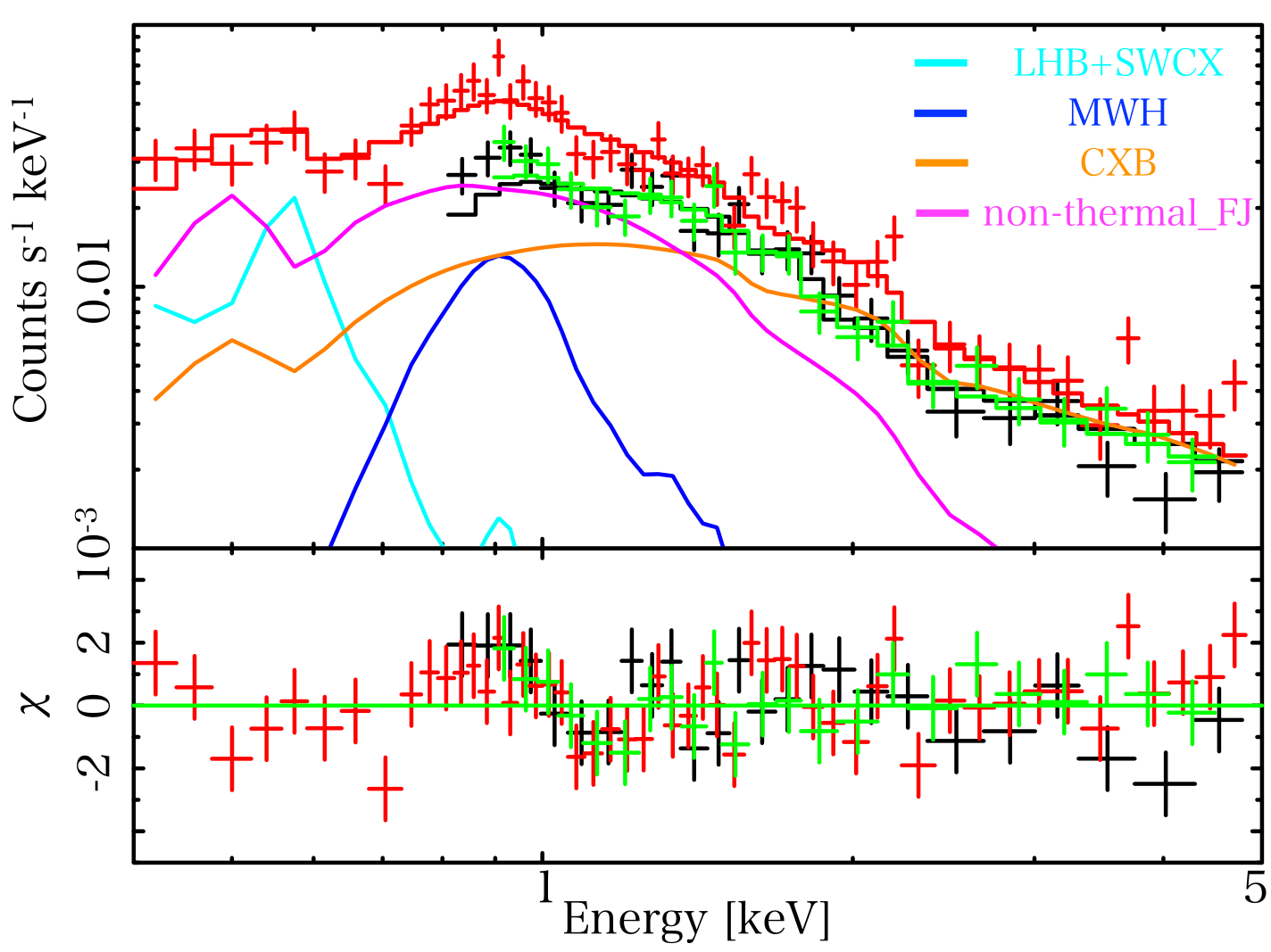}
\vspace{0.2cm}
\end{center}
\end{minipage}
\end{tabular}
\setcounter{figure}{3}
\vspace{-1cm}
  \caption{continued.}
  \label{fig:FJn-A2}
\end{figure*}
\begin{table*}[h!]
  \caption{The results of model fitting of the five FJn-A regions.}
\label{table:FJn-Aenter-result}
  \begin{center}
  \scriptsize
{\tabcolsep=0.6mm
    \begin{tabular}{lcccccc}
\hline\hline
 \multicolumn{7}{c}{Model: ${\it apec_\mathrm{SWCX+LHB}}$ + {\it phabs} $\times$ ({\it apec$_\mathrm{MWH}$} + {\it pow-law$_\mathrm{CXB}$} + {\it phabs$_\mathrm{FJ}$} $\times$ {\it apec$_\mathrm{FJ}$})}\\ \hline
 \multirow{2}*{Region}        & $f_{{\rm bgd}}$$^\ast$                                                         & $kT_\mathrm{SWCX+LHB}$ [keV] & Norm$^\dagger$$_\mathrm{SWCX+LHB}$& $kT_\mathrm{MWH}$ [keV]                             & Norm$^\dagger$$_\mathrm{MWH}$ & S$^\ddagger_{\rm CXB}$  \\ \cline{2-7}
                                                & $f_{{\rm FJ}}$$^\ast$                                                           & N$_H$$_\mathrm{:FJ}$ [$\times$10$^{21}$~cm$^{-2}$]& $kT_\mathrm{FJ}$ [keV]                      & $Z^\ddagger_\mathrm{FJ}$ [$Z_{\odot}$]    & Norm$^\dagger$$_\mathrm{FJ}$& $\chi^2/$d.o.f                \\  \hline
 \multirow{2}*{FJ1-A}      & (1.05$^{+0.09}_{-0.08}$, 1.14$\pm$0.09)                           & 0.09$^{+0.03}_{-0.06}$                    & 55$^{+72000}_{-34}$                                  & 0.22$^{+0.08}_{-0.06}$                                   & 4.2$^{+5.4}_{-2.3}$                                     & 7.5$\pm$0.4 \\ \cline{2-7}
                                                & (0.95$^{+0.10}_{-0.09}$, 0.96$^{+0.11}_{-0.10}$)        & $<$3                                                     & 0.61$^{+0.11}_{-0.10}$                                                             & 1 (fix)                                    & 11$^{+9}_{-6}$                                             & 191/210 \\ \hline
 \multirow{2}*{$\uparrow$} &(1.06$^{+0.09}_{-0.08}$, 1.14$\pm$0.09)                       & 0.09$^{+0.03}_{-0.06}$                    & 55$^{+50000}_{-33}$                                  & 0.23$^{+0.08}_{-0.05}$                                   & 4.1$^{+4.6}_{-1.3}$                                     & 7.5$\pm$0.4 \\ \cline{2-7}
                                                & (0.94$^{+0.10}_{-0.09}$, 0.96$^{+0.11}_{-0.10}$)        & $<$5                                                     & 0.62$^{+0.12}_{-0.11}$                                                             & 0.3 (fix)                                 & 22$^{+22}_{-5}$                                          & 190/210 \\ \hline
 \multirow{2}*{FJ2-A}      & (1.09$\pm$0.08, 1.09$\pm$0.08)                                         & 0.09$^{+0.01}_{-0.02}$                     & 75$^{+170}_{-30}$                                      & 0.28$\pm$0.05                                                  & 3.7$^{+2.3}_{-1.1}$                                     & 8.0$\pm$0.4 \\ \cline{2-7}
                                                & (1.07$\pm$0.06, 0.98$\pm$0.06)                                     & $<$2                                                     & 1.3$\pm$0.1                                                                                 & 0.2$\pm$0.1                       & 77$^{+15}_{-12}$                                        & 343/319 \\ \hline
 \multirow{2}*{FJ3-A}     & (1.11$^{+0.09}_{-0.08}$, 0.90$^{+0.08}_{-0.07}$)             & 0.12$^{+0.03}_{-0.02}$                     & 28$^{+33}_{-13}$                                        & 0.79$\pm$0.13                                                  & 1.1$\pm$0.3                                                 & 7.8$^{+0.2}_{-0.5}$ \\ \cline{2-7}
                                               & (1.13$\pm$0.12, 1.04$^{+0.15}_{-0.14}$)                       & $<$2                                                     & 0.76$^{+0.15}_{-0.16}$                                                              & 1 (fix)                                    & 4.7$^{+7.2}_{-1.4}$                                     & 176/146 \\ \hline
 \multirow{2}*{$\uparrow$}& (1.12$^{+0.09}_{-0.08}$, 0.90$^{+0.08}_{-0.07}$)        & 0.12$\pm$0.02                                   & 27$^{+29}_{-12}$                                        & 0.79$^{+0.12}_{-0.13}$                                    & 1.1$^{+0.2}_{-0.3}$                                     & 7.8$^{+0.3}_{-0.5}$ \\ \cline{2-7}
                                               &(1.11$^{+0.14}_{-0.13}$, 1.03$^{+0.16}_{-0.15}$)         & $<$2                                                     & 0.76$^{+0.14}_{-0.15}$                                                              & 0.3 (fix)                                 & 12$^{+12}_{-4}$                                          & 178/146 \\ \hline
 \multirow{2}*{FJ4-A}     & (1.31$^{+0.14}_{-0.13}$, 1.06$^{+0.12}_{-0.11}$)            & 0.09$^{+0.04}_{-0.05}$                     & 60$^{+800}_{-43}$                                      & 0.28$^{+0.12}_{-0.09}$                                    & 3.2$^{+5.7}_{-1.6}$                                     & 8.7$^{+0.5}_{-0.7}$ \\ \cline{2-7}
                                               & (1.33$^{+0.23}_{-0.21}$, 0.92$^{+0.19}_{-0.18}$)        & $<$2                                                     & 0.69$^{+0.24}_{-0.39}$                                                               & 1 (fix)                                   & 6.2$^{+48}_{-3}$                                          & 111/92 \\ \hline
 \multirow{2}*{$\uparrow$}& (1.32$^{+0.14}_{-0.13}$, 1.06$^{+0.12}_{-0.11}$)        & 0.09$\pm$0.04                                   & 60$^{+790}_{-42}$                                     & 0.28$^{+0.12}_{-0.08}$                                     & 3.2$^{+5.4}_{-1.5}$                                     & 8.6$\pm$0.7 \\ \cline{2-7}
                                               &(1.32$^{+0.23}_{-0.21}$, 0.91$^{+0.19}_{-0.18}$)         & $<$3                                                     & 0.69$^{+0.23}_{-0.45}$                                                               & 0.3 (fix)                                & 15$^{+110}_{-6}$                                        & 112/92 \\ \hline
 \multirow{2}*{FJ5-A}     & (1.25$^{+0.09}_{-0.08}$, 0.93$^{+0.08}_{-0.07}$)            & 0.11$\pm$0.02                                   & 39$^{+57}_{-16}$                                         & 0.88$^{+0.09}_{-0.10}$                                    & 2.2$\pm$0.5                                                 & 12$^{+0.7}_{-0.6}$ \\ \cline{2-7}
                                               & (1.16$\pm$0.08, 0.91$\pm$0.07)                                     & $<$1                                                     & 0.99$^{+0.08}_{-0.09}$                               & $<$0.1                                                                 & 86$^{+21}_{-12}$                                        & 278/224 \\
 \hline\hline
 \multicolumn{6}{c}{Model: ${\it apec_\mathrm{SWCX+LHB}}$ + {\it phabs} $\times$ ({\it apec$_\mathrm{MWH}$} + {\it pow-law$_\mathrm{CXB}$} + {\it phabs$_\mathrm{FJ}$} $\times$ {\it pow$_\mathrm{FJ}$})}\\ 
\hline
 \multirow{2}*{Region}    & $f_{{\rm bgd}}$$^\ast$                                                    & $kT_\mathrm{SWCX+LHB}$ [keV] & Norm$^\dagger$$_\mathrm{SWCX+LHB}$ & $kT_\mathrm{MWH}$ [keV] & Norm$^\dagger$$_\mathrm{MWH}$ & $S^\ddagger_\mathrm{CXB}$  \\ \cline{2-7}
                                            & $f_{{\rm FJ}}$$^\ast$                                                      & N$_H$$_\mathrm{:FJ}$ [$\times$10$^{21}$~cm$^{-2}$] & $\Gamma_\mathrm{FJ}$                          & $z$$_\mathrm{FJ}$ & $\chi^2/$d.o.f                              &                           \\ \hline
 \multirow{2}*{FJ1-A}  & (1.05$\pm$0.09, 1.13$\pm$0.09)                                     & 0.09$^{+0.03}_{-0.06}$                    & 50$^{+31000}_{-33}$                                        & 0.23$^{+0.07}_{-0.05}$        & 4.6$^{+6.3}_{-2.5}$                               & 7.4$\pm$0.4                                 \\ \cline{2-7}
                                            & (0.94$^{+0.11}_{-0.10}$, 0.95$^{+0.11}_{-0.10}$)   & 6$\pm$4                                                                                     & $>$4.7                                                         & 0 (fix)                      & 208/210 \\ \hline
 \multirow{2}*{FJ2-A}  & (1.14$^{+0.09}_{-0.08}$, 1.09$^{+0.09}_{-0.08}$)        & 0.10$^{+0.03}_{-0.01}$                    & 54$^{+28}_{-33}$                                                & 0.76$^{+0.07}_{-0.07}$      & 1.6$\pm$0.3                                            & 7.4$\pm$0.5   \\ \cline{2-7}
                                            & (1.03$^{+0.07}_{-0.06}$, 0.97$^{+0.07}_{-0.06}$)   & 2$\pm$1                                                                                     & 3.3$^{+0.4}_{-0.3}$                                                                          & 0 (fix)                      & 411/320 \\ \hline
 \multirow{2}*{FJ3-A} & (1.10$^{+0.09}_{-0.08}$, 0.89$^{+0.08}_{-0.07}$)         & 0.14$^{+0.01}_{-0.03}$                    & 18$^{+13}_{-4}$                                                  & 0.79$\pm$0.13                      & 1.2$^{+0.4}_{-0.3}$ & 7.8$\pm$0.5     \\ \cline{2-7}
                                           & (1.18$^{+0.15}_{-0.14}$, 1.07$\pm$0.15)                   & $<$11                                                  & $>$2.5                                                                                                 & 0 (fix)                      & 199/146 \\ \hline
 \multirow{2}*{FJ4-A} & (1.30$^{+0.14}_{-0.13}$, 1.06$^{+0.12}_{-0.11}$)         & 0.09$\pm$0.03                                  & 61$^{+550}_{-43}$                                              & 0.29$^{+0.11}_{-0.07}$       & 3.3$^{+2.9}_{-1.5}$                               & 8.6$\pm$0.7  \\ \cline{2-7}
                                           & (1.39$^{+0.25}_{-0.23}$, 0.93$^{+0.21}_{-0.19}$)    & $<$12                                                  & $>$2.7                                                                                                 & 0 (fix)                      & 121/92 \\ \hline
 \multirow{2}*{FJ5-A} & (1.40$^{+0.11}_{-0.10}$, 1.03$^{+0.09}_{-0.08}$)        & 0.11$^{+0.04}_{-0.02}$                    & 35$^{+47}_{-17}$                                                 & 0.87$^{+0.07}_{-0.06}$       & 2.5$^{+0.5}_{-0.4}$                               & 11$\pm$1      \\ \cline{2-7}
                                           & (1.14$^{+0.09}_{-0.08}$, 0.90$^{+0.08}_{-0.07}$)    & $<$3                                                    & 3.6$^{+0.7}_{-0.5}$                                                                           & 0 (fix)                      & 305/225 \\ \hline
 \hline
    \end{tabular}}
  \end{center}
\begin{flushleft}
\footnotesize{
$\ast$ Constant factors of XIS1 (left) and XIS3 (right) detectors relative to the XIS0 detector. \\
$\dagger$ Normalization of the $apec$ models divided by a solid angle $\Omega$, 
 assumed in a uniform-sky ARF calculation (20$'$ radius), 
 i.e. $\it{Norm} = (1/\Omega) \int n_e n_H dV / (4\pi(1+z)^2)D^2_A)$ cm$^{-5}$ sr$^{-1}$ 
in unit of 10$^{-14}$, where $D_A$ is the angular diameter distance.\\
 $\ddagger$ Surface brightness of the {\it power-law } model in the unit of $10^{-8}$ erg cm$^{-2}$ s$^{-1}$ sr$^{-1}$ in 2--10 keV.\\
 $\ddagger$ Abundance table tabulated in \citet{solar-abundance-table-anders} was adopted.}
 \end{flushleft}
\end{table*}
\newpage
\begin{figure*}[h!]
\begin{tabular}{cc}
\begin{minipage}{0.5\hsize}
\hspace{0.6cm}
FJ1-B: + absorbed thin-thermal plasma 
\begin{center}
\vspace{-3.2cm}
    \includegraphics[bb=0 0 1424 1067, width=10cm]{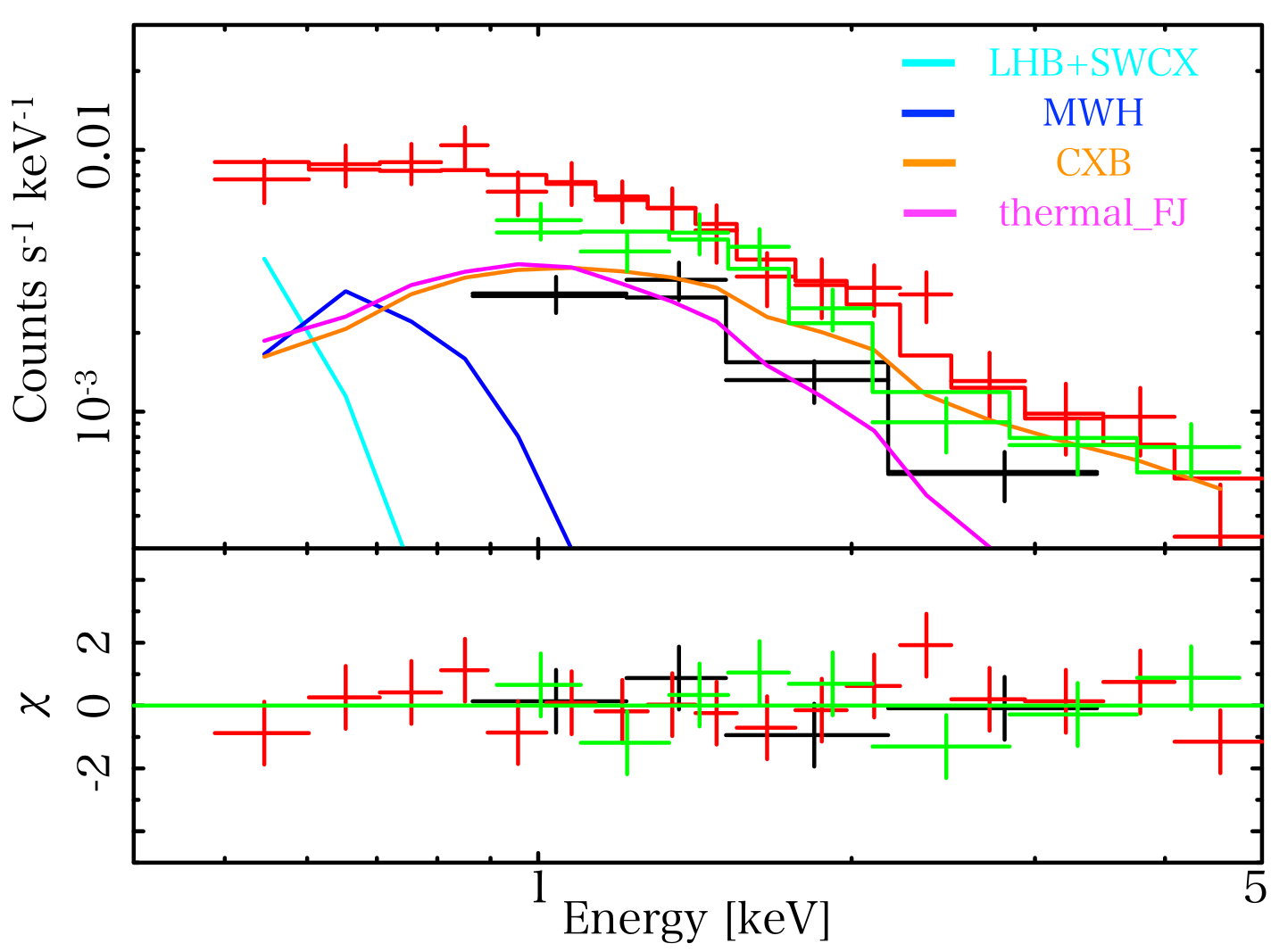}
\vspace{0.2cm}
\end{center}
\end{minipage}
\begin{minipage}{0.5\hsize}
\hspace{0.6cm}
FJ1-B: + absorbed non-thermal plasma 
\begin{center}
\vspace{-3.2cm}
    \includegraphics[bb=0 0 1424 1067, width=10cm]{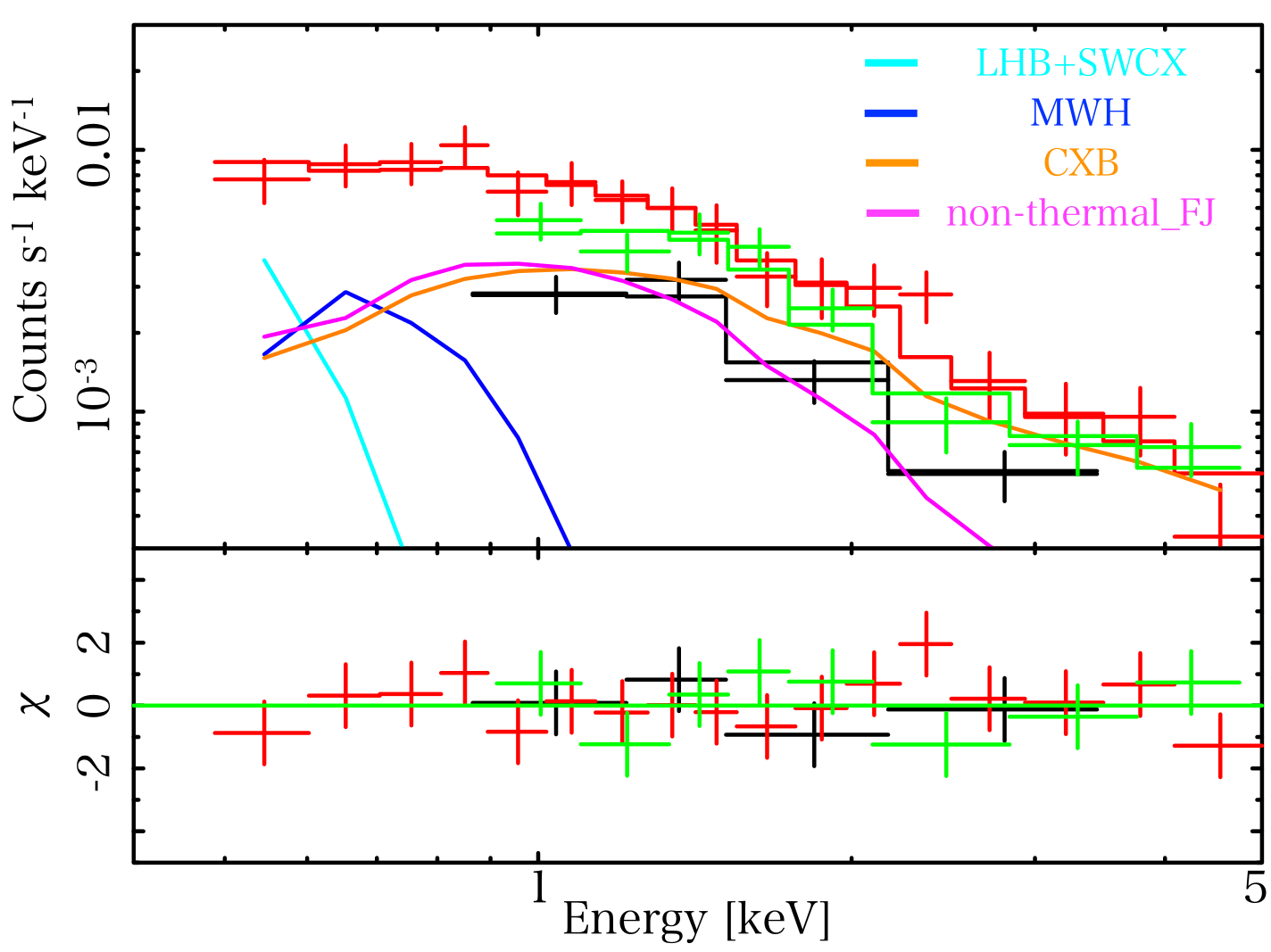}
\vspace{0.2cm}
\end{center}
\end{minipage}
\\
\begin{minipage}{0.5\hsize}
\hspace{0.6cm}
FJ2-B: + absorbed thin-thermal plasma 
\begin{center}
\vspace{-3.2cm}
    \includegraphics[bb=0 0 1424 1067, width=10cm]{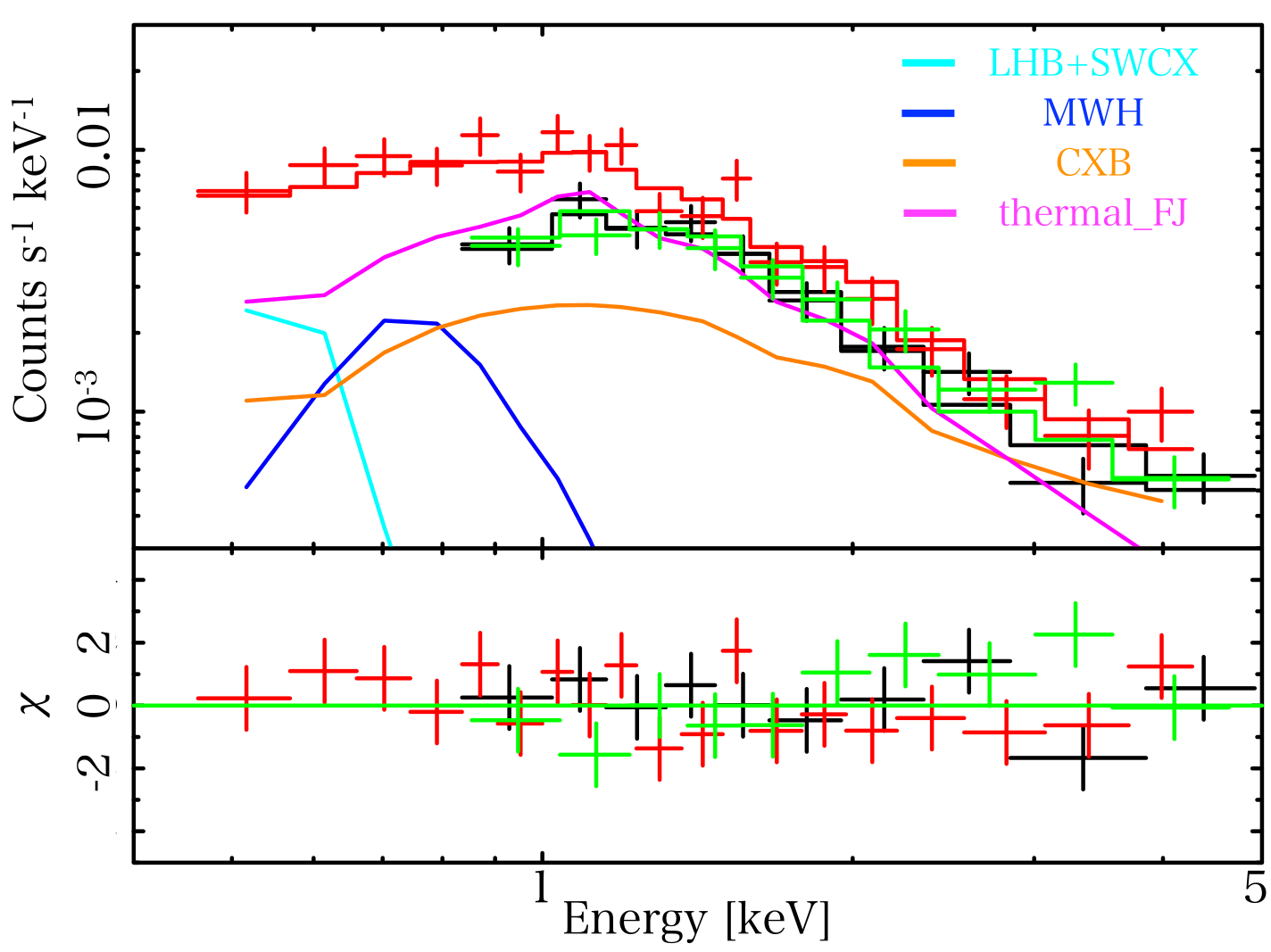}
\vspace{0.2cm}
\end{center}
\end{minipage}
\begin{minipage}{0.5\hsize}
\hspace{0.6cm}
FJ2-B: + absorbed non-thermal plasma 
\begin{center}
\vspace{-3.2cm}
    \includegraphics[bb=0 0 1424 1067, width=10cm]{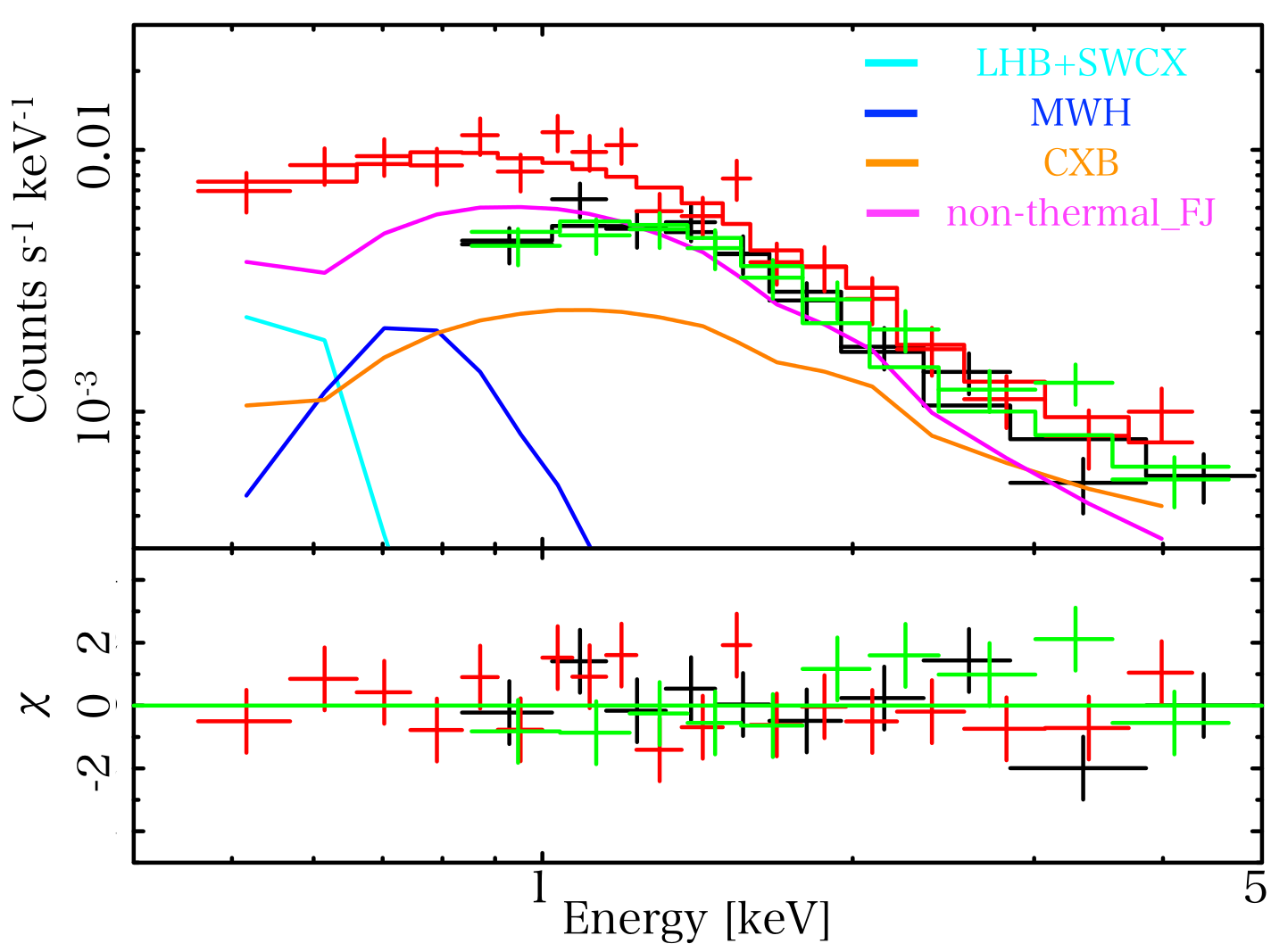}
\vspace{0.2cm}
\end{center}
\end{minipage}
\\
\begin{minipage}{0.5\hsize}
\hspace{0.6cm}
FJ3-B: + absorbed thin-thermal plasma 
\begin{center}
\vspace{-3.2cm}
    \includegraphics[bb=0 0 1424 1067, width=10cm]{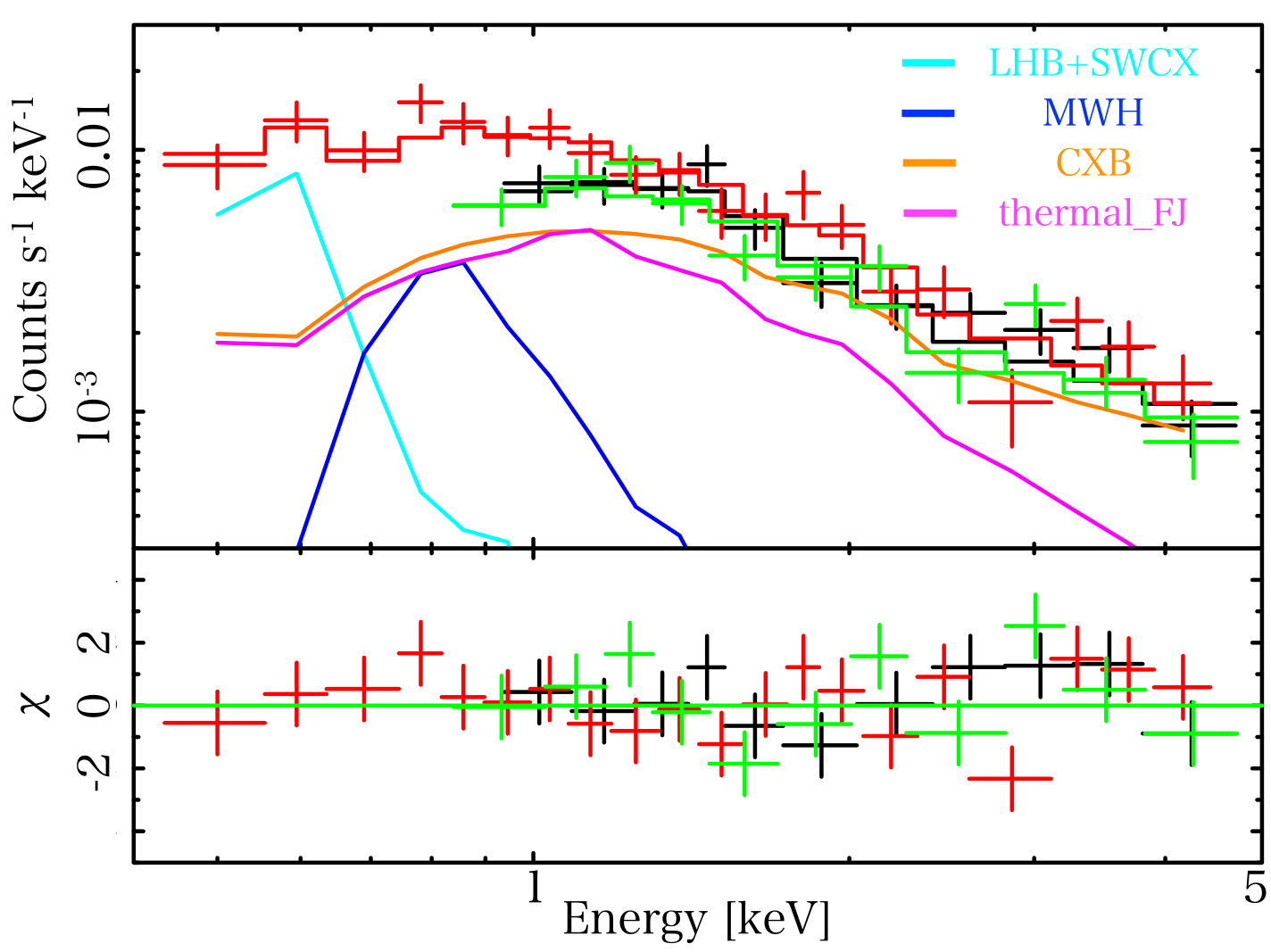}
\vspace{0.2cm}
\end{center}
\end{minipage}
\begin{minipage}{0.5\hsize}
\hspace{0.6cm}
FJ3-B: + absorbed non-thermal plasma 
\begin{center}
\vspace{-3.2cm}
    \includegraphics[bb=0 0 1424 1067, width=10cm]{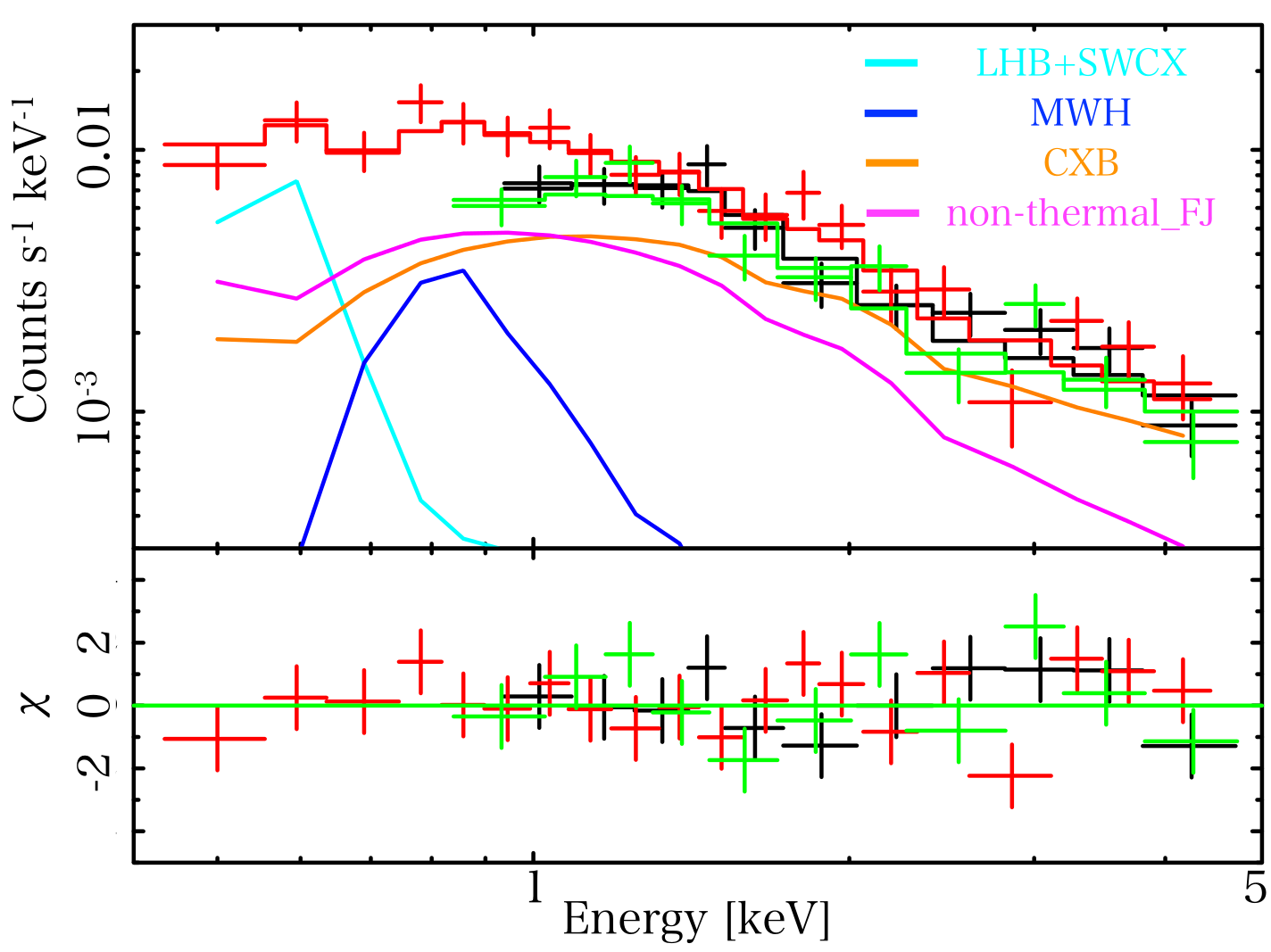}
\vspace{0.2cm}
\end{center}
\end{minipage}
\end{tabular}
\vspace{-1cm}
  \caption{The same as Figure \ref{fig:FJn-A1} but spectra obtained from FJn-B regions.}
  \label{fig:FJn-B1}
\end{figure*}
\begin{figure*}[h!]
\begin{tabular}{cc}
\begin{minipage}{0.5\hsize}
\hspace{0.6cm}
FJ4-B: + absorbed thin-thermal plasma 
\begin{center}
\vspace{-3.2cm}
    \includegraphics[bb=0 0 1424 1067, width=10cm]{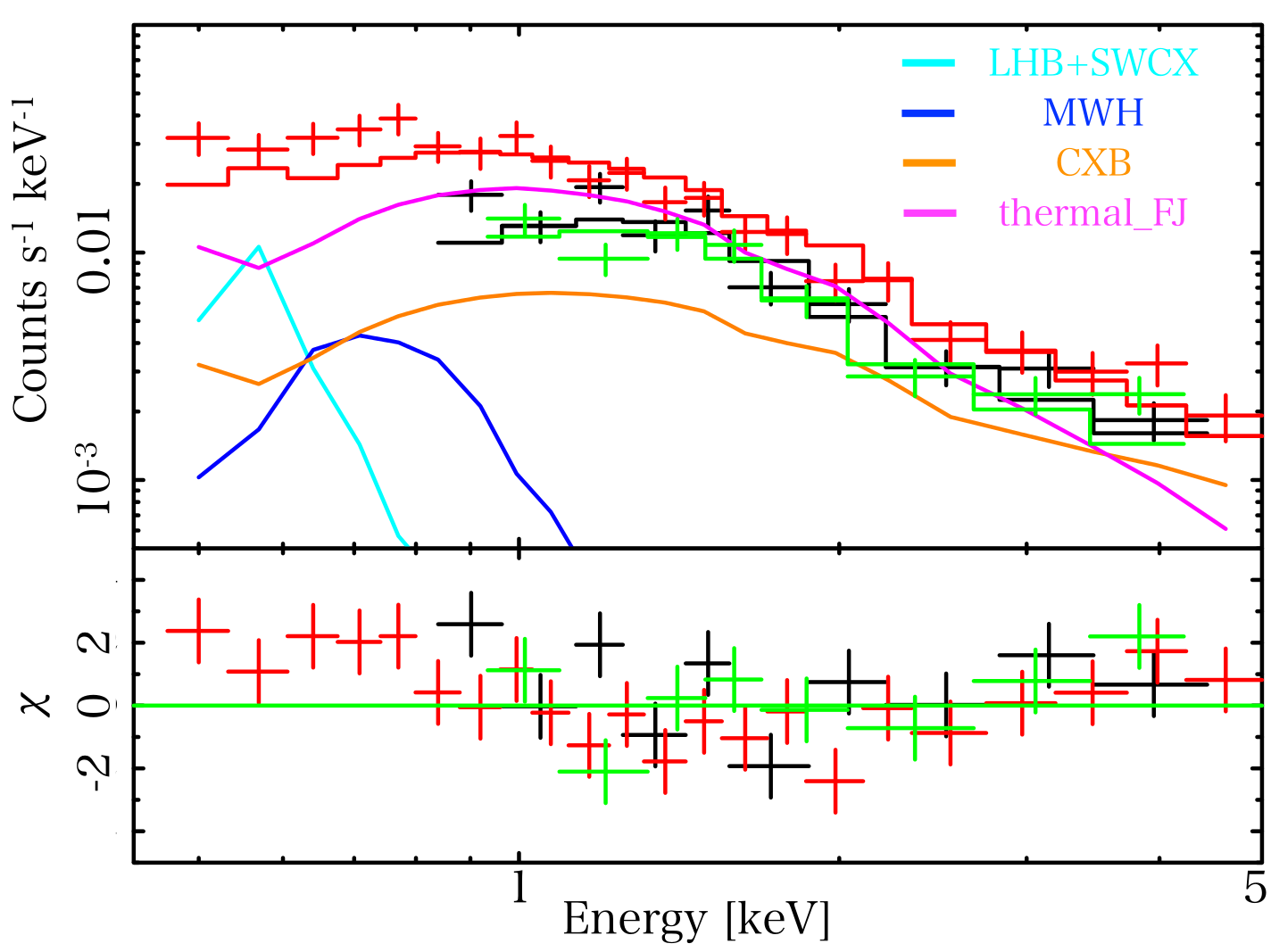}
\vspace{0.2cm}
\end{center}
\end{minipage}
\begin{minipage}{0.5\hsize}
\hspace{0.6cm}
FJ4-B: + absorbed non-thermal plasma 
\begin{center}
\vspace{-3.2cm}
    \includegraphics[bb=0 0 1424 1067, width=10cm]{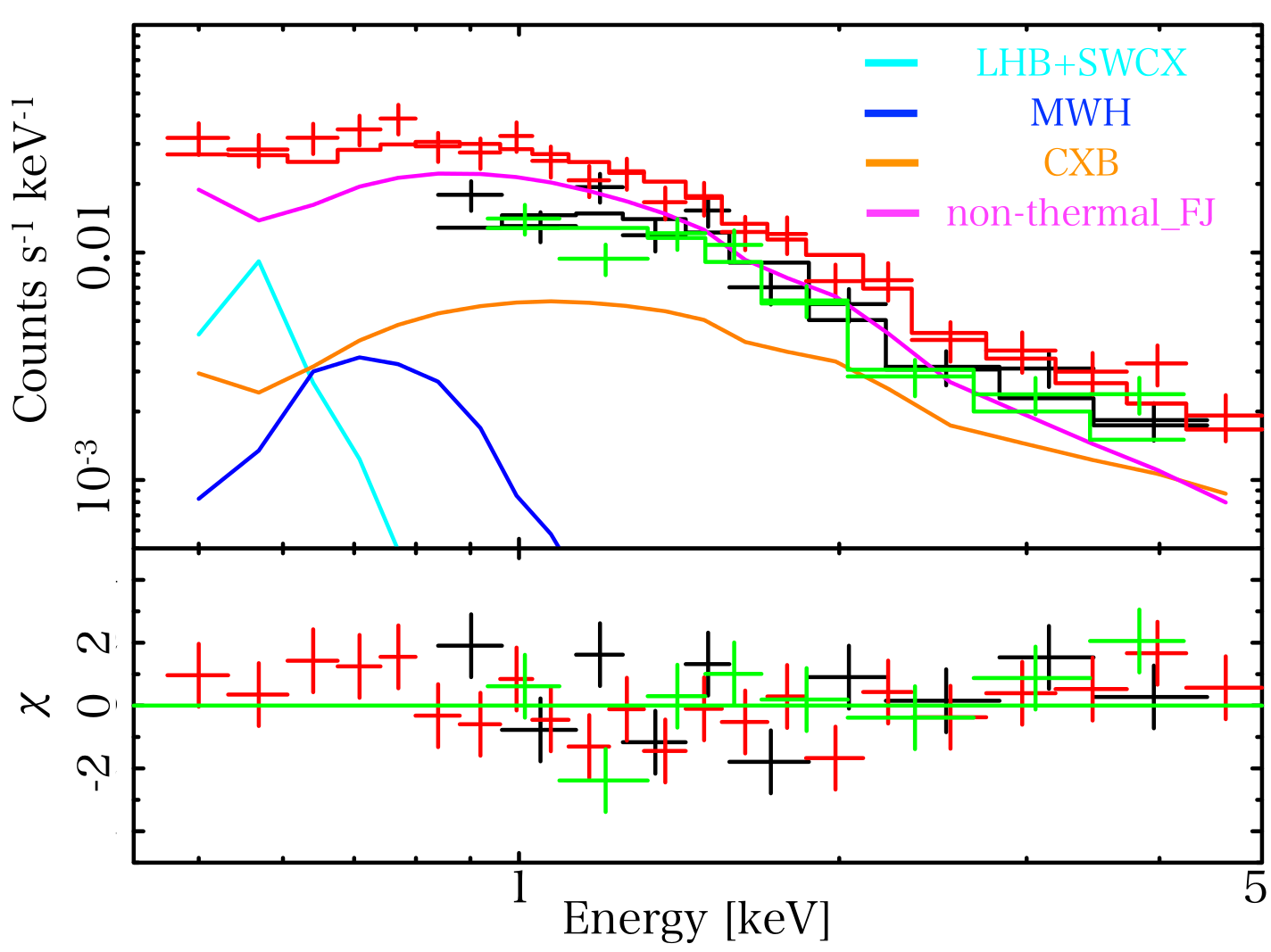}
\vspace{0.2cm}
\end{center}
\end{minipage}
\\
\begin{minipage}{0.5\hsize}
\hspace{0.6cm}
FJ5-B: + absorbed thin-thermal plasma 
\begin{center}
\vspace{-3.2cm}
    \includegraphics[bb=0 0 1424 1067, width=10cm]{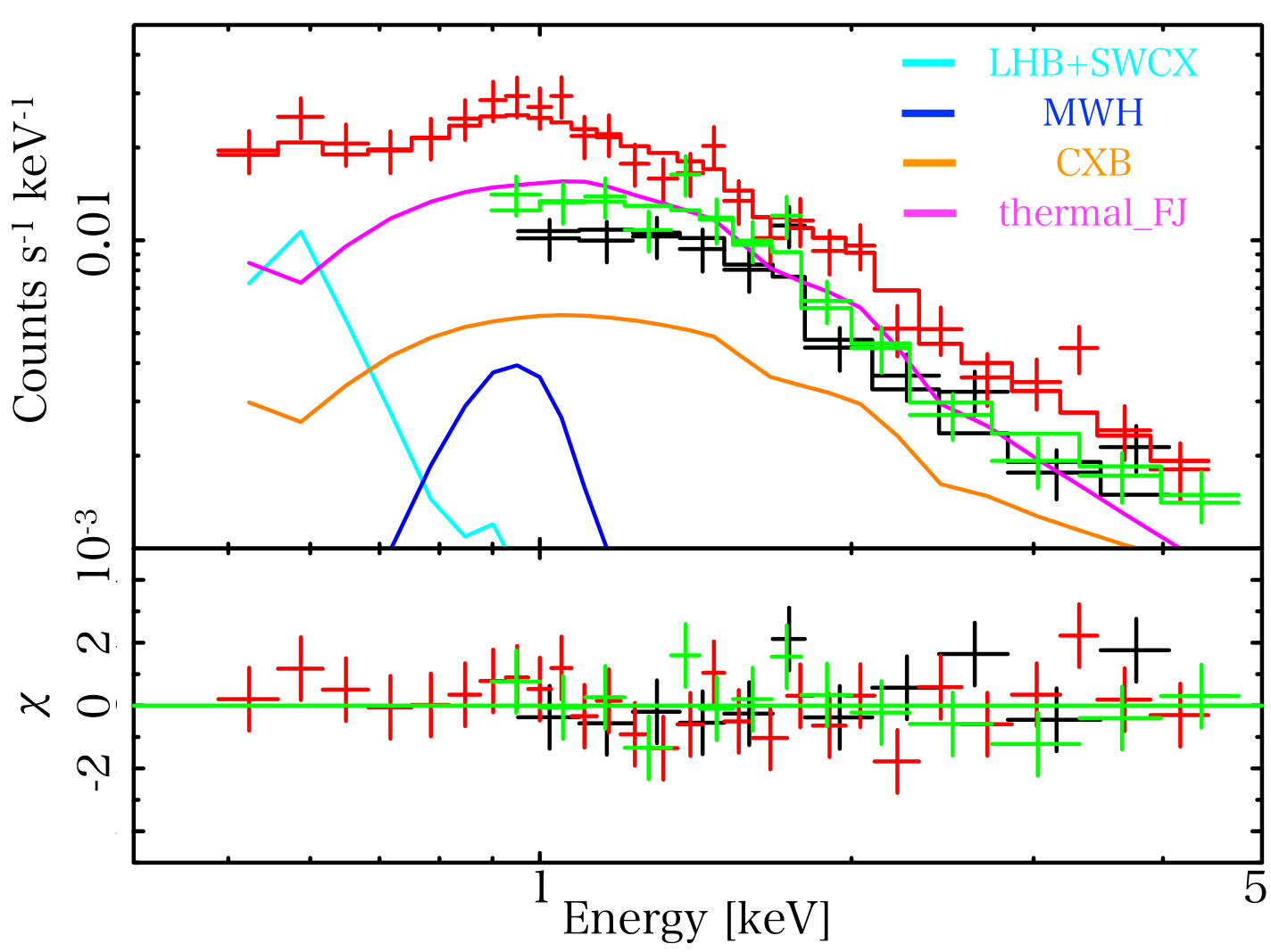}
\vspace{0.2cm}
\end{center}
\end{minipage}
\begin{minipage}{0.5\hsize}
\hspace{0.6cm}
FJ5-B: + absorbed non-thermal plasma 
\begin{center}
\vspace{-3.2cm}
    \includegraphics[bb=0 0 1424 1067, width=10cm]{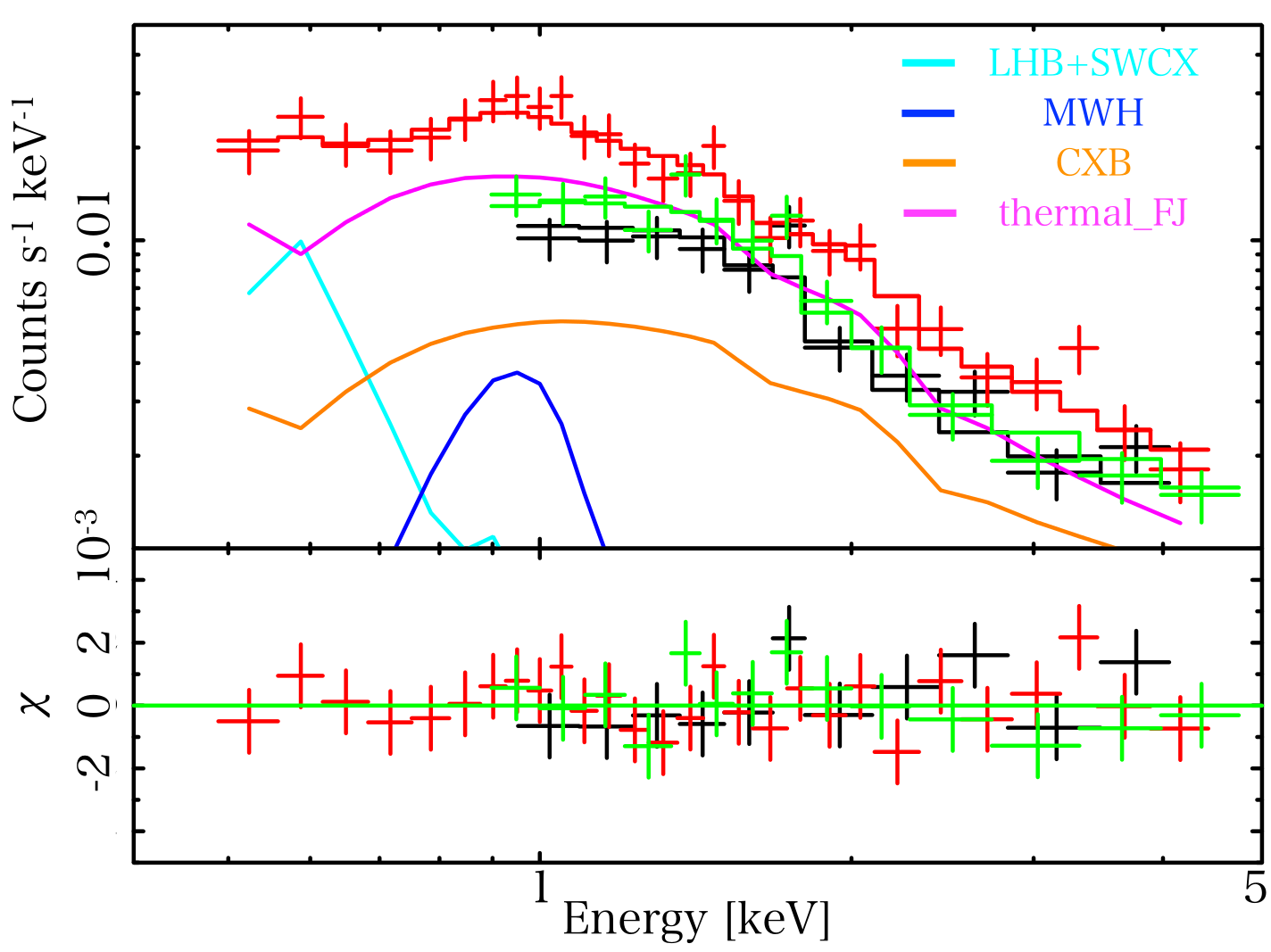}
\vspace{0.2cm}
\end{center}
\end{minipage}
\end{tabular}
\setcounter{figure}{4}
\vspace{-1cm}
  \caption{continued.}
  \label{fig:FJn-B2}
\end{figure*}
\begin{table*}[h!]
\vspace{-0.7cm}
  \caption{The results of model fitting of the five FJn-B regions.}
\label{table:FJn-B-result}
  \begin{center}
  \scriptsize
{\tabcolsep=0.6mm
    \begin{tabular}{lccccccc}
\hline\hline
 \multicolumn{7}{c}{Model: ${\it apec_\mathrm{SWCX+LHB}}$ + {\it phabs} $\times$ ({\it apec$_\mathrm{MWH}$} + {\it pow-law$_\mathrm{CXB}$} + {\it phabs$_\mathrm{FJ}$} $\times$ {\it apec$_\mathrm{FJ}$})}\\ \hline
 \multirow{2}*{Region}        & $f_{{\rm bgd}}$$^\ast$                                                         & $kT_\mathrm{SWCX+LHB}$ [keV]                  & Norm$^\dagger$$_\mathrm{SWCX+LHB}$& $kT_\mathrm{MWH}$ [keV]  & Norm$^\dagger$$_\mathrm{MWH}$ & S$^\ddagger_{\rm CXB}$ &   \\ \cline{2-8}
                                                & $f_{{\rm FJ}}$$^\ast$                                                           & N$_H$$_\mathrm{:FJ}$ [$\times$10$^{21}$~cm$^{-2}$]         & $kT_\mathrm{FJ}$ [keV]                                  & $z$$_\mathrm{FJ}$  & $Z^\ddagger_\mathrm{FJ}$ [$Z_{\odot}$]    & Norm$^\dagger$$_\mathrm{FJ}$& $\chi^2/$d.o.f   \\  \hline
 \multirow{2}*{FJ1-B}           & (1.05$\pm$0.09, 1.12$^{+0.10}_{-0.09}$)                      & 0.09$^{+0.03}_{-0.06}$                                      & 57$^{+520}_{-36}$                                           & 0.23$^{+0.08}_{-0.05}$        & 4.3$^{+6.4}_{-2.2}$                               & 7.6$\pm$0.5 \\ \cline{2-8}
                                                & (1.31$^{+0.25}_{-0.19}$, 1.23$^{+0.25}_{-0.16}$)        & $<$8                                                                                                   & 1.4$^{+1.2}_{-0.9}$                                           & 0 (fix)                      & $<$0.4                                                                 & 38$^{+150}_{-13}$ & 149/186 \\ \hline
 \multirow{2}*{FJ2-B}           & (1.11$^{+0.09}_{-0.08}$, 1.10$^{+0.04}_{-0.08}$)        & 0.07$\pm$0.02                                                    & 220$^{+810}_{-140}$                                      & 0.30$^{+0.09}_{-0.05}$        & 3.5$^{+1.7}_{-1.1}$                               & 7.8$\pm$0.5 \\ \cline{2-8}
                                                & (1.20$^{+0.14}_{-0.12}$, 1.00$^{+0.14}_{-0.12}$)        & $<$1                                                                                                   & 2.1$^{+0.7}_{-0.4}$                                           & 0 (fix)                      & $<$0.4                                                                 & 71$^{+20}_{-18}$ & 198/182 \\ \hline
 \multirow{2}*{FJ3-B}           & (1.13$^{+0.10}_{-0.09}$, 0.92$^{+0.09}_{-0.08}$)        & 0.12$^{+0.03}_{-0.02}$                                     & 29$^{+42}_{-14}$                                             & 0.79$^{+0.13}_{-0.11}$        & 1.2$\pm$0.3                                            & 7.5$\pm$0.5  \\ \cline{2-8}
                                               & (0.88$^{+0.11}_{-0.10}$, 0.90$^{+0.12}_{-0.11}$)         & $<$1                                                                                                   & 2.6$^{+2.3}_{-1.0}$                                           & 0 (fix)                      & $<$1.4                                                                 & 31$^{+14}_{-13}$ & 183/150 \\ \hline
 \multirow{2}*{FJ4-B}          &  (1.18$^{+0.13}_{-0.12}$, 1.00$^{+0.11}_{-0.10}$)       & 0.09$^{+0.02}_{-0.04}$                                      & 95$^{+1500}_{-59}$                                        & 0.27$\pm$0.08                       & 4.3$^{+5.6}_{-1.7}$                                & 9.1$\pm$0.7  \\ \cline{2-8}
                                               & (1.33$^{+0.16}_{-0.14}$, 0.88$^{+0.13}_{-0.11}$)         & $<$1                                                                                                   & 1.9$^{+0.7}_{-0.5}$                                           & 0 (fix)                      & $<$0.1                                                                 & 110$^{+28}_{-20}$ & 186/117 \\ \hline
% \multirow{2}*{$\uparrow$}& \\  \cline{2-8}
%                                               &\\  \hline
 \multirow{2}*{FJ5-B}          & (1.39$^{+0.11}_{-0.10}$, 1.03$^{+0.09}_{-0.08}$)         &  0.15$^{+0.02}_{-0.01}$                                    & 18$\pm$4                                                          & 0.97$^{+0.09}_{-0.11}$        & 2.2$\pm$0.5                                             & 11$\pm$1 \\ \cline{2-8}
                                               & (1.39$^{+0.14}_{-0.12}$, 1.16$^{+0.13}_{-0.11}$)         & $<$1                                                                                                   & 3.2$^{+1.0}_{-0.7}$                                           & 0 (fix)                      & $<$0.3                                          & 120$^{+21}_{-23}$     & 213/177 \\ \hline
 \multirow{2}*{$\uparrow$}& (1.39$^{+0.11}_{-0.10}$, 1.03$^{+0.09}_{-0.08}$)         & 0.15$^{+0.02}_{-0.01}$                                     & 18$\pm$4                                                          & 0.97$^{+0.09}_{-0.11}$         & 2.2$\pm$0.5                                            & 11$\pm$1                    \\ \cline{2-8}
                                               & (1.39$^{+0.14}_{-0.12}$, 1.16$^{+0.13}_{-0.12}$)         & $<$1                                                                                                   & 3.5$^{+1.1}_{-0.7}$                                           & 0.11 (fix)                & $<0.7$                                           & 140$\pm$28                & 211/177 \\ \hline
 \multirow{2}*{$\uparrow$}& (1.25$^{+0.11}_{-0.10}$, 0.94$^{+0.09}_{-0.08}$)        & 0.15$^{+0.02}_{-0.01}$                                      & 18$\pm$4                                                          & 0.97$^{+0.09}_{-0.11}$         & 2.2$\pm$0.5                                            & 11$\pm$1                    \\ \cline{2-8}
                                               & (1.39$^{+0.14}_{-0.12}$, 1.15$^{+0.13}_{-0.12}$)        & $<$1                                                                                                    & 3.7$^{+1.1}_{-0.7}$                                           & 0.16 (fix)                & $<$0.8                                          & 150$\pm$30                 & 211/177 \\ \hline
 \multirow{2}*{$\uparrow$}& (1.39$^{+0.11}_{-0.10}$, 1.03$^{+0.09}_{-0.08}$)        & 0.15$^{+0.02}_{-0.01}$                                      & 18$\pm$4                                                          & 0.97$^{+0.09}_{-0.11}$         & 2.2$\pm$0.5                                            & 11$\pm$                       \\ \cline{2-8}
                                               & (1.39$^{+0.14}_{-0.12}$, 1.14$^{+0.13}_{-0.11}$)        & $<$1                                                                                                    & 3.8$^{+1.0}_{-0.8}$                                           & 0.17 (fix)                & $<$0.9                                           & 150$^{+30}_{-32}$     & 211/177 \\ \hline
 \hline
 \multicolumn{6}{c}{Model: ${\it apec_\mathrm{SWCX+LHB}}$ + {\it phabs} $\times$ ({\it apec$_\mathrm{MWH}$} + {\it pow-law$_\mathrm{CXB}$} + {\it phabs$_\mathrm{FJ}$} $\times$ {\it pow$_\mathrm{FJ}$})}\\ 
\hline
 \multirow{2}*{Region}    & $f_{{\rm bgd}}$$^\ast$                                                    & $kT_\mathrm{SWCX+LHB}$ [keV] & Norm$^\dagger$$_\mathrm{SWCX+LHB}$ & $kT_\mathrm{MWH}$ [keV] & Norm$^\dagger$$_\mathrm{MWH}$ & $S^\ddagger_\mathrm{CXB}$  \\ \cline{2-7}
                                            & $f_{{\rm FJ}}$$^\ast$                                                      & N$_H$$_\mathrm{:FJ}$ [$\times$10$^{21}$~cm$^{-2}$] & $\Gamma_\mathrm{FJ}$                          & $z$$_\mathrm{FJ}$ & $\chi^2/$d.o.f                                  &                           \\ \hline
 \multirow{2}*{FJ1-B}  & (1.06$^{+0.10}_{-0.09}$, 1.13$^{+0.10}_{-0.09}$)       & 0.09$^{+0.03}_{-0.06}$                     & 57$^{+550}_{-36}$                                            & 0.23$^{+0.08}_{-0.05}$        & 4.3$^{+7.2}_{-2.2}$                               & 7.6$\pm$0.5 \\ \cline{2-7}
                                            & (1.30$^{+0.24}_{-0.20}$, 1.21$^{+0.23}_{-0.19}$)   &$<$6                                                                                            & 2.9$^{+1.7}_{-0.8}$                                    & 0 (fix)                     & 149/187                         \\ \hline
 \multirow{2}*{FJ2-B}  & (1.15$^{+0.09}_{-0.08}$, 1.11$^{+0.09}_{-0.08}$)        & 0.09$^{+0.01}_{-0.02}$                    & 94$^{+210}_{-39}$                                            & 0.65$^{+0.09}_{-0.08}$        & 1.5$\pm$0.3                                           & 7.5$\pm$0.5  \\ \cline{2-7}
                                            & (1.17$^{+0.15}_{-0.13}$, 1.00$^{+0.14}_{-0.12}$)   & $<$1                                                                                           & 2.3$^{+0.4}_{-0.2}$                                    & 0 (fix)                     & 195/183                         \\ \hline
 \multirow{2}*{FJ3-B} & (1.15$^{+0.10}_{-0.09}$, 0.93$^{+0.09}_{-0.08}$)        & 0.12$^{+0.03}_{-0.02}$                     & 29$^{+44}_{-14}$                                               & 0.79$^{+0.14}_{-0.11}$        & 1.2$\pm$0.3                                          & 7.5$\pm$0.5    \\ \cline{2-7}
                                           & (0.84$^{+0.11}_{-0.10}$, 0.88$^{+0.12}_{-0.11}$)    &$<$1                                                                                            & 2.1$\pm$0.3                                                & 0 (fix)                     & 181/151                          \\ \hline
 \multirow{2}*{FJ4-B} & (1.25$^{+0.14}_{-0.12}$, 1.02$^{+0.12}_{-0.10}$)        & 0.09$^{+0.03}_{-0.05}$                     & 75$^{+840}_{-50}$                                             & 0.27$^{+0.09}_{-0.07}$       & 3.6$^{+7.0}_{-1.6}$                              & 8.9$\pm$0.7     \\ \cline{2-7}
                                           & (1.23$^{+0.14}_{-0.12}$, 0.85$^{+0.12}_{-0.11}$)    &$<$1                                                                                            & 2.4$\pm$0.2                                                & 0 (fix)                     & 156/118                          \\ \hline
 \multirow{2}*{$\uparrow$} & (1.47$^{+0.17}_{-0.14}$, 1.10$^{+0.12}_{-0.11}$)&0.09$^{+0.03}_{-0.05}$                    & 75$^{+590}_{-49}$                                             & 0.27$^{+0.06}_{-0.09}$       & 3.6$^{+4.8}_{-1.4}$                              & 8.9$\pm$0.7     \\ \cline{2-7} 
                                           & (1.21$^{+0.14}_{-0.12}$, 0.88$^{+0.12}_{-0.11}$)    & $<$1                                                                                           & 2.4$\pm$0.2                                               & 0.37 (fix)               & 156/118                           \\ \hline
 \multirow{2}*{FJ5-B} & (1.41$^{+0.12}_{-0.11}$, 1.04$\pm$0.09)                       & 0.15$^{+0.02}_{-0.01}$                     & 18$^{+7}_{-4}$                                                    & 0.98$^{+0.09}_{-0.11}$         & 2.2$\pm$0.5                                        & 11$\pm$1          \\ \cline{2-7}
                                           & (1.34$^{+0.14}_{-0.11}$, 1.13$^{+0.13}_{-0.11}$)    & $<$1                                                                                           & 2.0$^{+0.3}_{-0.1}$                                     & 0 (fix)                                       & 211/178  \\ \hline
 \hline
    \end{tabular}}
  \end{center}
\begin{flushleft}
\footnotesize{
$\ast$ Constant factors of XIS1 (left) and XIS3 (right) detectors relative to the XIS0 detector. \\
$\dagger$ Normalization of the $apec$ models divided by a solid angle $\Omega$, 
 assumed in a uniform-sky ARF calculation (20$'$ radius), 
 i.e. $\it{Norm} = (1/\Omega) \int n_e n_H dV / (4\pi(1+z)^2)D^2_A)$ cm$^{-5}$ sr$^{-1}$ 
in unit of 10$^{-14}$, where $D_A$ is the angular diameter distance.\\
 $\ddagger$ Surface brightness of the {\it power-law } model in the unit of $10^{-8}$ erg cm$^{-2}$ s$^{-1}$ sr$^{-1}$ in 2--10 keV.\\
 $\ddagger$ Abundance table tabulated in \citet{solar-abundance-table-anders} was adopted.}
 \end{flushleft}
\end{table*}
\newpage
\section{Discussion}
\subsection{Corresponding Object and Origin of FJ sources}
\label{label:origin}
In this section, we discuss the origins of the excess sources 
by searching for corresponding objects in NED.
$L_{X}$--$kT$ relation is also utilized to support our discussions.
Redshift known galaxies and point sources with/without redshift information are 
taken into account as supposed candidates of the origin.
The spectral analysis as indicated in \S\ref{sec:spectral-analysis} demonstrates that 
FJn-A and FJ4-B prefer thermal and non-thermal origins, respectively,  
while both thermal and non-thermal components are acceptable for the others.
Thus, the preferred component is considered in this discussion.
\paragraph{FJ1 (A $\&$ B):}
For FJ1-A, 
The metal abundance was not constrained and 
thus 0.3 or 1 solar abundance is assumed for the thin thermal emission.
These values correspond to typical values of groups of galaxies \citep[e.g.,][]{2003ApJS..145...39M} 
and X-ray luminous (L$_{{\rm X}}$ $>$ 10$^{41}$ erg s$^{-1}$) early-type galaxies 
\citep{2000PASJ...52..685M}.
The derived temperatures of the thin thermal emissions 
with the abundances of 0.3 and 1 solar are $\sim$0.6 keV and 
consistent with those of typical elliptical galaxies and 
groups of galaxies \citep{1996ApJ...470L..35D,2003ApJS..145...39M}. 
The observed bolometric X-ray luminosities are 
0.6$^{+0.6}_{-0.1}$$\times$10$^{42}$ erg s$^{-1}$.
Thus, the observed $L_{X}$--$kT$ relation also supports the ISM or ICM origin.

For FJ1-B, 
no difference between the thin thermal plasma model and the power-law component was seen 
statistically.
The resultant temperature is in good agreement 
with typical values of elliptical galaxies and groups of galaxies.
However, no corresponding galaxies are present within a radius of 1.5 arcmin 
in all redshift planes around the peak in 0.5--2 keV 
while the nearest radio source (NVSS J133623+434405) is located 
$\sim$30 and 70 arcsec away from the peaks in 0.5--2 keV and 2--5 keV, respectively.
Considering the {\it Suzaku} spatial resolution ($\sim$1.8$'$ in HPD), 
we cannot reject the possibility of the radio source as the origin.
Thus, we concluded that the origin of FJ1-B is possibly the nearest radio source 
or X-ray source(s) without any identifications in previous radio and optical surveys 
in the line of sight.
\paragraph{FJ2 (A $\&$ B):}
For FJ2-A, the abundance of the excess thin thermal plasma is well constrained and 
the resulting abundance is 0.2$\pm$0.1 solar, which indicates the ICM origin.
The bolometric X-ray luminosity is 3.5$^{+0.7}_{-0.6}$$\times$10$^{42}$ erg s$^{-1}$ and 
the observed $L_{X}$--$kT$ relation is also in accordance with those 
obtained from groups of galaxies.
In conclusion, 
the observed X-ray halo possibly originates from the hot diffuse halo 
associated with the group of galaxies, i.e., ShCG 188 where the central optically bright galaxy belongs.
Physical properties of this group are discussed in the next section.

As for FJ2-B, 
the thin thermal plasma and the power-law component are both acceptable statistically.
However, we cannot find any spectroscopically-identified galaxies 
within a radius of 3 arcmin from the peak position in 0.5--2 keV in the line of sight.
Meanwhile, the QSO (SDSS J095733.33+260636.5) lies $\sim$50 arcsec  
away from the peak position in 0.5--2 keV and 2--5 keV and 
the resultant photon index of $\sim$2.3 is consistent with that of QSO 
\citep{2000ApJ...531...52G}.
The observed X-ray emission in FJ2-B likely comes from the QSO or 
X-ray source(s) without any detections in radio and optical in the line of sight.
\paragraph{FJ3 (A $\&$ B):}
%%
%%
%%
%Spectral analysis prefers the thin thermal plasma to the power-law component 
%to explain the observed emission in 0.8--1 keV statistically.
%
Because no constraint is obtained on the abundance,  
both the ISM and the ICM are candidates as the origin of FJ3-A.
The observed luminosity of 0.6$^{+0.6}_{-0.2}$$\times$10$^{42}$ erg s$^{-1}$ 
is in good agreement with the expected value from the $L_{X}$--$kT$ relation 
of early-type galaxies and groups of galaxies.

In the case of FJ3-B, 
considering the fact that no corresponding galaxies exist 
within a radius of 1 arcmin from the peak in 0.5--2 keV 
in the line of sight and 
even the nearest radio sources (FIRST J100539.1+394327 \& FIRST J100520.7+394128) are  
$>$90 arcsec apart from the peaks in 0.5--2 keV and 2--5 keV 
X-ray source(s) in the line of sight is possibly the origin.
\paragraph{FJ4 (A $\&$ B):}
%%
%%
%%
%As with the case of FJ1-A and FJ3-A, 
%the power-law component cannot reproduce the emission feature in 0.7--1 keV and 
%therefore the fit with the thin thermal plasma model significantly improves.
%%
%Also in terms of its physical characteristic, 
%we can rule out the possibility of the power-law component 
%since the derived photon index of $>$2.6 is much higher than the typical value of AGN.
%
As the origin is not distinguishable based on the abundance parameter 
in spectral analysis   
due to the poor statistics, 
ISM and ICM are possible origins of FJ4-A.
We confirmed that the observed $L_{X}$--$kT$ relation 
of 0.3$^{+2.2}_{-0.1}$$\times$10$^{42}$ erg s$^{-1}$ and $\sim$0.7 keV 
also reach the same conclusion.

In contrast to FJ4-A, 
the power-law component with the photon index of $\sim$2.3 
is favorable rather than the thermal plasma 
to explain the soft band below 0.8 keV in its energy spectrum. 
Two QSOs (2MASSi J1103075+291230 \& SDSS J110307.58+291230.0) lie 
very close ($\sim$10 arcsec) to the peaks in both 0.5--2 keV and 2--5 keV and 
are spatially overlapped at almost the same redshift plane of $\sim$0.37.
We conducted spectral analysis using the power-law component with the redshift of 0.37 
as shown in Table \ref{table:FJn-B-result} and the derived photon index and luminosity 
of $\sim$2.4 and $\sim$1$\times$10$^{44}$ erg s$^{-1}$ 
are in good agreement with those of QSO \citep{2000ApJ...531...52G}.
Thus, these observational results lead to a conclusion that 
background QSO(s) is responsible for the excess emission detected in FJ4-B.
\paragraph{FJ5 (A $\&$ B):}
%%
%%
%%
%The X-ray image for FJ5-A shows the truly extended structure and 
%moreover the power-law component is not acceptable statistically and physically 
%in spectral analysis owing to the residual around 0.9 keV and the large photon index of $>$3.1.
%
Resultant temperature, abundance and luminosity of $\sim$1 keV, $<$0.1 and 
3.6$^{+0.5}_{-0.9}$ $\times$10$^{42}$ erg s$^{-1}$ indicate the ICM origin.
Details on FJ5-A are discussed in the next section.

Finally, we discuss the origin of FJ5-B.
No statistical significant difference is found in spectral analysis 
adopting the power-law component or the thin thermal plasma as the excess emission.
The photon index of $\sim$2.0 is slightly higher than that of the typical value of AGN and 
the X-ray source (1RXS J113534.0+210228) is present $\sim$50$"$ and 120$"$ 
away from the peaks in 0.5--2 keV and 2--5 keV, respectively.
Thus, the known X-ray source is not necessarily the origin of FJ5-B 
because the peak positions in both the soft and the hard bands are distant from the X-ray source 
even though the large {\it Suzaku} spatial resolution (1.8$'$ in HPD) is taken into account.
Next, we study the possibility of the thin thermal plasma origin.
Considering the resulting temperature of $\sim$3.2 keV, 
FJ5-B is considered to be a halo associated with a low temperature cluster of galaxies.
Actually, three optically-identified clusters of galaxies, 
i.e., WHL J113535.6+210333, GMBCG J173.89411+21.04810 and NSC J113528+210301, 
are located $\sim$30, 60 and 60 arcsec apart from the peak in 0.5--2 keV 
at the redshifts of $\sim$0.16, 0.17 and 0.11, respectively.
Elongations of 60 arcsec correspond to physical distances of $\sim$100-200 kpc 
at the redshift planes of 0.11, 0.16 and 0.17 
which are much smaller than a Mpc-scale typical virial radius of clusters of galaxies.
We confirmed that another filamentary junction surely exists in the FJ5-B region 
at the redshift plane between 0.16 and 0.17.
A filamentary junction at the redshift of $\sim$0.11 was not found 
presumably due to a small number of spectroscopic galaxies.
Spectral fittings assuming their redshifts to be 0.11, 0.16 and 0.17 are performed.
Resultant temperatures, abundance and luminosities are 
($\sim$3.5 keV, $\sim$3.7 keV, $\sim$3.8 keV), ($<$0.7 solar, $<$0.8 solar, $<$0.9 solar) and 
(7.9$\pm$1.5, 15$\pm$3, 15$\pm$3)$\times$10$^{42}$ erg s$^{-1}$, respectively.
The abundances are not well constrained 
but do not deny the possibility of the hot halo origin 
associated with the background cluster(s) of galaxies.
The observed luminosities are about an order of magnitude smaller than the expected values 
from the $L_{X}$--$kT$ relation for clusters of galaxies.
However, the {\it Suzaku} FOV does not cover the whole clusters of galaxies.
Thus, in conclusion, the origin of FJ5-B is 
likely the optically-identified background cluster(s) of galaxies.

\begin{figure*}[h!]
\begin{center}
    \includegraphics[bb=0 0 1424 1067, width=22cm]{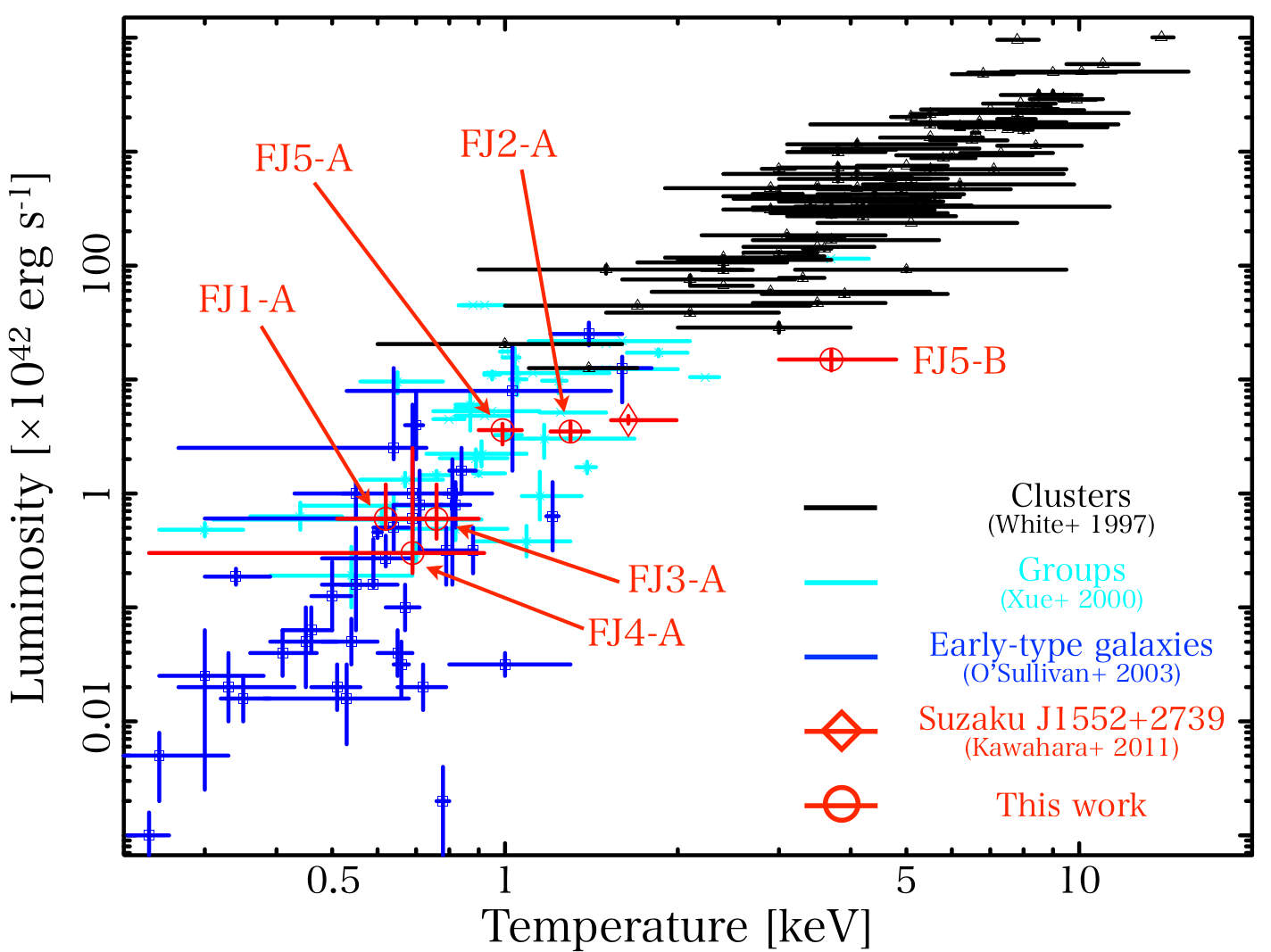}
\end{center}
\vspace{-0.5cm}
  \caption{The observed $L_{X}$--$kT$ relation of FJ sources including the result of Paper I 
  with previous studies for clusters of galaxies \citep{1997MNRAS.292..419W}, 
  groups of galaxies \citep{2000ApJ...538...65X} and early-type galaxies \citep{2003MNRAS.340.1375O}.}
  \label{fig:lx-kt-relation}
\end{figure*}

The observed $L_{X}$--$kT$ relation of FJ sources is shown in Figure \ref{fig:lx-kt-relation} 
with previous works obtained from early-type galaxies, groups of galaxies and clusters of galaxies 
\citep{1997MNRAS.292..419W,2000ApJ...538...65X,2003MNRAS.340.1375O} for comparison and 
the possible origins and corresponding objects for FJ sources discussed above is summarized 
in Table \ref{table:origins}.
Note that luminosities for FJ sources without redshift information were not calculated and 
therefore Figure \ref{fig:lx-kt-relation} does not include them.
\begin{table*}[h!]
  \caption{Summary of possible origins and corresponding objects for FJ sources with physical properties.}
\label{table:origins}
  \begin{center}
  \scriptsize
{\tabcolsep=0.6mm
    \begin{tabular}{l|cccccc}
\hline\hline
 \multicolumn{4}{c}{FJn-A}\\ \hline
Region    & $kT^{\ast}$ [keV]                                                            & L$_\mathrm{X}$ [$\times$10$^{42}$ erg s$^{-1}$] &Possible origin  \\ \hline
FJ1      & (0.62$^{+0.12}_{-0.11}$, 0.61$^{+0.11}_{-0.10}$)      & 0.6$^{+0.6}_{-0.1}$                                                       & ISM or ICM   \\ \hline
FJ2      & 1.3$\pm$0.1                                                                        & 3.5$^{+0.7}_{-0.6}$                                                       & ICM                        \\ \hline
FJ3      & (0.76$^{+0.14}_{-0.15}$, 0.76$^{+0.15}_{-0.16}$)      & 0.6$^{+0.6}_{-0.2}$                                                       & ISM or ICM   \\ \hline
FJ4      & (0.69$^{+0.23}_{-0.45}$, 0.69$^{+0.24}_{-0.39}$)      & 0.3$^{+2.2}_{-0.1}$                                                        & ISM or ICM   \\ \hline
FJ5      & 0.99$^{+0.08}_{-0.09}$                                                     & 3.6$^{+0.5}_{-0.9}$                                                       & ICM                        \\ \hline
% \multicolumn{6}{c}{FJn-B}\\ 
%\hline
% \hline
    \end{tabular}
        \vspace{0.5cm}
}
%  \end{center}
    \begin{tabular}{l|cccccc}
\hline\hline
 \multicolumn{6}{c}{FJn-B}\\  \hline
Region                                               & $kT$ [keV]                                                                                  & L$_\mathrm{X}$ [$\times$10$^{42}$ erg s$^{-1}$]           & $\Gamma$                                             & N$_H$ [$\times$10$^{21}$~cm$^{-2}$]   & Possible origin   \\ \hline
\multirow{2}*{FJ1}                            &                                                                                                      &                                                                                                     & \multirow{2}*{2.9$^{+1.7}_{-0.8}$}     &  \multirow{2}*{$<$6}                                     & radio source (NVSS J133623+434405)  \\
                                                            &                                                                                                      &                                                                                                     &                                                                   &                                                                          & or X-ray source(s) in the line of sight \\ \hline  
\multirow{2}*{FJ2}                           &                                                                                                      &                                                                                                     &\multirow{2}*{2.3$^{+0.4}_{-0.2}$}     & \multirow{2}*{$<$1}                                        & QSO (SDSS J095733.33+260636.5) \\ 
                                                            &                                                                                                      &                                                                                                     &                                                                   &                                                                          & or X-ray source(s) in the line of sight \\ \hline
FJ3-B                                                 &                                                                                                      &                                                                                                     &2.1$\pm$0.3                                            & $<$1                                                                & X-ray source(s) in the line of sight \\ \hline
\multirow{2}*{FJ4}                           &                                                                                                       & \multirow{2}*{120 ($z$ = 0.37)}                                            & \multirow{2}*{2.4$\pm$0.2}                 & \multirow{2}*{$<$1}                                      & QSOs (2MASSi J1103075+291230  \\ 
                                                           &                                                                                                       &                                                                                                     &                                                                   &                                                                          & and/or SDSS J110307.58+291230.0) \\ \hline
\multirow{3}*{FJ5}                          & 3.5$^{+1.1}_{-0.7}$                                                                   &7.9$\pm$1.5               ($z$ = 0.11)                                           &                                                                   &                                                                          & clusters of galaxies (NSC J113528+210301  \\ 
                                                           & 3.7$^{+1.1}_{-0.7}$                                                                   & 15$\pm$3                  ($z$ = 0.16)                                           &                                                                   &                                                                          & and/or WHL J113535.6+210333   \\
                                                           & 3.8$^{+1.0}_{-0.8}$                                                                   & 15$\pm$3                  ($z$ = 0.17)                                           &                                                                   &                                                                          & and/or GMBCG J173.89411+21.04810)   \\ \hline
    \end{tabular}
  \end{center}
\begin{flushleft}
\footnotesize{
$\ast$ Values in parentheses correspond to temperatures obtained 
by assuming the abundances to be 0.3 (left) and 1 (right) solar, respectively. \\
}
 \end{flushleft}
\end{table*}
%\newpage
%%
%%
%%
%%
%%

%%
%%
\subsection{Physical Condition in Galaxy Filamentary Junction}
Here, we discuss environmental specificity, i.e., galaxy filamentary junctions, 
in terms of structure formation which we can straightforwardly expect its speciality 
compared to other fields.

Among five fields, 
two active interactions are confirmed in FJ2-A and FJ5-A.
In FJ2-A, 
the enlarged halos of the two brightest galaxies in optical R band  
including the central S0 galaxy in the group ShCG 188 are observed and 
one of the halo is directed towards another as shown in \citet{2005RMxAA..41....3T}, 
which suggests that the group is experiencing an ongoing major merger event 
because a single massive galaxy can be the responsible perturber in groups.
It is also suggested that 
one more member galaxy is involved in the interaction 
because the bulge of the galaxy is shifted 
towards the interacting galaxies.
This elongated feature is confirmed also in its X-ray image 
as shown in Figure \ref{fig:FJ-xis1-05-2keV-images} and 
its direction is consistent with that observed in optical.
Therefore, in combination with X-ray spectroscopic analysis, 
the observed X-ray hot gas may originate from ICM associated with the merging event 
as observed in optical 
although further detailed analysis with higher spatial resolution is needed.

As for FJ5-A, 
it holds the multiple peaks and irregular morphology in its surface brightness and 
no corresponding optically bright elliptical galaxies are present in the peaks.
These observational features are seen 
in groups of galaxies and clusters of galaxies experiencing ongoing mergers 
\citep[e.g.,][]{2011ApJ...727L..38K,2004ApJ...613..831D}.
Thus, we conclude that the observed X-ray emission 
derives from the ICM associated with the ongoing merger event.
Although as is the case with Paper I, 
we searched for a hot spot based on a hardness ratio map, 
no hot spot was found.
X-ray observatories with higher spatial resolution have the potential 
to detect such a local structure.

As discussed above and in \S\ref{label:origin}, 
%ten X-ray sources associated with six diffuse halos (ISM or ICM) and 
%four compact objects.
%%
three of six diffuse sources are possibly group- and cluster-scale X-ray halo origins and 
four such samples in the total six pointing observations are discovered 
from the junctions of galaxy filaments 
(Paper I and this work).
Their fluxes in 0.5--2 keV are (2--4)$\times$10$^{-13}$ erg s$^{-1}$ cm$^{-2}$.
According to an integral log N--log S relation for X-ray selected groups of galaxies and 
clusters of galaxies \citep[e.g.,][]{2002MNRAS.334..219J}, 
$\sim$0.04 targets in the {\it Suzaku} FOV are expected 
assuming the detection limit of 2$\times$10$^{-13}$ erg s$^{-1}$ cm$^{-2}$ in 0.5--2 keV.
Thus, it is indicated that our method can extract an X-ray diffuse halo 
associated with a large scale structure efficiently.
Similarly, because resultant fluxes for four compact objects are 
(0.2--6)$\times$10$^{-13}$ erg s$^{-1}$ cm$^{-2}$ in 2--10 keV, 
the expected number of X-ray point sources within the {\it Suzaku} FOV 
is $\sim$0.9 when the typical detection limit is assumed to be 10$^{-13}$ erg s$^{-1}$ cm$^{-2}$ in 2--10 keV 
\citep{2002PASJ...54..327K}.
Our samples includes two ICMs experiencing ongoing mergers and 
thus we discovered three ongoing merging groups of galaxies in total by adding the result of Paper I  
in spite of rare events observationally.
X-ray hot gas associated with an optically-bright elliptical galaxy will be detected 
with a deep X-ray observation 
while an X-ray halo involved in a large scale structure such as ICM  
is not necessarily observed from surroundings of an elliptical galaxy.
In summary, our results indicate that 
a galaxy filamentary junction is more than likely to possess X-ray emitting hot gas 
associated with the large scale structure and have high chances to occur 
for a merger phenomenon.

X-ray selected survey provides us with new insights in a sequence of a merging phenomena 
such as fossil groups which is considered to be the end result evolutionally.
Hence, 
we hope that our results on the basis of this optically-selected survey play a complementary role 
in the structure evolution towards clusters of galaxies from groups 
in conjunction with future X-ray selected surveys, 
such as eROSITA on board Spectrum-Roentgen-Gamma observatory \citep{2011MSAIS..17..159C} and 
$DIOS$ \citep{2010SPIE.7732E..54O}.
\section{Conclusions}
Five X-ray pointing observations were conducted with $Suzaku$. 
We selected five regions (FJ1, FJ2, FJ3, FJ4 and FJ5) located on junctions of filamentary structures 
in the galaxy distribution based on SDSS identified by the filament extractor, DisPerSE.
Significant X-ray signals were successfully detected in both images and energy spectra and 
we performed spectroscopic analysis for two bright sources in each field.
Spectral analysis demonstrated that their emissions from six sources originate from diffuse halos 
such as ISM and ICM while the others are compact object origins.
Among six diffuse emissions, 
two of them are group-scale diffuse X-ray halo origins and 
one is serendipitously-detected background cluster(s) of galaxies.
ISM and ICM are both acceptable for other three sources 
due to the poor statistics.
The observed two ICMs are associated with ongoing group-scale merging events.
Thus, we conclude that 
galaxy filamentary junctions have tendency to hold X-ray emitting hot gas 
involved in the large scale structure and merger events are more likely to happen 
in such active fields.
\acknowledgments
This work was supported by Grant-in-Aid for JSPS Fellows 12J07045 and 12J10673 and 
JSPS KAKENHI Grant Number 25800106.

\end{document}